\documentclass[english,pra,lengthcheck,final,superscriptaddress,longbibliography]{revtex4-2}
\usepackage[T1]{fontenc}
\usepackage[latin9]{inputenc}
\usepackage{xcolor}
\usepackage{babel}
\usepackage{amsmath}
\usepackage{amssymb}
\usepackage{graphicx}
\PassOptionsToPackage{normalem}{ulem}
\usepackage{ulem}
\usepackage[unicode=true,pdfusetitle,
 bookmarks=true,bookmarksnumbered=false,bookmarksopen=false,
 breaklinks=false,pdfborder={0 0 1},backref=false,colorlinks=true]
 {hyperref}

\makeatletter

\providecolor{lyxadded}{rgb}{0,0,1}
\providecolor{lyxdeleted}{rgb}{1,0,0}

\DeclareRobustCommand{\lyxdeleted}[3]{{\texorpdfstring{\color{lyxdeleted}\lyxsout{#3}}{}}}
\DeclareRobustCommand{\lyxsout}[1]{\ifx\\#1\else\sout{#1}\fi}

\makeatother

\begin{document}
\title{Dynamics of quantum information scrambling under decoherence effects
measured via active spins clusters}
\author{Federico D. Domínguez}
\email{federico.dominguez@cab.cnea.gov.ar}

\affiliation{Centro Atómico Bariloche, CONICET, CNEA, S. C. de Bariloche, Argentina.}
\author{Gonzalo A. Álvarez}
\email{gonzalo.alvarez@cab.cnea.gov.ar}

\affiliation{Centro Atómico Bariloche, CONICET, CNEA, S. C. de Bariloche, Argentina.}
\affiliation{Instituto Balseiro, CNEA, Universidad Nacional de Cuyo, S. C. de Bariloche,
Argentina.}
\affiliation{Instituto de Nanociencia y Nanotecnologia, CNEA, CONICET, S. C. de
Bariloche, 8400, Argentina}
\begin{abstract}
Developing quantum technologies requires the control and understanding
of the non-equilibrium dynamics of quantum information in many-body
systems. Local information propagates in the system by creating complex
correlations known as information scrambling, as this process prevents
extracting the information from local measurements. In this work,
we develop a model adapted from solid-state NMR methods, to quantify
the information scrambling. The scrambling is measured via time-reversal
Loschmidt echoes (LE) and Multiple Quantum Coherences experiments
that intrinsically contain imperfections. Considering these imperfections,
we derive expressions for out-of-time-order correlators (OTOCs) to
quantify the observable information scrambling based on measuring
the number of active spins where the information was spread. Based
on the OTOC expressions, decoherence effects arise naturally by the
effects of the nonreverted terms in the LE experiment. Decoherence
induces localization of the measurable degree of information scrambling.
These effects define a localization cluster size for the observable
number of active spins that determines a dynamical equilibrium. We
contrast the model's predictions with quantum simulations performed
with solid-state NMR experiments, that measure the information scrambling
with time-reversal echoes with controlled imperfections. An excellent
quantitative agreement is found with the dynamics of quantum information
scrambling and its localization effects determined from the experimental
data. The presented model and derived OTOCs set tools for quantifying
the quantum information dynamics of large quantum systems (more than
$10^{4}$ spins) consistent with experimental implementations that
intrinsically contain imperfections.
\end{abstract}
\maketitle

\section{introduction}

In a quantum many-body system, local information can propagate into
many degrees of freedom, creating complex correlations as entanglement
that prevents extracting the information from local measurements.
This propagation process of information is known as \emph{scrambling
}\citep{Sekino2008,Lashkari2013}. Characterizing and understanding
the scrambling dynamics is an outstanding problem that connects different
fields of physics such as quantum statistical mechanics, cosmology
and quantum information processing \citep{Martinez2016,Friis2018,Swingle2018,Lewis-Swan2019}.
The complexity of information scrambling limits our ability to study
many-body quantum systems and employ them in technological developments
\citep{Eisert2015,abanin_colloquium_2019}. As many-body systems generate
high-order quantum correlations that spread over the system's degrees
of freedom \citep{Alvarez2015,Schweigler2017,Friis2018,Lukin2019a,Landsman2019,Brydges2019a},
a high degree of scrambling is produced that make these large quantum
states more sensitive to perturbations \citep{Krojanski2004a,alvarez2010nmr,sanchez2014,Alvarez2015,niknam_sensitivity_2020,Dominguez}.

An accepted measure of scrambling is the tripartite information, which
is near maximally negative for quantum channels that scramble the
information \citep{Hosur2016}. This measure proves that information
scrambling is manifested through the decay of out-of-time order correlators
(OTOCs). OTOCs are special correlators 
\begin{align}
F_{AB}(t) & =\frac{1}{|A|\;|B|}\langle A^{\dagger}(t)B^{\dagger}A(t)B\rangle\label{eq:OTOC}
\end{align}
that quantify the degree of noncommutativity between two local,
initially commutating operators $A$ and $B$, where $A(t)=e^{-i\mathcal{H}t}Ae^{i\mathcal{H}t}$
and $\mathcal{H}$ is the Hamiltonian of the interacting system \citep{Garttner2017,Li2017,Swingle2018}.
The decay of the correlator $F_{AB}$ captures the essence of the
quantum butterfly effect: a time-evolved local operator fails to commute
with almost all the other local operators \citep{Roberts2015}. The
dynamics of $F_{AB}$ is a probe for quantum chaos \citep{larkin1969quasiclassical,Shenker2014,maldacena2016bound,garcia2018_chaos}.

The OTOC functions are more accessible experimentally than the tripartite
information, therefore these correlators are employed for observing
information scrambling in a wide variety of quantum systems using
time-reversal of quantum evolutions with Loschmidt echo (LE) experiments
\citep{alvarez2010nmr,Alvarez2015,Garttner2017,Li2017,Wei2018,Wei2019,Dominguez,niknam_sensitivity_2020,Sanchez2020,Joshi2020,Nie2020,Mi2021}.
A LE experiment is performed by first evolving forward in time with
a Hamiltonian $\mathcal{H}$ and then evolving backward in time with
the Hamiltonian $-\mathcal{H}$. A ubiquitous problem of LE experiments
is that the forward and backward Hamiltonians are not identical, therefore
leading to the existence of nonreverted interactions, such as imperfections
or environmental interactions, that introduce nonunitary decays \citep{peres1984stability,Jalabert2001,Jacquod2009a,Gorin2006,Goussev:2012,Suter2016}.
These decoherence effects add an additional decay to the OTOC functions
that should not be confused with the information scrambling generated
by the system interactions $\mathcal{H}$. Therefore, it is necessary
to model the information scrambling in open quantum systems to correctly
interpret scrambling measurements obtained from LE experiments \citep{alvarez2010nmr,Alvarez2015,Swingle2018a,Syzranov2018_out,GonzalezAlonso2019,Tuziemski2019,Landsman2019,Dominguez,zanardi2021information}.

In this work, we study the dynamics of quantum information scrambling
via LE experiments that intrinsically contain imperfections. We show
that the scrambling degree can be interpreted as the average number
of active spins $K$ \citep{Sorensen1983,Griesinger1986} \textendash equivalent
to a mean Hamming distance\textendash{} where the information was
spread by the evolution that survived the perturbation effects. We
derive a more general OTOC than the one described in Eq. (\ref{eq:OTOC}),
that allows modeling decoherence effects induced by the nonreverted
terms that affect the outcome of information scrambling measurements.
The decoherence effects arise naturally as a leakage of the unitary
dynamics of the ideal echo experiment with a rate that depends on
the number of active spins. This allows one to model the dynamics of the
effective cluster size of correlated spins $K$ that provides a measurable
degree of scrambling of information into the system under decoherence
effects. We model the OTOCs and the effects of the nonreversed terms
in the LE experiments within the framework of the multiple quantum
coherences (MQC) technique \citep{alvarez2010nmr,Alvarez2015,Garttner2017,Wei2018,Wei2019,Sanchez2020,Dominguez,niknam_sensitivity_2020}.
The MQC framework was originally developed in solid-state NMR for addressing the dynamics of large interacting
quantum systems\citep{Baum1985c,Munowitz1987}. Under specific dynamical
models, the MQC experiments provide the number of correlated spins
$K$ \citep{Lacelle1991,Lacelle1993a,Hughes2004}, which is an alternative
way of interpreting the information scrambling within the spin system
\citep{Wei2018,niknam_sensitivity_2020,Dominguez}. Since solving
exactly the dynamics of general and large spin-systems is not possible
with present technologies, a variety of models were developed to describe
the dynamics of $K$ during the quantum evolution with MQC experiments
\citep{Murdoch1984,Baum1985c,Munowitz1987,Levy1992c,Zobov2006,Zobov2008,Mogami2013a,Zobov2011,Lundin2016}.
We revisit one of these models developed by Levy and Gleason \citep{Levy1992c}
that successfully explained the growth of the cluster size of correlated
spins in different crystalline samples \citep{Levy1992c,Cho1993,Cho1996,Cho2005}.
We adapt it to quantify the information scrambling of a local excitation
using LE and MQC experiments, including also time-reversal imperfections.
The complex dynamics of a multispin system is then simplified, by
describing the system quantum state as a mixture of average operators
of $L$ active -correlated- spins that have a decoherence decay rate
depending on $L$.

We contrast the model's predictions with quantum simulations performed
with solid-state NMR experiments, that measure the information scrambling
with time-reversal echoes with controlled imperfections. The presented
experiments are based on previous methods and results, where a spin
system is quenched by a control Hamiltonian that induces the scrambling
of a magnetization that plays the role of a localized initial information
\citep{alvarez2010nmr,Alvarez2011,Alvarez2015,Dominguez}. We accurately
predict the growth of the cluster size of correlated spins of the
experimental data. In particular, our model predicts a localization
cluster size of the scrambling dynamics that behaves as a dynamical
equilibrium state showing excellent quantitative agreement with the
experimental data and previous findings \citep{alvarez2010nmr,Alvarez2011}.
The model also manifests a transition from a localized to a delocalized
dynamics as a function of the perturbation strength in finite systems
that might also be related with previous experiments \citep{Alvarez2015}.
The quantitative agreement between the model predictions and the experimental
data is consistent with a scaling law transition of the decoherence
effects of the clusters of active spins where the information is scrambled
\citep{Dominguez}. \lyxdeleted{Propietario}{Thu Nov 18 14:27:13 2021}{ }Therefore
the presented model combined with the derivation of OTOCs can be useful
tools to predict the information scrambling dynamics of large quantum
systems, and address the effect of imperfections on the control Hamiltonian
that drives the quantum evolutions.

Our article is organized as follows. In Sec. \ref{sec:Information-scrambling-in},
we introduce the considered spin system, the OTOCs used and how they
can be determined with LE and MQC experiments. In Sec. \ref{sec:Number-of-active},
we provide a measure for information scrambling based on a cluster
size of active \textendash correlated\textendash{} spins. We also
introduce the effects of imperfect echo experiments to quantify the
scrambling dynamics. In Sec. \ref{sec:Revisited-Levy-Gleason-model},
we first introduce the original Levy and Gleason model, and then we
adapt it to describe quantum information scrambling dynamics. Based
on the derived OTOCs, we introduce the decoherence effects induced
by imperfection on the time-reversal procedure. In Sec. \ref{sec:Cluster-size-evolution-with},
we analyze our model to show how scrambling is modified in the presence
of decoherence effects. In Sec. \ref{sec:Experimental-test-of}, we
contrast our model with experimental results, and show the consistency
on the predictions of the observable information scrambling bounds
measured with NMR quantum simulations. Finally, in Sec. \ref{sec:Conclusion}
we give the conclusions.

\section{Quantum information scrambling in spin systems\label{sec:Information-scrambling-in}}

\subsection{The system out of equilibrium \label{subsec:The-system-out}}

We consider a system of $N$ interacting $1/2$ spins in the presence
of a strong magnetic field along the $z$ direction. We assume the
Larmor frequency $\omega_{z}\gg d_{ij}$, with $d_{ij}$ the dipole-dipole
interaction strength between the $i$ and $j$ spins. In a frame of
reference rotating at the Larmor frequency, the Hamiltonian of the
system is given by the truncated dipolar interaction

\begin{align}
\mathcal{H} & =\mathcal{H}_{dd}\nonumber \\
 & =\sum_{i<j}d_{ij}\left[2I_{z}^{i}I_{z}^{j}-(I_{x}^{i}I_{x}^{j}+I_{y}^{i}I_{y}^{j})\right],\label{eq:hdd}
\end{align}
where we have neglected the nonsecular terms of the dipolar Hamiltonian
\citep{slichter2013principles}. Therefore the Hamiltonian $\mathcal{H}_{dd}$
conserves the total magnetization of the system as $[\mathcal{H}_{dd},I_{z}]=0$,
with $I_{z}=\sum_{j}I_{z}^{j}$ the total spin operator on the $z$
direction. The operators $I_{v}^{i}$ are the angular momentum of
the $i$-th spin in the $v$ direction. We assume the initial-state
of the system at a thermal equilibrium with the Zeeman interaction.
Its density matrix is $\rho(0)=e^{-\omega_{z}\hbar I_{z}\beta}/\mathrm{Tr}\left(e^{-\omega_{z}\hbar I_{z}\beta}\right)$,
where we have neglected the dipolar interaction due to $\omega_{z}\gg d_{ij}$.
In the high-temperature limit $\beta^{-1}=k_{B}T\gg\hbar\omega_{z}$,
the thermal state is approximated to \citep{slichter2013principles}
\begin{equation}
\rho(0)\sim(\mathbb{I}-\beta\omega_{z}\hbar I_{z})/\mathrm{Tr}\left(\mathbb{I}\right).
\end{equation}
 The identity operator $\mathbb{I}$ does not evolve over time and
does not contribute to the expectation value of an observable magnetization,
since $\mathrm{Tr}(I_{v}\mathbb{I})=0$ for any direction $v$. This
initial state thus represents an ensemble of local operators $I_{z}^{j}$,
that plays the role of an initial local information that will be scrambled
into the system degrees of freedom. To induce the spreading of the
quantum information from the initial state operator $\rho(0)$, we
drive the system away from equilibrium by quenching the Hamiltonian
\citep{alvarez2010nmr,Alvarez2015,Dominguez}. In NMR, this is typically
done by applying rotations to the spins with electromagnetic-field
pulses that engineer an average Hamiltonian that does not commute
with the initial state \citep{Haeberlen1968}. In particular, we consider
the effective interaction given by a double-quantum Hamiltonian

\begin{equation}
\mathcal{H}_{0}=\frac{1}{2}\sum_{i,j}d_{ij}(I_{+}^{i}I_{+}^{j}+I_{-}^{i}I_{-}^{j}),\label{eq:double quantum-1}
\end{equation}
which can be engineered from the dipolar interaction of Eq. (\ref{eq:hdd})
via electromagnetic pulse sequences \citep{Baum1985c}. The Hamiltonian
$\mathcal{H}_{0}$ allows one to probe the growth of the number of correlated
spins as a function of time \citep{Baum1985c,alvarez2010nmr,Alvarez2015}
as a measure of the degree of scrambling of information into the system
\citep{Garttner2017,Garttner2018,Wei2018,Wei2019,Dominguez,Sanchez2020,niknam_sensitivity_2020}.
Here, $I_{\pm}^{i}$ are the ladder operators of the $i$-th spin. 

In the Zeeman basis, the quantum states $|\vec{m}\rangle=|m_{1},...,m_{N}\rangle$
are characterized by the magnetization numbers $m_{i}=\pm\tfrac{1}{2}$
associated with the local spin operators $I_{z}^{i}$. The double-quantum
Hamiltonian flips simultaneously two spins with the same orientation
$m_{i}$, therefore inducing transitions from a state $|\vec{m}\rangle$
to a state $|\vec{n}\rangle$ that change the coherence order $M=\sum_{i}m_{i}-n_{i}$
by $\Delta M=\pm2$. Here $h\left(\vec{m},\vec{n}\right)=M$ is the
Hamming distance between the states $|\vec{m}\rangle$ and $|\vec{n}\rangle$.
Since the initial state $\rho(0)$ is diagonal, it only has nonvanishing
elements for $M=0$ and the double-quantum Hamiltonian only creates
even coherence orders. 

We consider as our observable the magnetization operator $I_{z}$.
The observable signal is then $\mathrm{Tr}\left[I_{z}\rho(t)\right]\ensuremath{\propto\mathrm{Tr}}\left[I_{z}I_{z}(t)\right]$,
which is proportional to the time evolution of the $I_{z}$ operator
as the identity term of the initial state gives null trace. Therefore,
we consider the evolution of the $I_{z}$ operator. At the evolution
time $t$, the scrambled state $I_{z}^{0}(t)=U_{0}(t)I_{z}U_{0}^{\dagger}(t)$
can be expanded in coherence orders as
\begin{align}
I_{z}^{0}(t) & =\sum_{M}\sum_{h\left(\vec{m},\vec{n}\right)=M}\langle\vec{m}|I_{z}^{0}(t)|\vec{n}\rangle\;|\vec{m}\rangle\langle\vec{n}|\nonumber \\
 & =\sum_{M}I_{z,M}^{0}(t),\label{eq:order_decomp}
\end{align}

where the operator $I_{z,M}^{0}=\sum_{h\left(\vec{m},\vec{n}\right)=M}\langle\vec{m}|I_{z}^{0}(t)|\vec{n}\rangle\;|\vec{m}\rangle\langle\vec{n}|$
contains all the elements of the density operator involving coherences
of order $M$. The superindex zero indicates evolution under the double-quantum
Hamiltonian $\mathcal{H}_{0}$, i.e., $U_{0}(t)=e^{-i\mathcal{H}_{0}t}$.
The amplitude of each coherence order is $f_{M}^{0}=\mathrm{Tr}\left(I_{z}^{2}\right)^{-1}\mathrm{Tr}\left[I_{z,M}^{0}(t)I_{z,M}^{0\dagger}(t)\right]$,
that defines the MQC spectrum \citep{Baum1985c}.

\subsection{Measuring information scrambling with Multiple Quantum Coherences\label{subsec:Measuring-information-scrambling}}

The many-body Hamiltonian $\mathcal{H}_{0}$ scrambles the initial
local state into the degrees of freedom of the system. The dynamics
of the information spreading is contained in the evolved state $I_{z}^{0}(t)=U_{0}(t)I_{z}U_{0}^{\dagger}(t)$.
Usually, the degree of information scrambling is quantified between
distant local operators $A$ and $B$ in Eq. (\ref{eq:OTOC}) \citep{Hosur2016}.
In our case, we consider the observation of scrambling based on monitoring
the spreading of an ensemble of local operators $I_{z}$ by its own
time evolution $I_{z}^{0}(t)$ via the commutator $[I_{z}^{0}(t),I_{z}]$,
as it is accessible by time-reversal echoes common in NMR experiments
\citep{Garttner2018,Wei2018,Wei2019,Dominguez,Sanchez2020}. Based
on Eq. (\ref{eq:OTOC}), the OTO commutator 
\begin{multline}
\frac{1}{\mathrm{Tr}\left(I_{z}^{2}\right)}\left\langle \left[I_{z}^{0}(t),I_{z}\right]\left[I_{z}^{0}(t),I_{z}\right]^{\dagger}\right\rangle =\\
=\frac{2}{\mathrm{Tr}\left(I_{z}^{2}\right)}\left\langle I_{z}I_{z}^{0}(t)I_{z}^{0\dagger}(t)I_{z}^{\dagger}\right\rangle -2\mathrm{Re}\left(F_{I_{z},I_{z}}\right)
\end{multline}
is related to the OTOC function $F_{I_{z},I_{z}}=\mathrm{Tr}\left(I_{z}^{2}\right)^{-1}\left\langle I_{z}^{0\dagger}(t)I_{z}^{\dagger}I_{z}^{0}(t)I_{z}\right\rangle $,
where $\left\langle \mathcal{O}\right\rangle =\mathrm{Tr}\left(\rho_{\beta}\mathcal{O}\right)$
is the expectation value of an operator $\mathcal{O}$ considering
the system on the density-matrix state $\rho_{\beta}$. If the system
state is in the infinite temperature limit $\rho_{\beta}=\mathbb{I}$,
then $\left\langle \mathcal{O}\right\rangle _{\beta=0}=\mathrm{Tr}\left(\mathcal{O}\right)$.
Therefore 
\begin{multline}
\lim_{\beta\rightarrow0}\left[\frac{2}{\mathrm{Tr}\left(I_{z}^{2}\right)}\left\langle I_{z}I_{z}^{0}(t)I_{z}^{0\dagger}(t)I_{z}^{\dagger}\right\rangle -2\mathrm{Re}\left(F_{I_{z},I_{z}}\right)\right]=\\
\frac{1}{\mathrm{Tr}\left(I_{z}^{2}\right)}\left\langle \left[I_{z}^{0}(t),I_{z}\right]\left[I_{z}^{0}(t),I_{z}\right]^{\dagger}\right\rangle _{\beta=0}=\\
\frac{1}{\mathrm{Tr}\left(I_{z}^{2}\right)}\mathrm{Tr}\left\{ \left[I_{z}^{0}(t),I_{z}\right]\left[I_{z}^{0}(t),I_{z}\right]^{\dagger}\right\} \label{eq:OTOC_IZ_t_infinita-1-1}
\end{multline}
quantifies the scrambling of local information on $I_{z}$ due to
the evolution driven by $\mathcal{H}_{0}$ in the system at infinite
temperature \citep{Garttner2018,Dominguez,Sanchez2020}.

\begin{figure}
\includegraphics[width=1\columnwidth]{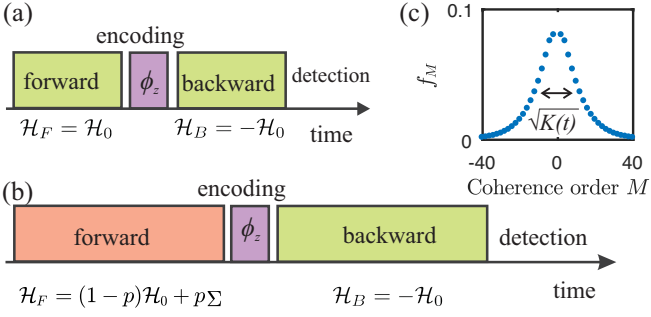}

\caption{\label{fig:(a)-The-MQC-protocol}Protocol for measuring multiple quantum
coherences and monitoring the quantum information scrambling. (a)
The MQC protocol consists of a sequence of quantum evolutions. First,
a forward in time evolution is driven by the double quantum Hamiltonian
$\mathcal{H}_{F}=\mathcal{H}_{0}$ that pumps coherences of different
orders $M$. The rotation $\phi_{z}=e^{iI_{z}\phi}$ encodes the coherence
order with a phase $\phi M$. Then a backward in time evolution is
driven by the Hamiltonian $\mathcal{H}_{B}=-\mathcal{H}_{0}$ to allow
the detection of the MQC spectrum $f_{M}^{0}$ from the signal $f_{0}(t,\phi)=\mathrm{Tr}\left(I_{z}^{2}\right)^{-1}\sum_{M}e^{i\phi M}f_{M}^{0}$.
The signal results from measuring the magnetization $I_{z}$, starting
from the initial state $\rho(0)=(\mathbb{I}-\beta\omega_{z}\hbar I_{z})/\mathrm{Tr}\left(\mathbb{I}\right)$.
(b) Modified MQC protocol to assess the sensitivity of a quantum dynamics
to perturbations in the Hamiltonian. The forward Hamiltonian $\mathcal{H}_{F}=(1-p)\mathcal{H}_{0}+p\Sigma$
differs from the backward Hamiltonian $\mathcal{H}_{B}=-\mathcal{H}_{0}$.
The resulting MQC spectrum $f_{M}(t,p)=\mathrm{Tr}\left(I_{z}^{2}\right)^{-1}\mathrm{Tr}\left[I_{z,M}^{0}(t)I_{z,M}^{\dagger}(t)\right]$
quantifies the inner product of the coherence orders between the ideal
$I_{z}^{0}(t)$ and the perturbed evolution $I_{z}(t)$. (c) Typical
form of the MQC spectrum. Its second moment is determined from the
squared width of the MQC distribution, and provides a measure for
the average number of active spins $K(t)$ where the information was
scrambled, shared by the perturbed and unperturbed dynamics.}
\end{figure}

Determining the MQC spectrum allows one to
measure the OTOC functions of Eq. (\ref{eq:OTOC_IZ_t_infinita-1-1})
\citep{Garttner2017}. A scheme of an experimental implementation
of a concatenation of quantum evolutions used to extract the MQC spectrum
is shown in Fig. \ref{fig:(a)-The-MQC-protocol}(a) \citep{Baum1985c}.
First, the system is at its initial state, then it is quenched by
suddenly turning on the double-quantum Hamiltonian $\mathcal{H}_{0}$.
The system evolves forward in time into the scrambled state $I_{z}^{0}(t)=U_{0}(t)I_{z}U_{0}^{\dagger}(t)$.
A rotation along the $z$ direction $\phi_{z}=e^{-i\phi I_{z}}$ is
applied to the state $I_{z}^{0}(t)$ to label each coherence term
$I_{z,M}^{0}(t)$ of Eq. (\ref{eq:order_decomp}), as they acquire
a phase $M\phi$. The $I_{z}^{0}(t)$ operator thus becomes 
\begin{align}
I_{z}^{0}\left(\phi,t\right) & =\phi_{z}I_{z}^{0}(t)\phi_{z}^{-1}\nonumber \\
 & =\sum_{M}e^{iM\phi}I_{z,M}^{0}(t).
\end{align}
Finally, the dynamics driven by the double-quantum Hamiltonian $\mathcal{H}_{0}$
is reversed in time by changing the sign of $\mathcal{H}_{0}$ by
a pulse sequence design \citep{Baum1985c}. This backward evolution
in time is driven by the evolution operator $U_{0}^{-1}(t)=U_{0}^{\dagger}(t)$.
This leads to a many-body Loschmidt echo \citep{peres1984stability,Jalabert2001,Goussev:2012}
on the resulting magnetization along $I_{z}$, where we obtain the
normalized signal at the end of the evolution:

\begin{align}
f_{0}(\phi,t) & =\frac{1}{\mathrm{Tr}\left(I_{z}^{2}\right)}\mathrm{Tr}\left[I_{z}\ U_{0}^{\dagger}(t)\phi_{z}U_{0}(t)I_{z}U_{0}^{\dagger}(t)\phi_{z}^{\dagger}U_{0}(t)\right]\nonumber \\
 & =\frac{1}{\mathrm{Tr}\left(I_{z}^{2}\right)}\mathrm{Tr}\left[I_{z}^{0}\left(t\right)I_{z}^{0\dagger}\left(\phi,t\right)\right]\nonumber \\
 & =\frac{1}{\mathrm{Tr}\left(I_{z}^{2}\right)}\sum_{MM'}e^{iM\phi}\mathrm{Tr}\left[I_{z,M}^{0}(t)I_{z,M'}^{0\dagger}(t)\right]\delta_{MM'}.\label{eq:global_fidelity_double quantum}
\end{align}
We have used the cyclic property of the trace, and the normalization
factor $\mathrm{Tr}\left(I_{z}^{2}\right)^{-1}$ to ensure the normalization
of the signal $f_{0}(\phi,t=0)=1$. This time-reversal of the quantum
dynamics combined with the rotation $\phi_{z}$ allows quantifying
the contribution of the different coherence orders $f_{M}^{0}=\mathrm{Tr}\left(I_{z}^{2}\right)^{-1}\mathrm{Tr}\left[I_{z,M}^{0}(t)I_{z,M}^{0\dagger}(t)\right]$
to the global spin state by performing a Fourier transform on $\phi$
\citep{Baum1985c,Dominguez}. Therefore the scrambling of information
into multispin states can be quantified.

As the relevant observable term in the initial state $\rho(0)=(\mathbb{I}-\beta\omega_{z}\hbar I_{z})/\mathrm{Tr}\left(\mathbb{I}\right)$
is proportional to the observable magnetization $I_{z}$, the Loschmidt
Echo $f_{0}(\phi,t)$ can be considered as a fidelity between the
rotated state $I_{z}^{0}\left(\phi,t\right)$ and the original one
$I_{z}^{0}\left(t\right)$ quantified by the inner product $\mathrm{Tr}\left[I_{z}^{0}\left(t\right)I_{z}^{0\dagger}\left(\phi,t\right)\right]$.
Moreover, $f_{0}(\phi,t)$ also defines an OTOC function $F_{\phi_{z},I_{z}}$
according to Eq. (\ref{eq:OTOC_IZ_t_infinita-1-1})
\begin{align}
f_{0}(t,\phi) & =\frac{1}{\mathrm{Tr}\left(I_{z}^{2}\right)}\mathrm{Tr}\left[I_{z}^{0}\left(t\right)\phi_{z}I_{z}^{0}\left(t\right)\phi_{z}^{\dagger}\right]\nonumber \\
 & =F_{\phi_{z},I_{z}}\nonumber \\
 & =1-\frac{1}{2\mathrm{Tr}\left(I_{z}^{2}\right)}\left\langle \left|\left[I_{z}^{0}(t),\phi_{z}\right]\right|^{2}\right\rangle _{\beta=0},\label{eq:fidelity and otoc}
\end{align}
where again $\mathrm{Tr}(\mathcal{O})=\left\langle \mathcal{O}\right\rangle _{\beta=0}$
is the expectation value of $\mathcal{O}$ with the system state at
infinite temperature $\rho_{\beta}=\mathbb{I}$ \citep{Garttner2018,Dominguez,Sanchez2020,Yan2020}
and we have used that $\left\langle \phi_{z}I_{z}^{0}(t)I_{z}^{0\dagger}(t)\phi_{z}^{\dagger}\right\rangle _{\beta=0}=\mathrm{Tr}\left(I_{z}^{2}\right)$.
Using a Taylor expansion as a function of $\phi$ in Eq. (\ref{eq:fidelity and otoc}),
the second moment of the MQC spectrum $K_{0}$ is determined by \citep{Khitrin1997,Garttner2018,Wei2019,Dominguez}
\begin{multline}
K_{0}=\sum_{M}M^{2}f_{M}^{0}(t)\\
=\frac{1}{\mathrm{Tr}(I_{z}^{2})}\mathrm{Tr}\left(\left[I_{z},I_{z}^{0}(t)\right]\left[I_{z},I_{z}^{0}(t)\right]^{\dagger}\right),\label{eq:OTOConmuta}
\end{multline}
which is related to an OTOC function following Eq. (\ref{eq:OTOC_IZ_t_infinita-1-1}).
The second moment $K_{0}$ quantifies therefore the information scrambling
into the system driven by the double-quantum Hamiltonian $\mathcal{H}_{0}$
starting from the localized information at $I_{z}$.

\section{Number of active spins as a measure of information scrambling\label{sec:Number-of-active}}

\subsection{Cluster size of correlated spins created by the information scrambling\label{subsec:Cluster-size-of}}

In the NMR community, the second moment $K_{0}$ is typically used
for spin counting of the number of correlated spins by the dynamics
induced by a many-body Hamiltonian \citep{Baum1985c,Lacelle1991,Hughes2004}.
The second moment quantifies the cluster size of spins where an initial
local state was spread into the system \citep{alvarez2010nmr,Alvarez2015}.
To relate the second moment of the MQC spectrum $K_{0}$ to a number
of correlated spins by the information scrambling, it is necessary
to know the propagation model of excitations within the system. Baum
\emph{et. al. }proposed a simple model by assuming that all coherence
orders are equally excited for a given system size \citep{Baum1985c,Munowitz1987}.
The corresponding MQC spectrum has a Gaussian distribution, the
second moment of which determines the system size $K_{0}$ which corresponds
to the instantaneous number of correlated spins. While this simple
assumption works reasonably well in several solid-state systems \citep{Munowitz86_principles,Lacelle1991,Levy1992c},
the MQC distribution is not always Gaussian \citep{Lacelle1993a},
and other models are required to give a quantitative number of correlated
spins \citep{Lacelle1993a,Khitrin1997,Wei2018}. There is no general
method to define the number of correlated spins independently of the
system dynamics.

We here derive a formal definition for the \emph{cluster size of correlated
spins} from the second moment \emph{$K_{0}$ }created by the information
scrambling\emph{, }that does not require assumptions regarding the
dynamics\emph{ }of the system. This definition for the cluster size
is used then in Sec. \ref{sec:Revisited-Levy-Gleason-model} to describe
the information scrambling by OTOC functions determined from an imperfect
echo experiment.

We consider the product basis $\left\{ P_{\vec{u}}=\left(\sqrt{2}\right)^{N}\bigotimes_{i=1}^{N}I_{u_{i}}^{(1)}\right\} $
of the composite Hilbert space for $N$ 1/2-spins, where the index
$i$ labels the spins, $u_{i}\in\{0,x,y,z\}$ and $\{I_{u_{i}}^{(1)}\}$
is the set of spin operators of a single $1/2$-spin, including the
identity operator $I_{0}^{(1)}\equiv\mathbb{I}$. As these single
spin operators satisfy the orthogonal relation $\mathrm{Tr}\left(I_{u_{i}}^{(1)}I_{v_{j}}^{(1)}\right)=\tfrac{1}{2}\delta_{u_{i},v_{j}}$,
the operators set $\left\{ P_{\vec{u}}\right\} $ is an orthonormal
basis of the complete Hilbert space. The evolved quantum state $I_{z}^{0}(t)$
can then be expanded on this basis as
\begin{equation}
I_{z}^{0}(t)=\sum_{\vec{u}}C_{\vec{u}}^{0}(t)P_{\vec{u}},\label{eq:exp_iz0_pa}
\end{equation}
where $C_{\vec{u}}^{0}(t)$ are time-dependent complex coefficients.
Using this expansion, the cluster size of correlated spins is

\begin{equation}
K_{0}(t)=\sum_{\vec{u},\vec{v}}C_{\vec{u}}^{0}(t)C_{\vec{v}}^{0*}(t)\mathcal{L}(\vec{u},\vec{v}),\label{eq:K0_exp_pa}
\end{equation}
where $^{*}$ is the complex conjugate. The function $\mathcal{L}(\vec{u},\vec{v})$
has a functional dependence on $\vec{u}$ and $\vec{v}$ defined by
(see Appendix \ref{sec:Appendix-A} for a demonstration)

\begin{equation}
\mathcal{L}(\vec{u},\vec{v})=\begin{cases}
L(\vec{u}) & \vec{u}=\vec{v},\\
2 & h_{z0}(\vec{u},\vec{v})=0,\,h_{xy}(\vec{u},\vec{v})=2\\
 & \mathrm{and}\,\vec{v}=\Pi(\vec{u}),\\
-2 & h_{z0}(\vec{u},\vec{v})=0,\,h_{xy}(\vec{u},\vec{v})=2\\
 & \mathrm{and}\,\vec{v}\neq\Pi(\vec{u}),\\
0 & \mathrm{Other\,cases.}
\end{cases}\label{eq:funcion_F}
\end{equation}
The diagonal terms when $\vec{u}=\vec{v}$ give the number $L(\vec{u})$
of elements in $\vec{u}$ that are equal to $x$ or $y$. For the
nondiagonal terms $\vec{u}\ne\vec{v}$ , the conditions on the right
hand side have to be satisfied simultaneously. The function $h_{z0}(\vec{u},\vec{v})$
is the Hamming distance between $\vec{u}$ and $\vec{v}$ only considering
the elements that are equal to zero and $z$. In other words, the
condition $h_{z0}(\vec{u},\vec{v})=0$ implies that $u_{i}=v{}_{i}$
for every $u_{i}=0,z$. Similarly, $h_{xy}(\vec{u},\vec{v})$ is the
Hamming distance between $\vec{u}$ and $\vec{v}$ only considering
the elements that are equal to $x$ and $y$. The condition $\vec{v}=\Pi(\vec{u})$
indicates that the vector $\vec{v}$ is a permutation of vector $\vec{u}$.\lyxdeleted{Propietario}{Thu Nov 18 14:27:13 2021}{ }

The cross terms $\vec{u}\ne\vec{v}$ in Eq. (\ref{eq:K0_exp_pa})
have complex numbers with phases that interfere destructively when
they are summed, and therefore their contribution goes to zero as
the cluster size $K_{0}(t)$ increases \citep{Alvarez2008}. Therefore
the cross terms contribution is negligible for large $K_{0}(t)$,
obtaining the cluster size of correlated spins
\begin{equation}
K_{0}(t)\sim\sum_{\vec{u}}\left|C_{\vec{u}}^{0}(t)\right|^{2}L(\vec{u}).\label{eq:K0_hamming}
\end{equation}
Here the magnitude $L(\vec{u})$ quantifies the number of \emph{active
spins} associated to a coherence transfer process $\langle\vec{n}|P_{\vec{u}}|\vec{m}\rangle$
\citep{Sorensen1983,Griesinger1986}. The transition element $\langle\vec{n}|P_{\vec{u}}|\vec{m}\rangle\neq0$
if and only if there are $L(\vec{u})$ spins that flip their state
during the transition $|\vec{m}\rangle\rightarrow|\vec{n}\rangle$
or, equivalently $L=h(\vec{m},\vec{n})$ with $h$ the Hamming distance.
This implies that $K_{0}$ is the average number of active spins in
the state $I_{z}^{0}(t)$ weighted by the coefficients $\left|C_{\vec{u}}^{0}(t)\right|^{2}$
that depends on the quantum dynamics, thus providing an interpretation
for the information scrambling in spin systems. The cluster size $K_{0}$
provides the average number of spins correlated by quantum superpositions
generated by the information scrambling dynamics. The expression of
Eq. (\ref{eq:K0_hamming}) for $K_{0}$ is similar to the ``average
correlation length'' introduced by Wei \emph{et. al.} in Ref. \citep{Wei2018},
but in their work $K_{0}$ is determined by the average number of
spins that does not contain identity operators on the system state.
A formal connection between OTOC functions and ``average correlation
length'' is also provided in Ref. \citep{Wei2018}, but only for
a particular noninteracting system defined by a spin-chain network
topology. In our case, $L(\vec{u})$ quantifies the number of nonidentity
operators and non-$I_{z}$ operators in $P_{\vec{u}}$. Based on Eqs.
(\ref{eq:K0_exp_pa}) and (\ref{eq:K0_hamming}), we provide an average
Hamming distance as a way of quantifying a correlation length which
is directly connected with the quantum information spreading derived
from the OTOC function in Eq. (\ref{eq:OTOConmuta}). Moreover, our
expression to determine $K_{0}$ and the corresponding OTOC is independent
of the spin-network topology, and it does not require assumptions
on the MQC dynamics.

\subsection{Imperfect echo effects on estimating the information scrambling dynamics}

Implementations of echo experiments for measuring the information
scrambling from the MQC spectrum, always contain imperfections. They
thus lead to a time reversion that is not fully performed, altering
the OTOC quantification. Different sources of imperfections might
occur induced by i) nonidealities of the control operations and ii)
the existence of external degrees of freedom considered as an environment
\citep{Suter2016}. Both imperfection terms in the Hamiltonian cannot
be typically reversed. As a paradigmatic model of these imperfections,
we consider a perturbation term $p\Sigma$ in the forward Hamiltonian
as generally considered within the Loschmidt echo formalism \citep{Gorin2006,Jacquod2009a,Goussev:2012}
{[}see Fig. \ref{fig:(a)-The-MQC-protocol}(b){]}. The perturbation
strength $p$ is a dimensionless parameter and $\Sigma$ is a perturbation
Hamiltonian. This perturbation spoils the time reversion process and
thus induces decoherence effects \citep{alvarez2010nmr,Alvarez2015,Dominguez,Suter2016}.
As shown in Fig. \ref{fig:(a)-The-MQC-protocol}(b), the forward evolution
is driven by the evolution operator $U_{P}(t)=e^{-i\mathcal{H}(p)t}$,
where the forward Hamiltonian is $\mathcal{H}_{F}=\mathcal{H}(p)=(1-p)\mathcal{H}_{0}+p\Sigma.$
After the forward evolution, again the global rotation $\phi_{z}=e^{-i\phi I_{z}}$
encodes the coherence orders as $\phi_{z}I_{z}(t)\phi_{z}^{\dagger}=\sum_{M}e^{-iM\phi}I_{z,M}(t)$,
and then the system is evolved backward in time with an ideal evolution
operator $U_{0}^{-1}=e^{it\mathcal{H}_{0}}$. The resulting state
is then projected on $I_{z}$, which is the observable magnetization,
leading to the signal
\begin{multline}
f(\phi,t,p)=\\
\frac{1}{\mathrm{Tr}\left(I_{z}^{2}\right)}\mathrm{Tr}\left[I_{z}\ U_{0}^{\dagger}(t)\phi_{z}U_{p}(t)I_{z}U_{p}^{\dagger}(t)\phi_{z}^{\dagger}U_{0}(t)\right]\\
=\frac{1}{\mathrm{Tr}\left(I_{z}^{2}\right)}\sum_{M}e^{-iM\phi}\mathrm{Tr}\left[I_{z,M}^{0}(t)I_{z,M}^{\dagger}(t)\right].\label{eq:fidelity_perturbada-1}
\end{multline}
The MQC spectrum is now defined by the inner product between the coherence
orders of the forward and backward density matrix terms 
\begin{equation}
f_{M}(t,p)=\mathrm{Tr}\left(I_{z}^{2}\right)^{-1}\mathrm{Tr}\left[I_{z,M}^{0}(t)I_{z,M}^{\dagger}(t)\right].
\end{equation}
 As the time reversion evolution is not ideal, therefore it compares
the two different information scrambling dynamics by comparing the
terms $I_{z,M}^{0}(t)$ and $I_{z,M}(t)$ based on the inner product
in $f_{M}$. As derived in Eq. (\ref{eq:fidelity and otoc}), the
Loschmidt echo of Eq. (\ref{eq:fidelity_perturbada-1}) can be recast
as
\begin{align}
f(\phi,t) & =\frac{1}{\mathrm{Tr}\left(I_{z}^{2}\right)}\mathrm{Tr}\left[I_{z}^{0}(t)I_{z}^{\dagger}(t)\right]-\nonumber \\
 & -\frac{1}{2\mathrm{Tr}\left(I_{z}^{2}\right)}\mathrm{Tr}\left(\left[\phi_{z},I_{z}^{0}(t)\right]\left[\phi_{z},I_{z}(t)\right]^{\dagger}\right)\nonumber \\
 & =f(\phi=0,t)-\nonumber \\
 & -\frac{1}{2\mathrm{Tr}\left(I_{z}^{2}\right)}\mathrm{Tr}\left(\left[\phi_{z},I_{z}^{0}(t)\right]\left[\phi_{z},I_{z}(t)\right]^{\dagger}\right).\label{eq:OTO comm pert-1}
\end{align}
Here, the Loschmidt echo is related to a more general OTO commutator
$\mathrm{Tr}\left(I_{z}^{2}\right)^{-1}\mathrm{Tr}\left(\left[\phi_{z},I_{z}^{0}(t)\right]\left[\phi_{z},I_{z}(t)\right]^{\dagger}\right)$
than the one described in Eq. (\ref{eq:OTOC}). This OTOC quantifies
the overlap between the commutators $\left[\phi_{z},I_{z}^{0}(t)\right]$
and $\left[\phi_{z},I_{z}(t)\right]$ via the inner product \citep{Dominguez}.
The second moment of the MQC spectrum $f_{M}(t,p)$ is now

\begin{align}
\sum_{M}M^{2}f_{M}(t) & =\frac{1}{\mathrm{Tr}(I_{z}^{2})}\mathrm{Tr}\left(\left[I_{z},I_{z}^{0}(t)\right]\left[I_{z},I_{z}(t)\right]^{\dagger}\right).\label{eq:otocom dec-1-1}
\end{align}
 In this imperfect echo experiment, the cluster size is then
\begin{equation}
K(t)=\tfrac{\sum_{M}M^{2}f_{M}(t)}{f(\phi=0,t,p)},\label{eq:K(t) y espectro MQC f_M}
\end{equation}
where the second moment of the MQC spectrum must be normalized to
the fidelity $f(\phi=0,t,p)=\mathrm{Tr}\left(I_{z}^{2}\right)^{-1}\mathrm{Tr}\left[I_{z}^{0}(t)I_{z}^{\dagger}(t)\right]$.
This normalization is required to extract the MQC distribution width,
as the fidelity decays as a function of time, and therefore the overall
amplitude of the MQC distribution. The second moment $K(t)$ is an
\emph{effective cluster size} that represents the common number of
correlated spins between the ideal and the perturbed information scrambling
dynamics, quantified by the commutators $\left[I_{z},I_{z}^{0}(t)\right]$
and $\left[I_{z},I_{z}(t)\right]$ respectively \citep{Dominguez}.

By expanding $I_{z}(t)$ in the basis of the multispin product operators
$\left\{ P_{\vec{u}}\right\} $, 
\begin{equation}
I_{z}(t)=U_{p}(t)I_{z}U_{p}^{\dagger}(t)=\sum_{\vec{u}}C_{\vec{u}}^{p}(t)P_{\vec{u}},
\end{equation}
the perturbed version of Eq. (\ref{eq:K0_exp_pa}) is

\begin{align}
K(t) & =\frac{1}{f(\phi=0,t,p)}\sum_{\vec{u},\vec{v}}C_{\vec{u}}^{0}(t)C_{\vec{v}}^{p*}(t)\mathcal{L}(\vec{u},\vec{v}),\label{eq: K(t) con decoherencia}
\end{align}
where $\mathcal{L}(\vec{u},\vec{v})$ is defined as in Eq. (\ref{eq:funcion_F}).
Again, the cross-terms in Eq. (\ref{eq: K(t) con decoherencia}) are
negligible when $K(t)$ is large \citep{Alvarez2008}, and we obtain
\begin{align}
K(t) & \simeq\frac{1}{f(\phi=0,t,p)}\sum_{\vec{u}}C_{\vec{u}}^{0}(t)C_{\vec{u}}^{p*}(t)L(\vec{u})\nonumber \\
 & =\frac{1}{\sum_{\vec{u}}C_{\vec{u}}^{0}(t)C_{\vec{u}}^{p*}(t)}\sum_{\vec{u}}C_{\vec{u}}^{0}(t)C_{\vec{u}}^{p*}(t)L(\vec{u}).\label{eq:K_hamming}
\end{align}
This expression demonstrates that $K(t)$ is an average Hamming distance
based on the active spins, weighted by the product between the forward
and backward coefficients $C_{\vec{u}}^{0}(t)C_{\vec{u}}^{p*}$ that
determine the respective dynamics evolutions. Therefore the expression
we derived here for $K(t)$ gives the average number of active spins
shared by the forward and backward dynamics. As experimental implementations
of time reversions have always a nonreverted interaction, our expression
for $K(t)$ provides a definition of what is actually measured as
information scrambling with echo experiments.

\section{Model for the dynamics of quantum information scrambling under decoherence\label{sec:Revisited-Levy-Gleason-model}}

\subsection{Revisiting the Levy-Gleason model: Quantum information scrambling
dynamics}

We develop a phenomenological model to describe the time evolution
of the information scrambling observed via the second moment of the
MQC distribution $K(t)$. Levy and Gleason developed a model to describe
the MQC dynamics in solid state systems \citep{Levy1992c}. We here
adapt this model for determining the information scrambling dynamics
in a spin system by using the expressions of Eqs. (\ref{eq:K0_hamming})
and (\ref{eq:K_hamming}).

As the complexity of many-spin system dynamics is an outstanding problem
in physics that only allows one to obtain exact solutions for $I_{z}(t)$
for very special cases \citep{Scruggs1992b,Doronin2001,Munowitz2006,Doronin2009,Doronin2011},
\lyxdeleted{Propietario}{Thu Nov 18 14:27:13 2021}{ }approximations
or phenomenological models need to be implemented \citep{Murdoch1984,Baum1985c,Levy1992c,Munowitz1987,de2004computational,Zobov2006,Zhang_2007,Alvarez2008,Zobov2008}.
The Levy-Gleason model describes the growth of the cluster size of
correlated spins $K_{0}(t)$ as a function of time by introducing
average operators that contains all the elements of the product operator
basis with the same number of nonidentity operators. Instead, based
on the result of Eq. (\ref{eq:K0_hamming}), which shows that the
active spins are the relevant ones for quantifying the information
scrambling dynamics, we adapt the Levy-Gleason model by defining an
average operator $P_{L}$ of the operators $P_{\vec{u}}$ that contain
$L$ active spins, defined by the number of nonidentity and non-$I_{z}$
operators, as described in Sec. \ref{subsec:Cluster-size-of}. The
time evolution of the $I_{z}$ operator is thus

\begin{equation}
I_{z}^{0}(t)=\sum_{\vec{u}}C_{\vec{u}}^{0}(t)P_{\vec{u}}\approx\sum_{L}C_{L}^{0}(t)P_{L}.\label{eq:rho_simp}
\end{equation}
Applying the Liouville-von Neumann equation to Eq. (\ref{eq:rho_simp}),
the dynamics of the coefficients $C_{L}^{0}(t)$ is determined by
the set of $N$ coupled linear differential equations

\begin{equation}
\frac{d}{dt}C_{L}^{0}=-\frac{i}{4}W_{L-1}C_{L-1}^{0}-\frac{i}{4}W_{L}C_{L+1}^{0},\label{eq:balance_lg}
\end{equation}
where $W_{L}$ and $W_{L-1}$ are the transition probabilities of
increasing or reducing the number of active spins $L(\vec{u})$ by
one spin respectively. Following the Levy and Gleason assumptions,
we consider that the operator $P_{L}$ represents a contiguous group
of $L$ spins in real space, and that the only relevant interactions
are between near neighbors. Therefore, the transition probability
$W_{L}$ is determined by an effective dipolar coupling strength $d$
multiplied by the number of spins at the cluster edge $L^{\delta}$,
where the exponent $\delta$ can be estimated from the the spatial
dimension $D$ of the system as $\delta\sim1-1/D$. The effective
coupling $d$ is of the same order of magnitude as the width of
the spins' resonance line and can be determined from a fit to the
experimental data \citep{Levy1992c}. Based on these assumptions,
the transition probability is 
\begin{equation}
W_{L}=dL^{\delta},\label{eq: rate}
\end{equation}
as a function of the number of active spins $L$.

Our results of Sec. \ref{subsec:Cluster-size-of} that derive Eq.
(\ref{eq:K0_hamming}), show that the second moment $K_{0}(t)$ and
hence the information scrambling of Eq. (\ref{eq:OTOConmuta}), are
determined by the average number of active spins weighted by the dynamics
coefficients $|C_{L}(t)|^{2}$

\begin{equation}
K_{0}(t)=\sum_{L}L|C_{L}^{0}(t)|^{2}.\label{eq:K(t)}
\end{equation}
Its time evolution can then be obtained by solving a set of $N$ differential
equations rather than $2^{N}$ as in the exact Liouville-von Neumann
equation for an $N$ spin system. Therefore, the connection between
the cluster size of correlated spins and the second moment of the
MQC spectrum arises naturally based on the average number of active
spins involved in the information scrambling correlations. The consistency
of this assumption relies on the fact that the Levy and Gleason model
reproduced well several experimental results in solid-state systems
where power-law growth of the cluster size evolution is seen \citep{Levy1992c}.
The predicted growth rates and power-law exponents were also consistent
with the spin-spin coupling network topologies.\lyxdeleted{Propietario}{Thu Nov 18 14:27:13 2021}{ }

\subsection{Model for the decoherent dynamics of quantum information scrambling}

\begin{figure}
\includegraphics[width=0.6\columnwidth]{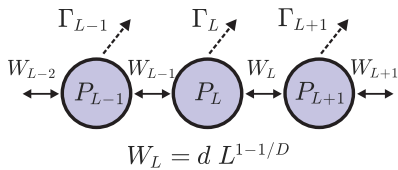}

\caption{Schematic representation of the model for the decoherent dynamics
of quantum information scrambling. It models the evolution of the
effective cluster size of active spins $K(t)$ in the presence of
decoherence effects induced by a perturbation on the echo experiment
described in Fig. \ref{fig:(a)-The-MQC-protocol}. The rate $W_{L}$
is the transition probability between $P_{L-1}$ and $P_{L}$, and
it is determined by the strength of the dipolar interaction $d$,
the size of the cluster of active spins $L$ and the lattice dimension
$D$. The spin clusters are affected by a decoherence process, characterized
by the decoherence rate $\Gamma_{L}$. \label{fig:Schematic-model-of}}
\end{figure}

We now consider that decoherence effects or perturbations to the ideal
Hamiltonian affect the time reversion of the protocol described in
Sec. \ref{subsec:Measuring-information-scrambling}. The product of
the two complex coefficients $C_{\vec{u}}^{p}$ and $C_{\vec{u}}^{0}$
of the forward and backward dynamics, respectively, in Eq. (\ref{eq:K_hamming}),
produces a reduction of the effective cluster size $K(t)$ compared
to the one determined by the ideal echo experiment case $K_{0}(t)$.
For weak perturbations, we consider that the coefficients $C_{\vec{u}}^{0}$
and $C_{\vec{u}}^{p}$ mainly differ by a phase, where $C_{\vec{u}}^{p}\sim|C_{\vec{u}}^{0}|\,e^{i\phi_{\vec{u}}}$.
Therefore the cluster size of Eq. (\ref{eq:K_hamming}) is

\begin{equation}
K(t)\sim\frac{1}{f(\phi=0,t,p)}\sum_{\vec{u}}\left|C_{\vec{u}}^{0}(t)\right|^{2}\cos(\phi_{\vec{u}})L(\vec{u}).\label{eq:K fenomeno}
\end{equation}
We here only consider the real part $\cos(\phi_{\vec{u}})$ of the
phase term, since $K(t)$ is real and the imaginary terms cancel out.
We then recast Eq. (\ref{eq:K fenomeno}) in terms of the number of
active spins $L$

\begin{equation}
K(t)\sim\frac{1}{f(\phi=0,t,p)}\sum_{L}L\sum_{\vec{u}_{L}}\left|C_{\vec{u}_{L}}^{0}(t)\right|^{2}\cos(\phi_{\vec{u}_{L}}).
\end{equation}
The coefficients $C_{L}^{0}(t)\equiv\sum_{\vec{u_{L}}}\left|C_{\vec{u}_{L}}^{0}(t)\right|^{2}$
of the average operator $P_{L}$, in Eq. (\ref{eq:rho_simp}), need
now to be modified as the phase $\cos(\phi_{\vec{u}_{L}})$ introduces
an attenuation factor that depends on $L$. The effective cluster
size $K$ is now described by the attenuated coefficients $C_{L}(t)=\sum_{\vec{u}_{L}}\left|C_{\vec{u}_{L}}^{0}(t)\right|^{2}\cos(\phi_{\vec{u}_{L}})$,
with $\left|C_{L}(t)\right|<\left|C_{L}^{0}(t)\right|$. The perturbation
term in the echo experiment is therefore modeled as a source of decoherence
in an open quantum system. We model this attenuation by adding to
Eq. (\ref{eq:balance_lg}) a leakage term with a rate $\Gamma_{L}$
that destroys the quantum superpositions given by the operator product
$P_{L}$,

\begin{equation}
\frac{d}{dt}C_{L}=-\frac{i}{4}W_{L-1}C_{L-1}-\frac{i}{4}W_{L}C_{L+1}-\Gamma_{L}C_{L}.\label{eq:lg_pert-1}
\end{equation}
The rate $\Gamma_{L}$ is the average decoherence rate for the product
of $L$ active spin operators. A schematic representation of the
model is shown in Fig. \ref{fig:Schematic-model-of}.

The matrix representation of Eq. (\ref{eq:lg_pert-1}) is

\[
\frac{d}{dt}\vec{C}=-\Upsilon\vec{C},
\]

\begin{equation}
\begin{bmatrix}\frac{d}{dt}C_{1}\\
\frac{d}{dt}C_{2}\\
\vdots\\
\frac{d}{dt}C_{N}
\end{bmatrix}=-i\begin{bmatrix}-i\Gamma_{1} & \frac{1}{4}W_{1} & 0 & \cdots & 0\\
\frac{1}{4}W_{1} & -i\Gamma_{2} & \frac{1}{4}W_{2} & \cdots & 0\\
\vdots & \vdots & \vdots & \ddots & \vdots\\
0 & 0 & \cdots & \frac{1}{4}W_{N-1} & -i\Gamma_{N}
\end{bmatrix}\begin{bmatrix}C_{1}\\
C_{2}\\
\vdots\\
C_{N}
\end{bmatrix},\label{eq:upsilon}
\end{equation}
and the effective cluster size $K$ in terms of $C_{L}$, is

\begin{equation}
K(t)=\left[\sum_{L}|C_{L}(t)|^{2}\right]^{-1}\sum_{L}L|C_{L}(t)|^{2}.\label{eq:K(t) con decoherencia}
\end{equation}
If the rate $\Gamma_{L}$ has no dependence on $L$, then all the
coefficients $C_{L}$ are affected by a global attenuation factor.
The effective cluster size $K(t)$ evolves then equally to the case
without perturbation, consistently with the predictions derived in
Ref. \citep{Garttner2018}. However, typically the decoherence effects
induced by perturbations increase with the number of active spins
that are correlated, and therefore $\Gamma_{L}$ generally depends
on $L$ \citep{Palma1997a,Duan1998,Reina2002,Krojanski2004a,Alvarez2011,Jing2015,Dominguez}.

\section{Effective cluster size evolution: decoherent dynamics of quantum
information scrambling\label{sec:Cluster-size-evolution-with}}

To analyze the effective cluster size evolution predicted by our model,
we consider the diagonal base for $\Upsilon$ to solve Eq. (\ref{eq:upsilon})

\begin{equation}
\frac{d}{dt}\tilde{C}_{i}=-\lambda_{i}\tilde{C}_{i},
\end{equation}
where $\tilde{C}_{i}$ is the $i$-th component of the populations
vector $\vec{C}$ in the eigenbasis of $\Upsilon$, and $\lambda_{i}$
is the the $i$-th eigenvalue. The solution is then
\begin{equation}
\tilde{C}_{i}(t)=\tilde{C}_{i}(0)e^{-\lambda_{i}t},\label{eq:sol_c}
\end{equation}
where $\tilde{C}_{i}(0)$ gives the initial condition $C(0)$ expressed
in the $\Upsilon$-eigenbasis. We consider that the decoherence rate
increases as a power law with the cluster size of active spins $\Gamma_{L}=\Gamma_{1}L^{\alpha}$,
where $\alpha$ is the scaling exponent, as reported in quantum simulations
in solid-state spin systems \citep{Krojanski2004a,Dominguez} and
this dependence is also expected for spin-boson models \citep{Palma1997a,Duan1998,Reina2002}.
We focus on three-dimensional (3D) systems, and thus we set the parameter
$\delta=0.66$ in Eq. (\ref{eq: rate}).

\begin{figure*}
\includegraphics[width=1\textwidth]{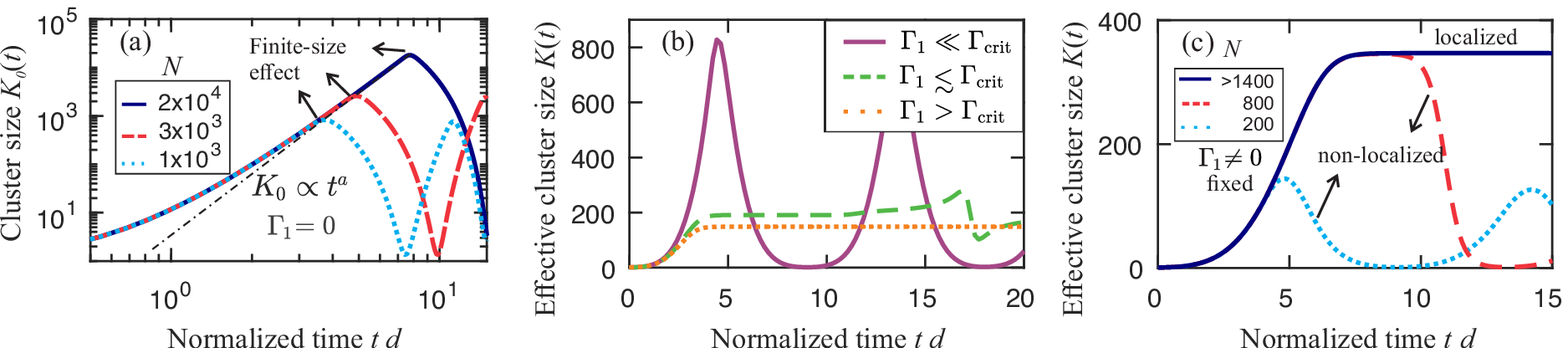}

\caption{Effective cluster size of correlated spins $K(t)$ that defines the
information scrambling predicted by our model. All simulations were
performed for a three-dimensional system {[}$\delta=0.66$, see Eq.
(\ref{eq: rate}){]}. (a) Cluster size dynamics $K_{0}(t)$ in the
ideal case, i.e., without considering decoherence effects ($\Gamma_{1}=0$).
Three different system sizes $N$ are shown in the legend. The cluster
size grows as $K_{0}(t)\propto t^{a}$ until finite-size effects distort
the evolution. All solutions are equivalent before the finite-size
effects appear independently of the value of $N$. (b) Effective cluster
size evolution $K(t)$, now including the decoherence effects ($\Gamma_{1}\protect\neq0$),
using $N=10^{3}$. When $\Gamma_{1}\ll\Gamma_{\mathrm{crit}}$, the
growth of $K(t)$ is delocalized until finite-size effects appear
as in the case of $\Gamma_{1}=0$. Near the critical decoherence strength
$\Gamma_{\mathrm{crit}}$, $K(t)$ tends to localize although it exhibits
low-frequency oscillations. For $\Gamma_{1}>\Gamma_{\mathrm{crit}}$,
$K(t)$ localizes indefinitely. (c) Finite-size effects on the observation
of the localization dynamics. If $\Gamma_{1}\protect\neq0$, $K(t)$
will localize only if $N$ is large enough. If not, finite-size effects
avoid the observation of the localization effects. \label{fig:finite size effects}}
\end{figure*}

We determine the evolution of the cluster size of correlated spins
for the case without perturbation. Figure \ref{fig:finite size effects}(a)
shows the cluster size evolution $K_{0}(t)$ for $\Gamma_{1}=0$ and
different system sizes $N$. The cluster size grows with a power law
as $K_{0}(t)=K_{1}t^{a}$, until it reaches a maximum value due to
finite-size effects determined by the system size $N$. The power-law
exponent $a$ and the constant $K_{1}$ are determined by the dimension
of the system $D$ and the average dipolar coupling $d$ respectively,
from Eq. (\ref{eq: rate}). After $K_{0}(t)$ reaches its maximum
value, it begins to oscillate indefinitely, again due to finite-size
effects. Therefore, this solution is only useful before finite-size
effects dominate the dynamical behavior.

We then introduce the decoherence effects. For $\Gamma_{1}>0$, an
initial power-law growth for the effective cluster size $K(t)$ is
observed similarly to the case of $\Gamma_{1}=0$ {[}Fig. \ref{fig:finite size effects}(b){]}.
However the exponent $a$ decreases slightly with increasing $\Gamma_{1}$.
Again, at long times when finite-size effects are reached, oscillations
are seen on the dynamics of $K(t)$. These oscillations decrease in
frequency and amplitude as $\Gamma_{1}$ increases. When the decoherence
strength $\Gamma_{1}$ increases above a critical value $\Gamma_{\mathrm{crit}}$,
the effective cluster size $K(t)$ reaches a plateau which is maintained
indefinitely in time. This plateau defines a localization size for
the observed information scrambling determined by the effective cluster
size. The predicted localization size is consistent with the experimental
observations of Refs. \citep{alvarez2010nmr,Alvarez2015,Dominguez}.
If the system size of the model $N\rightarrow\infty$, we observe
that $\Gamma_{\mathrm{crit}}\rightarrow0$. This implies that considering
power-law scalings for the leakage rates $\Gamma_{L}=\Gamma_{1}L^{\alpha}$ 
our model predicts that the effective cluster size dynamics localizes,
provided that the system size $N$ is large enough if $\Gamma_{1}\neq0$.
We have also observed that localization effects are manifested even
for slow-growing scaling laws as for $\Gamma_{L}\propto\log(L)$
or $L^{\alpha}$ with $\alpha<1$. Notice that,
if the decoherence rate is independent of $L$, localization effects
are not observed.

Figure \ref{fig:finite size effects}(c) shows the dynamics of $K(t)$
for different values of $N$ for a given $\Gamma_{1}$, manifesting
that localization effects are observed once $N$ is large enough.
All the curves behave equally before finite-size effects are significant.
Therefore if $N$ is large enough to manifest localization for a fixed
rate $\Gamma_{1}>\Gamma_{\mathrm{crit}}$, the predicted curve for
$K(t)$ is independent of $N$.

\begin{figure}
\centering \includegraphics[width=0.9\columnwidth]{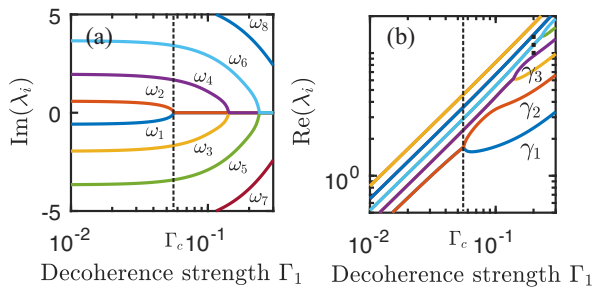} \caption{Eigenvalues of the transition rate operator $\Upsilon$ derived from
our model. (a) Imaginary and (b) real parts of the eigenvalues ${\lambda_{i}}$
as a function of the decoherence strength $\Gamma_{1}$. We use a
low number of $N=10$ to show clearer the functional behavior of the
eigenfrequencies, but the qualitative behavior is analogous for larger
$N$. For $\Gamma_{1}=0$, all the frequencies are pure imaginary
numbers, therefore the global amplitude $\sum_{L}\left|C_{L}\right|^{2}$
of the system is strictly conserved. For $\Gamma_{1}\protect\neq0$,
the solutions $\tilde{C}(t)$ become damped oscillations as ${\lambda_{i}}$
are complex numbers. There is a critical decoherence strength $\Gamma_{\mathrm{crit}}$
from which the smallest eigenvalue $\lambda_{1}$ becomes a real number.
Localization effects can be deduced from the eigenspectrum: the existence
of a nonoscillating, long-living solution $\tilde{C}_{1}(t)$ for
$\Gamma_{1}>\Gamma_{\mathrm{crit}}$ implies that observed scrambling
dynamics by $K(t)$ will eventually localize for long times.}

\label{fig:spectrum}
\end{figure}

The effective cluster size $K(t)$ of Eq. (\ref{eq:K(t) con decoherencia})
can be written in terms of the eigenbasis of $\Upsilon$ as

\begin{equation}
K(t)=\frac{\sum_{L}L|\sum_{i}c_{Li}\tilde{C}_{i}(t)|^{2}}{\sum_{L}|\sum_{i}c_{Li}\tilde{C}_{i}(t)|^{2}},
\end{equation}
where the populations $C_{L}(t)$ are in terms of a linear combination
of the evolution of $\tilde{C}_{i}(t)$, i.e., $C_{L}(t)=\sum_{i}c_{Li}\tilde{C}_{i}(t)$
with $c_{Li}$ the eigenvector coefficients. Using the solution for
$\tilde{C}_{i}(t)$ determined from Eq. (\ref{eq:sol_c}), we get

\begin{equation}
\begin{aligned}K(t) & =\frac{\sum_{L}L\left|\sum c_{Li}\tilde{C}_{i}(0)e^{-t\lambda_{i}}\right|^{2}}{\sum_{L}|\sum_{i}c_{Li}\tilde{C}_{i}(0)e^{-t\lambda_{i}}|^{2}}.\end{aligned}
\label{eq:Kt_eigenvectors-1}
\end{equation}
The transition from a delocalized to a localized scrambling dynamics
is evidenced in the behavior of the eigenvalues of $\Upsilon$ as
a function of $\Gamma_{1}$ {[}Fig. \ref{fig:spectrum}(a) and (b){]}.
The eigenvalues $\left\{ \lambda_{i}\right\} $ are purely imaginary
for $\Gamma_{1}=0$, which implies the conservation of $\sum_{L}\left|C_{L}\right|^{2}$.
However, if $\Gamma_{1}\neq0$, the eigenvalues $\left\{ \lambda_{i}\right\} $
are in general complex numbers $\lambda_{i}=\gamma_{i}+i\omega_{i}$,
with real $\gamma_{i}$ and imaginary $\omega_{i}$ components. We
consider the set $\left\{ \lambda_{i}\right\} $ sorted by its real
value, so as $\gamma_{i}\leq\gamma_{i+1}$. If $\Gamma_{1}<\Gamma_{\mathrm{crit}}$,
then $\lambda_{1}=\lambda_{2}^{*}$, which implies that two different
frequencies $\omega_{1}$ and $\omega_{2}=-\omega_{1}$ have the same
decay constant $\gamma_{1}=\gamma_{2}$. Hence, in the long time limit,
terms with decay constants $\gamma_{i}>\gamma_{1}$ become negligible
and we obtain
\begin{gather}
K(t\rightarrow\infty)=\frac{\sum_{L}L\left|e^{-t\lambda_{1}}\sum_{i}c_{Li}\tilde{C}_{i}(0)e^{-t(\lambda_{i}-\lambda_{1})}\right|^{2}}{\sum_{L}|e^{-t\lambda_{1}}\sum_{i}c_{Li}\tilde{C}_{i}(0)e^{-t(\lambda_{i}-\lambda_{1})}|^{2}}\nonumber \\
\sim\frac{\sum_{L}L\left|\left(c_{L1}\tilde{C}_{1}(0)+c_{L2}\tilde{C}_{2}(0)e^{-2it\omega_{1}}\right)\right|^{2}}{\sum_{L}\left|\left(c_{L1}\tilde{C}_{1}(0)+c_{L2}\tilde{C}_{2}(0)e^{-2it\omega_{1}}\right)\right|^{2}}.
\end{gather}
This solution provides the oscillations observed for $K(t)$ due to
the finite-size effects in Fig. \ref{fig:finite size effects}.

If $\Gamma_{1}>\Gamma_{\mathrm{crit}}$, then $\lambda_{1}$ becomes
a nondegenerate real eigenvalue ($\omega_{1}=0$ and $\gamma_{2}>\gamma_{1}$),
implying that the effective cluster size $K(t)$ attain a localization
value at long times

\begin{equation}
\lim_{t\rightarrow\infty}K(t)=K_{loc}=\frac{\sum_{L}L\left|c_{L1}\right|^{2}}{\sum_{L}\left|c_{L1}\right|^{2}}.\label{eq:Kloc}
\end{equation}
This demonstrates the existence of a localized regime for the observable
information scrambling determined by $K(t)$ when $\Gamma_{1}>\Gamma_{\mathrm{crit}}$.
Moreover, $K(t)$ converges always to the same stationary value, provided
that the initial condition $\vec{C}(0)$ has a nonzero contribution
of the $\tilde{C}_{1}$ eigenvector. This is because the coefficients
$c_{L1}$ are independent of the initial condition $\vec{C}(0)$ in
Eq. (\ref{eq:Kloc}). Therefore, independently of the initial cluster
size of correlated spins, the effective cluster size in the long time
limit will converge to the same localization size consistently with
experimental observations \citep{alvarez2010nmr,Alvarez2011,Alvarez2013}.

The delocalization-localization transition on the dynamical behavior
of the information scrambling manifested by the evolution of $K(t)$,
resembles the quantum dynamical phase transitions induced by decoherence
effects \citep{Alvarez2006,Danieli2007} that are connected with exceptional
points ubiquitous in non-Hermitian Hamiltonians \citep{Rotter2009,Rotter2010,Rotter2015,Martinez2018}.

\section{Model vs. experiments: evaluation of the decoherent dynamics of information
scrambling\label{sec:Experimental-test-of}}

\subsection{Information scrambling determined with NMR quantum simulations\label{subsec:Localized-spreading-in}}

We evaluate here the presented model as a framework to describe quantum
information scrambling dynamics with imperfect echo experiments. We
consider the MQC protocol based on imperfect time reversion echoes
as described in Fig. \ref{fig:(a)-The-MQC-protocol}(b), following
the technique introduced in Refs. \citep{alvarez2010nmr,Alvarez2011}.
In this experimental protocol, a controlled perturbation is introduced
to the forward Hamiltonian $\mathcal{H}_{F}=(1-p)\mathcal{H}_{0}+p\Sigma$,
where the nonreverted term $\Sigma$ is weighted by the dimensionless
parameter $p$ using average Hamiltonian techniques that can control
the perturbation strength $p$. This allows performing quantum simulations
to evaluate the effect of nonreverted interactions that are inherent
to any echo experiment.

\begin{figure}
\includegraphics[width=1\columnwidth]{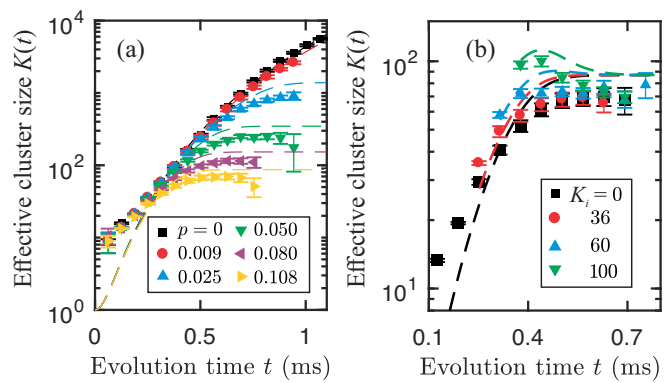}

\caption{Quantum information scrambling determined from MQC experiments with
imperfect time-reversals. (a) Effective cluster size $K(t)$ evolution
determined from quantum simulations with solid-state NMR experiments
on powdered adamantane for several perturbations strengths $p$ (symbols)
with the perturbation Hamiltonian $\Sigma=\mathcal{H}_{dd}$. The
corresponding predictions of our model are shown in dashed lines.
(b) Experimentally determined effective cluster size $K(t)$ evolution
starting from different initial conditions $K_{i}$ for $p=0.108$
(symbols). The effective cluster size $K(t)$ converges to a dynamical
equilibrium determined by the localization size $K_{loc}$ independently
of the initial cluster size $K_{i}$. The presented model correctly
predicts the effective cluster size evolution towards the dynamical
equilibrium (dashed lines). \label{fig:comp}}
\end{figure}

We perform the experimental quantum simulations on a Bruker Avance
III HD 9.4 T WB NMR spectrometer with a $^{1}\mathrm{H}$ resonance
frequency of $\omega_{z}=400.15$ MHz. We consider the nuclear spins
$^{1}\mathrm{H}$ of a powdered adamantane sample as the system, which
constitutes a dipolar interacting many-body system of equivalent $1/2$
spins with a 3D spin-spin coupling network topology. The imperfect
echo protocol of Fig. \ref{fig:(a)-The-MQC-protocol}(b) is implemented
using the perturbation Hamiltonian $\Sigma=\mathcal{H}_{dd}$ as the
raw dipolar interaction of the system of Eq. (\ref{eq:hdd}). The
perturbation Hamiltonian is introduced using the NMR sequence described
in Refs. \citep{alvarez2010nmr,Alvarez2011}. This protocol provides
a magnetization echo at the end of the sequence that is proportional
to the fidelity function $f(\phi,t,p)$ of Eq. (\ref{eq:fidelity_perturbada-1}),
from which we experimentally monitor the dynamics of the effective
cluster size $K(t,p)$ based on Eq. (\ref{eq:K(t) y espectro MQC f_M})
(see Fig. \ref{fig:comp}). 
We also calculate the instantaneous decay rate $\chi'(t)=\frac{d\chi(t)}{dt}$
with $\chi=\log(f)$ of the echo fidelity $f(\phi,t,p)$, which is
shown in Fig. \ref{Fig6: dec rate, alpha and chi(1)}(a).
\begin{figure}
\includegraphics[width=1\columnwidth]{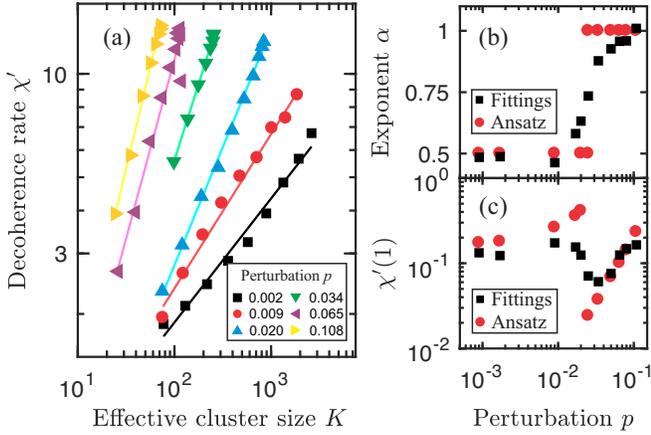}

\caption{Decoherence scaling as a function of the perturbation strength. (a)
Experimentally determined decoherence rates (symbols) and fitting
curves (solid lines) $\chi'(K)=\chi'(1)K^{\alpha}$ for different
perturbation strengths $p$. This power law behavior determines the
decay rates $\Gamma(L)$ of the presented model. The scaling exponent
$\alpha$ and the scaling factor $\chi'(1)$ are shown in panels (b)
and (c). Black squares are the fitting parameters of curves $\chi'(K)$,
and the red circles are the parameters determined by assuming the
single-parameter ansatz of Eq. (\ref{eq:chi_pc-2}). \label{Fig6: dec rate, alpha and chi(1)}}
\end{figure}
It is seen from the experiments that the decoherence rate scales with
a power-law function $\chi'(K)=\chi'(1)K^{\alpha}$. This scaling
behavior indicates the sensitivity of the controlled quantum dynamics
to the perturbation as a function of the instantaneous effective cluster
size $K$ \citep{Dominguez}.

For an unperturbed echo experiment ($p=0$), the cluster size $K_{0}(t)$
grows indefinitely following a power-law $K_{0}(t)\propto t^{4.3}$
up to the experimentally accessible timescales (black squares in Fig.
\ref{fig:comp}). If the perturbation to the control Hamiltonian $p\ne0$,
the effective cluster size $K(t)$ growth is reduced. The decay rate
$\chi'(K)$ and its power-law exponent $\alpha$ increase with the
perturbation strength $p$ (Fig. \ref{Fig6: dec rate, alpha and chi(1)}).
As seen in Fig. \ref{fig:comp}(a), $K(t)$ reaches a localization
value $K_{loc}$ that remains constant in time for large perturbations
$p>0.02$ \citep{alvarez2010nmr,Alvarez2015}. This localization size
reduces by increasing the perturbation strength $p$ consistently
with the predictions of Sec. \ref{sec:Cluster-size-evolution-with}.

The experimental results evidence that $K_{loc}$ is determined by
a dynamical equilibrium of $K(t)$ that converges to the same stationary
value, independently of the initial cluster size \citep{alvarez2010nmr,Alvarez2011,Alvarez2013}.
Figure \ref{fig:comp}(b) shows a series of experiments following
the protocol implemented in Refs. \citep{alvarez2010nmr,Alvarez2011,Alvarez2013}.
Here, an initial cluster size of correlated spins $K_{i}$ is prepared
by an unperturbed evolution with the propagator $U_{0}=e^{-it_{ini}\mathcal{H}_{0}}$,
where $t_{ini}$ is the initialization time required for preparing
the initial cluster size. Then, the cluster size $K_{i}$ is used
as the initial information state for the perturbed evolution $U_{p}=e^{-it[(1-p)\mathcal{H}_{0}+p\Sigma]}$.
As shown in Refs. \citep{alvarez2010nmr,Alvarez2011,Alvarez2013},
the cluster size converges to a localization size $K_{loc}(p)$ independently
of the initial value of $K_{i}$.

\subsection{Quantitative evaluation of the decoherent model for the information
scrambling dynamics}

We perform here a quantitative comparison between the predictions
of our model and the experimental results of the information scrambling
dynamics in imperfect echo experiments. We determine the parameters
$d$ and $\delta$ of Eq. (\ref{eq: rate}) to reproduce the cluster
size dynamics at $p=0$ shown in Fig. \ref{fig:comp}(a). We observe
from the experimental data that the cluster size is $K_{0}(t)\approx(7.41\,\frac{1}{\text{ms}}\,t)^{4.2}$.
We found that the power law exponent $a=4.2$ is obtained if $\delta=0.78$.
This value is slightly larger than the expected $1-1/D=0.66$
for a three-dimensional system as in our case, according to the original
Levy-Gleason model. This result is consistent with a cluster size
that keeps growing for a long time, at a rate that is faster than
in normal diffusion \citep{Alvarez2013}. This might be related with
a \textquotedblleft super-diffusion\textquotedblright{} mechanism
due to the complex long-range nature of the dipolar interaction in
our system \citep{METZLER20001,mercadier2009levy}. Setting $d=13$kHz
defined by the width of the resonance line of adamantane, we obtain
an excellent agreement with the experimental evolution of $K_{0}(t)$
as shown by black dashed lines in Fig. \ref{fig:comp}(a).

To predict the evolution of the effective cluster size $K(t)$ for
every perturbation strength $p$, we need to define the decay rates
$\Gamma(L)$ in Eq. (\ref{eq:lg_pert-1}) for the average $L$-spin
operators $P_{L}$. The experimentally observed decay rates $\chi'(K)=\chi'(1)K^{\alpha}$
{[}Fig. \ref{Fig6: dec rate, alpha and chi(1)} (a){]} of the fidelity
$f$ have a power law dependence on the instantaneous cluster size
$K(t)$ with a power law exponent $\alpha$ and proportionally constant
$\chi'(1)$ that depend on the perturbation strength {[}black squares
in Fig. \ref{Fig6: dec rate, alpha and chi(1)} (b) and (c) respectively{]}.
We assume that these decoherence rates determine the decay rates of
the model as $\Gamma(L)=\chi'(L)$ for each perturbation strength.
The exponent $\alpha$ presents a transition as a function of $p$,
between a low-scaling regimen with $\alpha\sim0.5$ for weak perturbations
and a high-scaling regimen with $\alpha\sim1$ for large perturbations
\citep{Dominguez}.

We calculate then $K(t)$, using a system size $N=2\times10^{4}$
large enough to avoid finite-size effects on the experimentally accessible
temporal scales of Fig. \ref{fig:comp}(a). We compare the experimental
results for $K(t)$ in Fig. \ref{fig:comp}(a) with their predictions
for several perturbation strengths. The calculated $K(t)$ correctly
predicts its time evolution and the achieved localization size $K_{loc}$
for large $p$, although $K_{loc}$ is slightly overestimated in all
cases. The model predicts with high accuracy the cluster size growth
for the long-time behavior of the weakest perturbation strengths.

Our model also quantitatively predicts well the dynamical equilibrium
localization size $K_{loc}$ predicted in subsection \ref{subsec:Localized-spreading-in}
as show in Fig. \ref{fig:comp}(b), for the perturbation strength
$p=0.108$. The dynamical equilibrium value for the localization size
$K_{loc}$ is determined from Eq. (\ref{eq:Kloc}) as the eigenvector
matrix $c_{Lk}$ is independent of the initial condition.

\begin{figure}
\includegraphics[width=1\columnwidth]{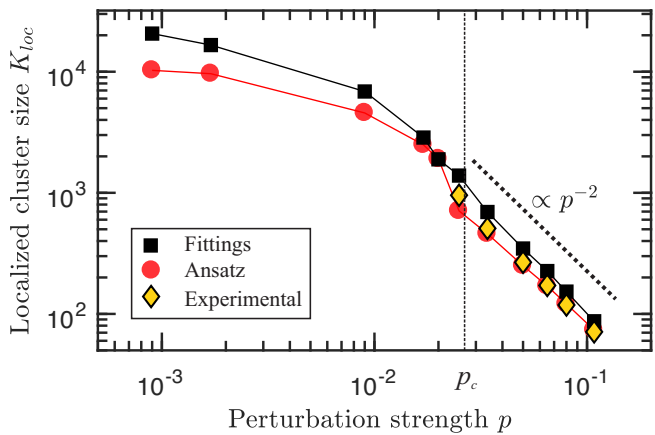}

\caption{Localization cluster size $K_{loc}$ as a function of the perturbation
strength for the perturbation Hamiltonian $\Sigma=\mathcal{H}_{dd}$.
Yellow diamonds show the experimentally determined localization sizes
only for the largest perturbation strengths where localization effects
are observed. We observe a power-law dependence $K_{loc}\propto p^{-2}$
shown with a dotted line as a guide to the eye. The black squares
are the localization size predicted by our model for every studied
perturbation using the measured parameters $\chi'(1)$ and $\alpha$
from the black symbols in Fig. \ref{Fig6: dec rate, alpha and chi(1)}
(b) and (c). The red circles are the predicted localization size determined
from our model assuming the single-parameter ansatz of Eq. (\ref{eq:chi_pc-2})
using the experimentally determined parameters $s,\nu,\alpha_{0}$,
and $\alpha_{\infty}$. The predicted localization size $K_{loc}$
exhibits two different scaling regimes as a function of the perturbation
strength, for weak and strong perturbations. The determined critical
perturbation $p_{c}$, where this transition occurs, is shown with
a vertical dashed line. \label{fig:Kloc}}
\end{figure}

Figure \ref{fig:Kloc} shows with black squares the prediction for
the localization size $K_{loc}$ as a function of the perturbation
strength $p$ compared with the ones determined from the experimental
data (yellow diamonds scatters). 
We determined $K_{loc}$ from Eq. (\ref{eq:Kloc}) which only depends
on the eigenvector coefficients $c_{L1}$ associated to the smallest
eigenvalue $\lambda_{1}$. The calculation of a single eigenvector
can be performed in shorter times compared with solving the full system
dynamics, allowing us to calculate $K_{loc}$ for systems with $N=2\times10^{5}$.
The computed values of $K_{loc}$ evidence two different regimes.
For large perturbations ($p>0.025$) our model predicts a scaling
for the localization size $K_{loc}(p)\propto p^{-2}$. For weak perturbations
($p<0.025$), we observe a weaker dependence of the localization size
as a function of the perturbation that can be described approximately
by $K_{loc}(p)\propto p^{-0.5}$. This transition is consistent with
the transition on the decoherence scaling exponent $\alpha$ shown
in Fig. \ref{Fig6: dec rate, alpha and chi(1)}(b). Experimental evidence
of localization effects is only observed for $p>0.026$, therefore
only these experimental data are shown in Fig. \ref{fig:Kloc}. Within
this perturbation regime the experimental values exhibit a functional
dependence $K_{loc}(p)\propto p^{-2}$ consistently with our predictions
from the model. We do not observe experimental evidence of localization
effects for weaker perturbations within the accessible time which
is limited by the decoherence decay of the NMR signal. The largest
cluster sizes observed experimentally for the weak perturbations are
always lower than the predicted localization sizes.

In a previous work \citep{Dominguez}, we showed experimental evidence
that the scaling behavior of decoherence as a function of the cluster
size is consistent with the single-parameter ansatz for the asymptotic
functional dependence at long times

\begin{equation}
\chi'(p,K)\sim\begin{cases}
(p_{c}-p)^{s}K^{\alpha_{0}} & p<p_{c}\\
(p-p_{c})^{-2\nu}K^{\alpha_{\infty}} & p>p_{c},
\end{cases}\label{eq:chi_pc-2}
\end{equation}
where we obtained the critical exponents $s=(-0.911\pm0.004$) and
$\nu=(-0.57\pm0.03)$. The scaling exponents are $\alpha_{\infty}=0.96\pm0.02$,
near to a linear scaling, and $\alpha_{0}=(0.48\pm0.03)$. The determined
critical perturbation is $p_{c}=(0.026\pm0.006)$. Due to the finite
evolution time of the experimental data, the transition from one regime
to another is smooth (see Fig. \ref{Fig6: dec rate, alpha and chi(1)}).
The experimental data give a smooth functional behavior also for the
decay rate $\Gamma(K)=\chi'(K)$ that we introduced in our model {[}black
squares in Fig. \ref{Fig6: dec rate, alpha and chi(1)}(b,c){]}. To
further evaluate the consistency of the scaling behavior determined
by Eq. (\ref{eq:chi_pc-2}) and extrapolate its predicted behavior
for the weak perturbations where we do not have experimental data,
we assume this functional behavior for $\Gamma(K)=\chi'(K)$ using
the extracted parameters $s,\nu,\alpha_{0}$ and $\alpha_{\infty}$.
Figure \ref{fig:Kloc} shows with red circles the prediction for the
localization size $K_{loc}$ as a function of the perturbation strength
$p$. The localization size $K_{loc}$ determined from Eq. (\ref{eq:Kloc})
now fit better the experimental data for the strong perturbations
($p>0.025$). Again for weak perturbations ($p<0.025$), we observe
a weaker dependence of the localization size as a function of the
perturbation strength. However, our model now predicts a finite localization
size when $p\rightarrow0$.

Both assumptions for $\Gamma(K)=\chi'(K)$ manifest different scaling
laws for $K_{loc}$ comparing the weak and strong perturbation regimes.
For weak perturbations, it seems from the predicted curves that there
might exist a limiting value for $K_{loc}$ for $p\rightarrow0$.
Further experimental designs need to be implemented to verify if the
experimental behavior matches these predictions, but we expect that
this limiting value is due to intrinsic perturbations on the experimental
implementation that are not accounted for in our microscopic model described
in Sec. \ref{sec:Information-scrambling-in} and in Refs. \citep{Alvarez2015,Dominguez}.

\section{Conclusion and discussions\label{sec:Conclusion}}

We developed a model for studying the quantum information scrambling
dynamics based on active spins clusters measured via out-of-time-order
correlators determined from Loschmidt echoes combined with multiple
quantum coherences experiments. Experimental implementations ubiquitously
contain imperfections in the quantum operations that lead to the
presence of nonreversed interactions in the LE procedure for measuring
the OTOC. Based on considering an imperfect experimental protocol,
we derived expressions for OTOC functions connected with the effective
number of correlated spins that are active on quantum superpositions
generated by coherence transfers of a local information, that survived
the perturbation effects. Decoherence effects induced by the time-reversal imperfection arise naturally by the derived OTOCs as a leakage
of the ideal unitary dynamics. The derived OTOCs quantify the observable
degree of scrambling of information based on an inner product between
the ideal and the perturbed scrambling dynamics. The main prediction
of our model is the existence of localization effects on the measurable
information scrambling that bound the effective cluster size $K(t)$
where the ideal information was propagated. We also found that whether
the initial information on $K(0)$ is local or not, the dynamics of
the effective cluster size $K(t)$ tends to a dynamical equilibrium
value $K_{loc}$. Our predictions were contrasted with quantum simulations
performed with NMR experiments on a solid-state adamantane sample,
showing excellent quantitative agreement with the experimental observations.

Levy and Gleason originally proposed a model to describe the dynamics
of the spin cluster of correlated spins in MQC experiments in solid-state
samples. The model was based on simplifying the spin quantum dynamics
to a phenomenological equation depending on two physical parameters:
the mean value of the spin-spin interactions and the number of dimensions
of the system. It correctly described how the number of correlated
spins grows in NMR solid-state experiments with only these parameters.
Our work revisits this model and adapts it to treat information scrambling.
We also introduced a decoherence process induced by imperfection on
the experimental implementations into the model. The decoherence process
is described by a leakage rate that depends on the number of active
spins involved in the states that describe the dynamics of the system.
Determining this rate from experimental data, we obtained accurate
predictions of the cluster size growth as a function of time and the
localization values that were experimentally observed. Our results
indicate that quantum information scrambling dynamics and its localization
effects due to perturbations are phenomena that can be predicted in
terms of few physical parameters by the presented model.

Treating many-body dynamics with exact numerical solutions is not
possible with present technology. Therefore, the results shown in
this work provide a framework for describing the quantum information
scrambling dynamics of many-body systems determined from experiments
affected by nonunitary decays, either by imperfections on the control
or from interaction with external degrees of freedom. In this article,
we have focused on modeling the dynamics of quantum information based
on OTOC functions that give the cluster size of correlated spins generated
by the scrambling dynamics. However, the model building block is based
on finding the density matrix evolution of the system by solving the
Liouville-von Neumann equation reduced to the average operator of the
active spins in the dynamics. Therefore, one can envisage that the
approach can be adapted to study the dynamic behavior of relevant
physical quantities such as other types of correlations in the presence
of decoherence effects \citep{Maziero2009_classical,Xu2010_experimental,Touil2021_information,Xu2021_thermofield,Styliaris2021_information}.
The framework presented here can be a useful tool to predict the quantum
information dynamics in large quantum systems and address the effect
of imperfections on the control Hamiltonian that drives the quantum
evolutions.
\begin{acknowledgments}
We thank M.C. Rodriguez for helpful discussions. This work was supported
by CNEA, ANPCyT-FONCyT PICT-2017-3447, PICT-2017-3699, PICT-2018-04333,
PIP-CONICET (11220170100486CO), UNCUYO SIIP Tipo I 2019-C028, and
Instituto Balseiro. F.D.D. and G.A.A. acknowledge support from CONICET.
\end{acknowledgments}

\appendix

\section{Quantum information scrambling as the cluster size of active spins\label{sec:Appendix-A}}

We here demonstrate that the norm of the commutator $\left[I_{z},\sigma^{0}(t)\right]$
quantifies the average number of active spins in the operator $\sigma^{0}(t)=U_{0}(t)\sigma U_{0}^{\dagger}(t)$
with $\sigma$ an arbitrary operator of the system. Then, when the
operator $\sigma=I_{z}$, the following demonstration relates the
cluster size of correlated spins of Eq. (\ref{eq:K0_exp_pa}) and
the OTO commutator of Eq. (\ref{eq:OTOConmuta}). We first calculate
the commutator $\left[I_{z},\sigma^{0}(t)\right]$ of the OTOC expression
of Eq. (\ref{eq:OTOConmuta}). We find that $\left[I_{z},\sigma^{0}(t)\right]=\sum_{j}^{N}\sum_{\vec{u}}C_{\vec{u}}^{0}(t)[I_{z}^{j},P_{\vec{u}}]$,
where we used that $\sigma^{0}(t)=\sum_{\vec{u}}C_{\vec{u}}^{0}(t)P_{\vec{u}}$
and $I_{z}=\sum_{j}^{N}I_{z}^{j}$. The operator $I_{z}$ is a mixture
of local spin operators $I_{z}^{j}$,

\begin{equation}
I_{z}^{j}=\underbrace{\mathbb{I}\otimes...}_{j-1}\otimes\underbrace{I_{z}^{(1)}}_{j}\otimes\underbrace{...\otimes\mathbb{I}}_{N-j}.\label{eq:Iz_j}
\end{equation}
The operators $P_{\vec{u}}$ are the elements of the orthonormal product
basis $\left\{ \left(\sqrt{2}\right)^{N}\bigotimes_{k=1}^{N}I_{u_{k}}^{^{(1)}}\right\} $,
where the coefficients of the vector $\vec{u}$ are $u_{i}\in\{x,y,z,0\}$
(zero for identity operator $I_{0}=\mathbb{I})$. By using the following
property of the Kronecker product

\begin{equation}
[A\otimes B,C\otimes D]=[A,C]\otimes BD+CA\otimes[B,D]
\end{equation}
and the expression for $I_{z}^{j}$ of Eq. (\ref{eq:Iz_j}), the commutator
$[I_{z}^{j},P_{\vec{u}}]$ is
\begin{align}
[I_{z}^{j},P_{\vec{u}}] & =\left(\sqrt{2}\right)^{N}\underbrace{I_{u_{1}}\otimes...}_{j-1}\otimes[I_{z}^{(1)},I_{u_{j}}]\otimes\underbrace{...\otimes I_{u_{N}}}_{N-j}\nonumber \\
 & =\begin{cases}
iP_{\vec{u}'(j)} & \text{\ensuremath{\mathrm{if}}\,\ensuremath{u_{j}=x}}\\
-iP_{\vec{u}'(j)} & \ensuremath{\mathrm{if}}\,\ensuremath{u_{j}=y}\\
0 & \ensuremath{\mathrm{if}}\,\ensuremath{u_{j}=0,z}
\end{cases}.
\end{align}
The operator $P_{\vec{u'}(j)}$ belongs to the product basis $\{P_{\vec{u}}\}$,
where the vector $\vec{u}'(j)$ is a vector identical to $\vec{u}$,
except for the $j$-th element (e.g. if $u_{j}=x$ then $u'(j)_{j}=y$
and vice versa). We have used that $[I_{z}^{(1)},I_{x}^{(1)}]=iI_{y}^{(1)}$,
$[I_{z}^{(1)},I_{y}^{(1)}]=-iI_{x}^{(1)}$ and $[I_{z}^{(1)},I_{z}^{(1)}]=[I_{z}^{(1)},\mathbb{I}]=0$.
This means that if $[I_{z}^{j},P_{\vec{u}}]\neq0$, then $[I_{z}^{j},P_{\vec{u}}]$
is proportional to the other element $P_{\vec{u'}(j)}$ of the product
basis $\{P_{\vec{u}}\}$. The sign of $[I_{z}^{j},P_{\vec{u}}]$ depends
on whether $u_{j}=x$ $(+)$ or $u_{j}=y$ $(-).$

We then find that the commutator 
\[
[I_{z},P_{\vec{u}}]=\sum_{j=1}^{N}[I_{z}^{j},P_{\vec{u}}]=\sum_{j\in\mathcal{A}(\vec{u})}\pm iP_{\vec{u}'(j)},
\]
where $\mathcal{A}(\vec{u})$ is the set of indeces $\{j\}$ corresponding
to active spins in $\vec{u}$, that is, those which satisfy $u_{j}=x,y$.
The set $\mathcal{A}(\vec{u})$ has $L(\vec{u})$ elements, where
$L(\vec{u})$ is the number of elements $u_{j}$ equal to $x$ and
$y$, which implies that $[I_{z},P_{\vec{u}}]$ has $L(\vec{u})$
nonzero terms $\pm iP_{\vec{u}'(j)}$. The resulting expression for
the OTOC and therefore for cluster size $K_{0}(t)$ is then
\begin{multline}
K_{0}(t)=\mathrm{Tr}\left\{ [I_{z},\sigma^{0}(t)][I_{z},\sigma^{0}(t)]{}^{\dagger}\right\} \\
=\sum_{\vec{u},\vec{v}}C_{\vec{u}}^{0}C_{\vec{v}}^{0*}\times\\
\times\sum_{i\in\mathcal{A}(\vec{u})}\sum_{j\in\mathcal{A}(\vec{v})}\mathrm{Tr}\left[\left(\pm iP_{\vec{u}'(i)}\right)\left(\pm iP_{\vec{v}'(j)}\right)^{\dagger}\right].
\end{multline}

Finally, we prove that the factor of the previous equation is equal
to the function $\mathcal{L}(\vec{u},\vec{v})$ introduced in Eq.
(\ref{eq:funcion_F})

\begin{equation}
\mathcal{L}(\vec{u},\vec{v})=\sum_{i\in\mathcal{A}(\vec{u})}\sum_{j\in\mathcal{A}(\vec{v})}\mathrm{Tr}\left[\left(\pm iP_{\vec{u}'(i)}\right)\left(\pm iP_{\vec{v}'(j)}\right)^{\dagger}\right].\label{eq:L definicion}
\end{equation}
To prove this, we divide the demonstration into five different propositions
as we discuss below. We introduce the notation $h_{0z}(\vec{u},\vec{v})$
for the Hamming distance between $\vec{u}$ and $\vec{v}$ considering
only the elements zero,$z$, and $h_{xy}(\vec{u},\vec{v})$ for the Hamming
distance between $\vec{u}$ and $\vec{v}$ considering only the elements
$x,y$. In Propositions 1 and 2, we deduce the necessary conditions
that $\vec{u}$ and $\vec{v}$ must satisfy for obtaining $\mathcal{L}(\vec{u},\vec{v})\neq0$.
Then, Propositions 3, 4 and 5 provide the values of $\mathcal{L}(\vec{u},\vec{v})$
when the conditions deduced in Propositions 1 and 2 are satisfied.

\paragraph{Proposition 1}

If $\mathcal{L}(\vec{u},\vec{v})\neq0$, then $h_{0z}(\vec{u},\vec{v})=0$.

\paragraph{Proof:}

Since $\{P_{\vec{u}}\}$ is an orthonormal base $\mathrm{Tr}\left(P_{\vec{u}'(i)}P_{\vec{v}'(j)}\right)=\delta_{\vec{u}'(i),\vec{v}'(j)}$,
to obtain $\mathcal{L}(\vec{u},\vec{v})\neq0$, there must be states
$\vec{u}'(i)=\vec{v}'(j)$ for some $i\in\mathcal{A}(\vec{u})$, $j\in\mathcal{A}(\vec{v})$.
Since the vectors $\vec{u}'(i)$ and $\vec{v}'(j)$ are identical
to $\vec{u}$ and $\vec{v}$ respectively, in those elements that
are equal to $z,$zero, it is necessary that $u_{l}=v_{l}$ if $l=0,z$
for $\vec{u}'(i)=\vec{v}'(j)$ to exist. Therefore, the Hamming distance
$h_{0z}(\vec{u},\vec{v})=0$.
\begin{flushright}
$\blacksquare$
\par\end{flushright}

\paragraph{Proposition 2}

If $\mathcal{L}(\vec{u},\vec{v})\neq0$, then $h_{xy}(\vec{u},\vec{v})=0$
or $2$.

\paragraph{Proof:}

Again, to obtain $\mathcal{L}(\vec{u},\vec{v})\neq0$, there must
be states $\vec{u}'(i)=\vec{v}'(j)$ for some $i\in\mathcal{A}(\vec{u})$,
$j\in\mathcal{A}(\vec{v})$. We prove Proposition 2 in two steps.

(i) We can see first that if $\mathcal{L}(\vec{u},\vec{v})\neq0$,
$h_{xy}(\vec{u},\vec{v})\le2$. From the definition of the vectors
$\vec{u'}(i)$, we know that $\vec{u'}(i)$ only differs from $\vec{u}$
in the $i$-th element, and that $u_{i}$ must be $x$ or $y$. We
thus deduce that $h_{xy}\left(\vec{u},\vec{u}'(i)\right)=h_{xy}\left(\vec{v},\vec{v}'(j)\right)=1$,
for every $i\in\mathcal{A}(\vec{u})$, $j\in\mathcal{A}(\vec{v})$.
If $\vec{u}'(i)=\vec{v}'(j)$ for some $i,j$, we then deduce from
the triangle inequality that

\begin{align}
h_{xy}(\vec{u},\vec{v}) & \le h_{xy}\left(\vec{u},\vec{u}'(i)\right)+h_{xy}\left(\vec{v},\vec{u}'(i)\right)\nonumber \\
 & =h_{xy}\left(\vec{u},\vec{u}'(i)\right)+h_{xy}\left(\vec{v},\vec{v}'(j)\right)\nonumber \\
 & =2.
\end{align}

(ii) We can also see that $h_{xy}(\vec{u},\vec{v})\neq1$. Let us
suppose that $h_{xy}(\vec{u},\vec{v})=1$, then there is a unique
index $l$ such that $u_{l}\neq v_{l}$. Therefore $\vec{v}=\vec{u}'(l)$,
and then it is impossible to find $\vec{v}'(j)$ such that $\vec{u}'(l)=\vec{v}'(j)$,
because $\vec{v}'(j)\neq\vec{v}=\vec{u}'(l)$. If we consider $\vec{u}'(i)$
(with $i\neq l$) then $h_{xy}\left(\vec{u}'(i),\vec{v}\right)=2$
and again it is impossible to find $\vec{v}'(j)$ such that $\vec{u}'(i)=\vec{v}'(j)$.
We have proven that if $h_{xy}(\vec{u},\vec{v})=1$ then $\mathcal{L}(\vec{u},\vec{v})=0$.
Therefore, if $\mathcal{L}(\vec{u},\vec{v})\neq0$, then $h_{xy}(\vec{u},\vec{v})\neq1$.
Therefore, if $\mathcal{L}(\vec{u},\vec{v})\neq0$, $h_{xy}(\vec{u},\vec{v})=0$
or $h_{xy}(\vec{u},\vec{v})=2$.
\begin{flushright}
$\blacksquare$
\par\end{flushright}

\paragraph{Proposition 3}

$\mathcal{L}(\vec{u},\vec{u})=L(\vec{u})$, where $L(\vec{u})$ is
the number of elements in $\vec{u}$ that are equal to $x,y$. Notice
that in this case when $\vec{u}=\vec{v}$, the Hamming distances are
$h_{xy}(\vec{u},\vec{v})=0$ and $h_{0z}(\vec{u},\vec{v})=0$.

\paragraph{Proof:}

Since $\{P_{\vec{u}}\}$ is an orthonormal basis, $\mathrm{Tr}\left(P_{\vec{u}'(i)}P_{\vec{v}'(j)}\right)=\delta_{\vec{u}'(i),\vec{v}'(j)}$.
As the number of elements in $\mathcal{A}(\vec{u})$ is $L(\vec{u})$,
we get

\begin{align}
\mathcal{L}(\vec{u},\vec{u}) & =\sum_{i\in\mathcal{A}(\vec{u})}\sum_{j\in\mathcal{A}(\vec{u})}\mathrm{Tr}\left[\left(\pm iP_{\vec{u}'(i)}\right)\left(\pm iP_{\vec{u}'(j)}\right)^{\dagger}\right]\nonumber \\
 & =-i^{2}\sum_{j\in\mathcal{A}(\vec{u})}1\nonumber \\
 & =L(\vec{u}).
\end{align}

\begin{flushright}
$\blacksquare$
\par\end{flushright}

\paragraph{Proposition 4\lyxdeleted{Propietario}{Thu Nov 18 14:27:13 2021}{ }}

If $h_{xy}(\vec{u},\vec{v})=2$, $h_{0z}(\vec{u},\vec{v})=0$ and
$\vec{u}$ is not a permutation of $\vec{v}$, i.e., $\vec{u}\neq\Pi(\vec{v})$,
then $\mathcal{L}(\vec{u},\vec{v})=-2$.

\paragraph{Proof:\lyxdeleted{Propietario}{Thu Nov 18 14:27:13 2021}{ }}

Since $h_{xy}(\vec{u},\vec{v})=2$ and $h_{0z}(\vec{u},\vec{v})=0$,
there are only two indices $l_{1}$ and $l_{2}$ for which $u_{l_{1}}\neq v_{l_{1}}$
and $u_{l_{2}}\neq v_{l_{2}}$. Since $l_{1},l_{2}\in\{x,y\}$ and
$\vec{u}\neq\Pi(\vec{v})$, there must be the following conditions
$u_{l_{1}}=x$, $u_{l_{2}}=x$ and $v_{l_{1}}=y$ , $v_{l_{2}}=y$,
or vice versa that are equivalent to exchange $\vec{u}$ and $\vec{v}$.
Then, we have that $\vec{u}'(l_{1})=\vec{v}'(l_{2})$ and $\vec{u}'(l_{2})=\vec{v}'(l_{1})$.
These are the only two terms different from zero in Eq. (\ref{eq:L definicion}).
If $u_{l_{1}}=u_{l_{2}}=x$, then the operators $P_{\vec{u}'(l_{1})}$
and $P_{\vec{u}'(l_{2})}$ have a multiplicative factor $i$. Analogously,
since $v_{l_{1}}=v_{l_{2}}=y$ , then the operators $P_{\vec{v}'(l_{1})}$
and $P_{\vec{v}'(l_{2})}$ have a multiplicative factor $-i$. Therefore,
the operator products $P_{\vec{u}'(l_{1})}P_{\vec{v}'(l_{2})}$ and
$P_{\vec{u}'(l_{2})}P_{\vec{v}'(l_{1})}$ are proportional to $i^{2}=-1$,
and the function $\mathcal{L}(\vec{u},\vec{v})$ is

\begin{multline}
\sum_{i\in\mathcal{A}(\vec{u})}\sum_{j\in\mathcal{A}(\vec{v})}\mathrm{Tr}\left[\left(\pm iP_{\vec{u}'(i)}\right)\left(\pm iP_{\vec{v}'(j)}\right)^{\dagger}\right]=\\
=i(-i)^{*}\mathrm{Tr}\left(P_{\vec{u}'(l_{1})}P_{\vec{v}'(l_{2})}\right)+\\
+(-i)^{*}i\mathrm{Tr}\left(P_{\vec{v}'(l_{1})}P_{\vec{u}'(l_{2})}\right)\\
=-2.
\end{multline}

\begin{flushright}
$\blacksquare$
\par\end{flushright}

\paragraph{Proposition 5}

If $h_{xy}(\vec{u},\vec{v})=2$, $h_{0z}(\vec{u},\vec{v})=0$ and
$\vec{u}$ is a permutation of $\vec{v}$, i.e. $\vec{u}=\Pi(\vec{v})$,
then $\mathcal{L}(\vec{u},\vec{v})=2$.

\paragraph{Proof:\lyxdeleted{Propietario}{Thu Nov 18 14:27:13 2021}{ }}

Since $h_{xy}(\vec{u},\vec{v})=2$ and $h_{0z}(\vec{u},\vec{v})=0$,
there are only two indices $l_{1}$ and $l_{2}$ for which $u_{l_{1}}\neq v_{l_{1}}$
and $u_{l_{2}}\neq v_{l_{2}}$. Since $l_{1},l_{2}\in\{x,y\}$ and
$\vec{u}=\Pi(\vec{v})$, it must hold that $u_{l_{1}}=x$ , $u_{l_{2}}=y$
and $v_{l_{1}}=y$ , $v_{l_{2}}=x$. Then, we have that $\vec{u}'(l_{1})=\vec{v}'(l_{2})$
and $\vec{u}'(l_{2})=\vec{v}'(l_{1})$. These are the only two terms
different from zero in Eq. (\ref{eq:L definicion}). Since $u_{l_{1}}=v_{l_{2}}=x$,
then the operators $P_{\vec{u}'(l_{1})}$ and $P_{\vec{v}'(l_{2})}$
have a multiplicative factor $i$. Analogously, since $v_{l_{1}}=u_{l_{2}}=y$
, then the operators $P_{\vec{v}'(l_{1})}$ and $P_{\vec{u}'(l_{2})}$
have a multiplicative factor $-i$. Therefore, the operator products
$P_{\vec{u}'(l_{1})}P_{\vec{v}'(l_{2})}$ and $P_{\vec{u}'(l_{2})}P_{\vec{v}'(l_{1})}$
are proportional to $-i^{2}=1$, and the function $\mathcal{L}(\vec{u},\vec{v})$
is
\begin{flushright}
\begin{align}
 & \sum_{i\in\mathcal{A}(\vec{u})}\sum_{j\in\mathcal{A}(\vec{v})}\mathrm{Tr}\left[\left(\pm iP_{\vec{u}'(i)}\right)\left(\pm iP_{\vec{v}'(j)}\right)^{\dagger}\right]=\nonumber \\
 & =i\cdot(i)^{*}\mathrm{Tr}\left(P_{\vec{u}'(l_{1})}P_{\vec{v}'(l_{2})}\right)+(i)^{*}i\mathrm{Tr}\left(P_{\vec{v}'(l_{1})}P_{\vec{u}'(l_{2})}\right)\nonumber \\
 & =2.
\end{align}
$\blacksquare$
\par\end{flushright}

These five propositions thus demonstrate the expression of Eq. (\ref{eq:funcion_F})
in the main text for the function $\mathcal{L}(\vec{u},\vec{v})$.

For the perturbed case the effective cluster size $K(t)$ of Eq. (\ref{eq: K(t) con decoherencia})
is derived directly from the previous demonstration. But, we now write
$\sigma^{0}(t)=\sum_{\vec{u}}C_{\vec{u}}^{0}(t)P_{\vec{u}}$ and $\sigma(t)=\sum_{\vec{u}}C_{\vec{u}}(t)P_{\vec{u}}$,
and then 

\begin{align}
 & K(t)=\mathrm{Tr}\left[[I_{z},\sigma^{0}(t)][I_{z},\sigma(t)]{}^{\dagger}\right]\nonumber \\
 & =\sum_{\vec{u},\vec{v}}C_{\vec{u}}^{0}(t)C_{\vec{v}}^{*}(t)\left\{ \sum_{i\in\mathcal{A}(\vec{u})}\sum_{j\in\mathcal{A}(\vec{v})}\mathrm{Tr}\left[\left(\pm iP_{\vec{u}'(i)}\right)\left(\pm iP_{\vec{v}'(j)}\right)^{\dagger}\right]\right\} ,
\end{align}
where the function $\mathcal{L}(\vec{u},\vec{v})=\sum_{i\in\mathcal{A}(\vec{u})}\sum_{j\in\mathcal{A}(\vec{v})}\mathrm{Tr}\left[\left(\pm iP_{\vec{u}'(i)}\right)\cdot\left(\pm iP_{\vec{v}'(j)}\right)^{\dagger}\right]$
is the same as Eq. (\ref{eq:funcion_F}) for the nonperturbed case.
The expression for $K(t)$ of Eq. (\ref{eq: K(t) con decoherencia})
is obtained when $\sigma^{0}(t)=I_{z}^{0}(t)$ and $\sigma(t)=I_{z}(t)$.


\begin{thebibliography}{93}%
	\makeatletter
	\providecommand \@ifxundefined [1]{%
		\@ifx{#1\undefined}
	}%
	\providecommand \@ifnum [1]{%
		\ifnum #1\expandafter \@firstoftwo
		\else \expandafter \@secondoftwo
		\fi
	}%
	\providecommand \@ifx [1]{%
		\ifx #1\expandafter \@firstoftwo
		\else \expandafter \@secondoftwo
		\fi
	}%
	\providecommand \natexlab [1]{#1}%
	\providecommand \enquote  [1]{``#1''}%
	\providecommand \bibnamefont  [1]{#1}%
	\providecommand \bibfnamefont [1]{#1}%
	\providecommand \citenamefont [1]{#1}%
	\providecommand \href@noop [0]{\@secondoftwo}%
	\providecommand \href [0]{\begingroup \@sanitize@url \@href}%
	\providecommand \@href[1]{\@@startlink{#1}\@@href}%
	\providecommand \@@href[1]{\endgroup#1\@@endlink}%
	\providecommand \@sanitize@url [0]{\catcode `\\12\catcode `\$12\catcode
		`\&12\catcode `\#12\catcode `\^12\catcode `\_12\catcode `\%12\relax}%
	\providecommand \@@startlink[1]{}%
	\providecommand \@@endlink[0]{}%
	\providecommand \url  [0]{\begingroup\@sanitize@url \@url }%
	\providecommand \@url [1]{\endgroup\@href {#1}{\urlprefix }}%
	\providecommand \urlprefix  [0]{URL }%
	\providecommand \Eprint [0]{\href }%
	\providecommand \doibase [0]{https://doi.org/}%
	\providecommand \selectlanguage [0]{\@gobble}%
	\providecommand \bibinfo  [0]{\@secondoftwo}%
	\providecommand \bibfield  [0]{\@secondoftwo}%
	\providecommand \translation [1]{[#1]}%
	\providecommand \BibitemOpen [0]{}%
	\providecommand \bibitemStop [0]{}%
	\providecommand \bibitemNoStop [0]{.\EOS\space}%
	\providecommand \EOS [0]{\spacefactor3000\relax}%
	\providecommand \BibitemShut  [1]{\csname bibitem#1\endcsname}%
	\let\auto@bib@innerbib\@empty
	\bibitem [{\citenamefont {Sekino}\ and\ \citenamefont
		{Susskind}(2008)}]{Sekino2008}%
	\BibitemOpen
	\bibfield  {author} {\bibinfo {author} {\bibfnamefont {Y.}~\bibnamefont
			{Sekino}}\ and\ \bibinfo {author} {\bibfnamefont {L.}~\bibnamefont
			{Susskind}},\ }\bibfield  {title} {\bibinfo {title} {{Fast scramblers}},\
	}\href {https://doi.org/10.1088/1126-6708/2008/10/065} {\bibfield  {journal}
		{\bibinfo  {journal} {J. High Energy Phys.}\ }\textbf {\bibinfo {volume}
			{2008}},\ \bibinfo {pages} {065}}\BibitemShut {NoStop}%
	\bibitem [{\citenamefont {Lashkari}\ \emph {et~al.}(2013)\citenamefont
		{Lashkari}, \citenamefont {Stanford}, \citenamefont {Hastings}, \citenamefont
		{Osborne},\ and\ \citenamefont {Hayden}}]{Lashkari2013}%
	\BibitemOpen
	\bibfield  {author} {\bibinfo {author} {\bibfnamefont {N.}~\bibnamefont
			{Lashkari}}, \bibinfo {author} {\bibfnamefont {D.}~\bibnamefont {Stanford}},
		\bibinfo {author} {\bibfnamefont {M.}~\bibnamefont {Hastings}}, \bibinfo
		{author} {\bibfnamefont {T.}~\bibnamefont {Osborne}},\ and\ \bibinfo {author}
		{\bibfnamefont {P.}~\bibnamefont {Hayden}},\ }\bibfield  {title} {\bibinfo
		{title} {{Towards the fast scrambling conjecture}},\ }\href
	{https://doi.org/10.1007/JHEP04(2013)022} {\bibfield  {journal} {\bibinfo
			{journal} {J. High Energy Phys.}\ }\textbf {\bibinfo {volume} {2013}},\
		\bibinfo {pages} {22}}\BibitemShut {NoStop}%
	\bibitem [{\citenamefont {Martinez}\ \emph {et~al.}(2016)\citenamefont
		{Martinez}, \citenamefont {Muschik}, \citenamefont {Schindler}, \citenamefont
		{Nigg}, \citenamefont {Erhard}, \citenamefont {Heyl}, \citenamefont {Hauke},
		\citenamefont {Dalmonte}, \citenamefont {Monz}, \citenamefont {Zoller},\ and\
		\citenamefont {Blatt}}]{Martinez2016}%
	\BibitemOpen
	\bibfield  {author} {\bibinfo {author} {\bibfnamefont {E.~A.}\ \bibnamefont
			{Martinez}}, \bibinfo {author} {\bibfnamefont {C.~A.}\ \bibnamefont
			{Muschik}}, \bibinfo {author} {\bibfnamefont {P.}~\bibnamefont {Schindler}},
		\bibinfo {author} {\bibfnamefont {D.}~\bibnamefont {Nigg}}, \bibinfo {author}
		{\bibfnamefont {A.}~\bibnamefont {Erhard}}, \bibinfo {author} {\bibfnamefont
			{M.}~\bibnamefont {Heyl}}, \bibinfo {author} {\bibfnamefont {P.}~\bibnamefont
			{Hauke}}, \bibinfo {author} {\bibfnamefont {M.}~\bibnamefont {Dalmonte}},
		\bibinfo {author} {\bibfnamefont {T.}~\bibnamefont {Monz}}, \bibinfo {author}
		{\bibfnamefont {P.}~\bibnamefont {Zoller}},\ and\ \bibinfo {author}
		{\bibfnamefont {R.}~\bibnamefont {Blatt}},\ }\bibfield  {title} {\bibinfo
		{title} {{Real-time dynamics of lattice gauge theories with a few-qubit
				quantum computer}},\ }\href {https://doi.org/10.1038/nature18318} {\bibfield
		{journal} {\bibinfo  {journal} {Nature}\ }\textbf {\bibinfo {volume} {534}},\
		\bibinfo {pages} {516} (\bibinfo {year} {2016})}\BibitemShut {NoStop}%
	\bibitem [{\citenamefont {Friis}\ \emph {et~al.}(2018)\citenamefont {Friis},
		\citenamefont {Marty}, \citenamefont {Maier}, \citenamefont {Hempel},
		\citenamefont {Holz{\"{a}}pfel}, \citenamefont {Jurcevic}, \citenamefont
		{Plenio}, \citenamefont {Huber}, \citenamefont {Roos}, \citenamefont
		{Blatt},\ and\ \citenamefont {Lanyon}}]{Friis2018}%
	\BibitemOpen
	\bibfield  {author} {\bibinfo {author} {\bibfnamefont {N.}~\bibnamefont
			{Friis}}, \bibinfo {author} {\bibfnamefont {O.}~\bibnamefont {Marty}},
		\bibinfo {author} {\bibfnamefont {C.}~\bibnamefont {Maier}}, \bibinfo
		{author} {\bibfnamefont {C.}~\bibnamefont {Hempel}}, \bibinfo {author}
		{\bibfnamefont {M.}~\bibnamefont {Holz{\"{a}}pfel}}, \bibinfo {author}
		{\bibfnamefont {P.}~\bibnamefont {Jurcevic}}, \bibinfo {author}
		{\bibfnamefont {M.~B.}\ \bibnamefont {Plenio}}, \bibinfo {author}
		{\bibfnamefont {M.}~\bibnamefont {Huber}}, \bibinfo {author} {\bibfnamefont
			{C.}~\bibnamefont {Roos}}, \bibinfo {author} {\bibfnamefont {R.}~\bibnamefont
			{Blatt}},\ and\ \bibinfo {author} {\bibfnamefont {B.}~\bibnamefont
			{Lanyon}},\ }\bibfield  {title} {\bibinfo {title} {{Observation of Entangled
				States of a Fully Controlled 20-Qubit System}},\ }\href
	{https://doi.org/10.1103/PhysRevX.8.021012} {\bibfield  {journal} {\bibinfo
			{journal} {Phys. Rev. X}\ }\textbf {\bibinfo {volume} {8}},\ \bibinfo {pages}
		{21012} (\bibinfo {year} {2018})}\BibitemShut {NoStop}%
	\bibitem [{\citenamefont {Swingle}(2018)}]{Swingle2018}%
	\BibitemOpen
	\bibfield  {author} {\bibinfo {author} {\bibfnamefont {B.}~\bibnamefont
			{Swingle}},\ }\bibfield  {title} {\bibinfo {title} {{Unscrambling the physics
				of out-of-time-order correlators}},\ }\href
	{https://doi.org/10.1038/s41567-018-0295-5} {\bibfield  {journal} {\bibinfo
			{journal} {Nat. Phys.}\ }\textbf {\bibinfo {volume} {14}},\ \bibinfo {pages}
		{988} (\bibinfo {year} {2018})}\BibitemShut {NoStop}%
	\bibitem [{\citenamefont {Lewis-Swan}\ \emph {et~al.}(2019)\citenamefont
		{Lewis-Swan}, \citenamefont {Safavi-Naini}, \citenamefont {Kaufman},\ and\
		\citenamefont {Rey}}]{Lewis-Swan2019}%
	\BibitemOpen
	\bibfield  {author} {\bibinfo {author} {\bibfnamefont {R.~J.}\ \bibnamefont
			{Lewis-Swan}}, \bibinfo {author} {\bibfnamefont {A.}~\bibnamefont
			{Safavi-Naini}}, \bibinfo {author} {\bibfnamefont {A.~M.}\ \bibnamefont
			{Kaufman}},\ and\ \bibinfo {author} {\bibfnamefont {A.~M.}\ \bibnamefont
			{Rey}},\ }\bibfield  {title} {\bibinfo {title} {{Dynamics of quantum
				information}},\ }\href {https://doi.org/10.1038/s42254-019-0090-y} {\bibfield
		{journal} {\bibinfo  {journal} {Nat. Rev. Phys.}\ }\textbf {\bibinfo
			{volume} {1}},\ \bibinfo {pages} {627} (\bibinfo {year} {2019})}\BibitemShut
	{NoStop}%
	\bibitem [{\citenamefont {Eisert}\ \emph {et~al.}(2015)\citenamefont {Eisert},
		\citenamefont {Friesdorf},\ and\ \citenamefont {Gogolin}}]{Eisert2015}%
	\BibitemOpen
	\bibfield  {author} {\bibinfo {author} {\bibfnamefont {J.}~\bibnamefont
			{Eisert}}, \bibinfo {author} {\bibfnamefont {M.}~\bibnamefont {Friesdorf}},\
		and\ \bibinfo {author} {\bibfnamefont {C.}~\bibnamefont {Gogolin}},\
	}\bibfield  {title} {\bibinfo {title} {{Quantum many-body systems out of
				equilibrium}},\ }\href {https://doi.org/10.1038/nphys3215} {\bibfield
		{journal} {\bibinfo  {journal} {Nat. Phys.}\ }\textbf {\bibinfo {volume}
			{11}},\ \bibinfo {pages} {124} (\bibinfo {year} {2015})}\BibitemShut
	{NoStop}%
	\bibitem [{\citenamefont {Abanin}\ \emph {et~al.}(2019)\citenamefont {Abanin},
		\citenamefont {Altman}, \citenamefont {Bloch},\ and\ \citenamefont
		{Serbyn}}]{abanin_colloquium_2019}%
	\BibitemOpen
	\bibfield  {author} {\bibinfo {author} {\bibfnamefont {D.~A.}\ \bibnamefont
			{Abanin}}, \bibinfo {author} {\bibfnamefont {E.}~\bibnamefont {Altman}},
		\bibinfo {author} {\bibfnamefont {I.}~\bibnamefont {Bloch}},\ and\ \bibinfo
		{author} {\bibfnamefont {M.}~\bibnamefont {Serbyn}},\ }\bibfield  {title}
	{\bibinfo {title} {Colloquium: Many-body localization, thermalization, and
			entanglement},\ }\href {https://doi.org/10.1103/RevModPhys.91.021001}
	{\bibfield  {journal} {\bibinfo  {journal} {Rev. Mod. Phys.}\ }\textbf
		{\bibinfo {volume} {91}},\ \bibinfo {pages} {21001} (\bibinfo {year}
		{2019})}\BibitemShut {NoStop}%
	\bibitem [{\citenamefont {{\'{A}}lvarez}\ \emph {et~al.}(2015)\citenamefont
		{{\'{A}}lvarez}, \citenamefont {Suter},\ and\ \citenamefont
		{Kaiser}}]{Alvarez2015}%
	\BibitemOpen
	\bibfield  {author} {\bibinfo {author} {\bibfnamefont {G.~A.}\ \bibnamefont
			{{\'{A}}lvarez}}, \bibinfo {author} {\bibfnamefont {D.}~\bibnamefont
			{Suter}},\ and\ \bibinfo {author} {\bibfnamefont {R.}~\bibnamefont
			{Kaiser}},\ }\bibfield  {title} {\bibinfo {title}
		{{Localization-delocalization transition in the dynamics of dipolar-coupled
				nuclear spins}},\ }\href {https://doi.org/10.1126/science.1261160} {\bibfield
		{journal} {\bibinfo  {journal} {Science}\ }\textbf {\bibinfo {volume}
			{349}},\ \bibinfo {pages} {846} (\bibinfo {year} {2015})}\BibitemShut
	{NoStop}%
	\bibitem [{\citenamefont {Schweigler}\ \emph {et~al.}(2017)\citenamefont
		{Schweigler}, \citenamefont {Kasper}, \citenamefont {Erne}, \citenamefont
		{Mazets}, \citenamefont {Rauer}, \citenamefont {Cataldini}, \citenamefont
		{Langen}, \citenamefont {Gasenzer}, \citenamefont {Berges},\ and\
		\citenamefont {Schmiedmayer}}]{Schweigler2017}%
	\BibitemOpen
	\bibfield  {author} {\bibinfo {author} {\bibfnamefont {T.}~\bibnamefont
			{Schweigler}}, \bibinfo {author} {\bibfnamefont {V.}~\bibnamefont {Kasper}},
		\bibinfo {author} {\bibfnamefont {S.}~\bibnamefont {Erne}}, \bibinfo {author}
		{\bibfnamefont {I.}~\bibnamefont {Mazets}}, \bibinfo {author} {\bibfnamefont
			{B.}~\bibnamefont {Rauer}}, \bibinfo {author} {\bibfnamefont
			{F.}~\bibnamefont {Cataldini}}, \bibinfo {author} {\bibfnamefont
			{T.}~\bibnamefont {Langen}}, \bibinfo {author} {\bibfnamefont
			{T.}~\bibnamefont {Gasenzer}}, \bibinfo {author} {\bibfnamefont
			{J.}~\bibnamefont {Berges}},\ and\ \bibinfo {author} {\bibfnamefont
			{J.}~\bibnamefont {Schmiedmayer}},\ }\bibfield  {title} {\bibinfo {title}
		{{Experimental characterization of a quantum many-body system via
				higher-order correlations}},\ }\href {https://doi.org/10.1038/nature22310}
	{\bibfield  {journal} {\bibinfo  {journal} {Nature}\ }\textbf {\bibinfo
			{volume} {545}},\ \bibinfo {pages} {323} (\bibinfo {year}
		{2017})}\BibitemShut {NoStop}%
	\bibitem [{\citenamefont {Lukin}\ \emph {et~al.}(2019)\citenamefont {Lukin},
		\citenamefont {Rispoli}, \citenamefont {Schittko}, \citenamefont {Tai},
		\citenamefont {Kaufman}, \citenamefont {Choi}, \citenamefont {Khemani},
		\citenamefont {L{\'{e}}onard},\ and\ \citenamefont {Greiner}}]{Lukin2019a}%
	\BibitemOpen
	\bibfield  {author} {\bibinfo {author} {\bibfnamefont {A.}~\bibnamefont
			{Lukin}}, \bibinfo {author} {\bibfnamefont {M.}~\bibnamefont {Rispoli}},
		\bibinfo {author} {\bibfnamefont {R.}~\bibnamefont {Schittko}}, \bibinfo
		{author} {\bibfnamefont {M.~E.}\ \bibnamefont {Tai}}, \bibinfo {author}
		{\bibfnamefont {A.~M.}\ \bibnamefont {Kaufman}}, \bibinfo {author}
		{\bibfnamefont {S.}~\bibnamefont {Choi}}, \bibinfo {author} {\bibfnamefont
			{V.}~\bibnamefont {Khemani}}, \bibinfo {author} {\bibfnamefont
			{J.}~\bibnamefont {L{\'{e}}onard}},\ and\ \bibinfo {author} {\bibfnamefont
			{M.}~\bibnamefont {Greiner}},\ }\bibfield  {title} {\bibinfo {title}
		{{Probing entanglement in a many-body-localized system}},\ }\href
	{https://doi.org/10.1126/science.aau0818} {\bibfield  {journal} {\bibinfo
			{journal} {Science}\ }\textbf {\bibinfo {volume} {364}},\ \bibinfo {pages}
		{256} (\bibinfo {year} {2019})}\BibitemShut {NoStop}%
	\bibitem [{\citenamefont {Landsman}\ \emph {et~al.}(2019)\citenamefont
		{Landsman}, \citenamefont {Figgatt}, \citenamefont {Schuster}, \citenamefont
		{Linke}, \citenamefont {Yoshida}, \citenamefont {Yao},\ and\ \citenamefont
		{Monroe}}]{Landsman2019}%
	\BibitemOpen
	\bibfield  {author} {\bibinfo {author} {\bibfnamefont {K.~A.}\ \bibnamefont
			{Landsman}}, \bibinfo {author} {\bibfnamefont {C.}~\bibnamefont {Figgatt}},
		\bibinfo {author} {\bibfnamefont {T.}~\bibnamefont {Schuster}}, \bibinfo
		{author} {\bibfnamefont {N.~M.}\ \bibnamefont {Linke}}, \bibinfo {author}
		{\bibfnamefont {B.}~\bibnamefont {Yoshida}}, \bibinfo {author} {\bibfnamefont
			{N.~Y.}\ \bibnamefont {Yao}},\ and\ \bibinfo {author} {\bibfnamefont
			{C.}~\bibnamefont {Monroe}},\ }\bibfield  {title} {\bibinfo {title}
		{{Verified quantum information scrambling}},\ }\href
	{https://doi.org/10.1038/s41586-019-0952-6} {\bibfield  {journal} {\bibinfo
			{journal} {Nature}\ }\textbf {\bibinfo {volume} {567}},\ \bibinfo {pages}
		{61} (\bibinfo {year} {2019})}\BibitemShut {NoStop}%
	\bibitem [{\citenamefont {Brydges}\ \emph {et~al.}(2019)\citenamefont
		{Brydges}, \citenamefont {Elben}, \citenamefont {Jurcevic}, \citenamefont
		{Vermersch}, \citenamefont {Maier}, \citenamefont {Lanyon}, \citenamefont
		{Zoller}, \citenamefont {Blatt},\ and\ \citenamefont {Roos}}]{Brydges2019a}%
	\BibitemOpen
	\bibfield  {author} {\bibinfo {author} {\bibfnamefont {T.}~\bibnamefont
			{Brydges}}, \bibinfo {author} {\bibfnamefont {A.}~\bibnamefont {Elben}},
		\bibinfo {author} {\bibfnamefont {P.}~\bibnamefont {Jurcevic}}, \bibinfo
		{author} {\bibfnamefont {B.}~\bibnamefont {Vermersch}}, \bibinfo {author}
		{\bibfnamefont {C.}~\bibnamefont {Maier}}, \bibinfo {author} {\bibfnamefont
			{B.~P.}\ \bibnamefont {Lanyon}}, \bibinfo {author} {\bibfnamefont
			{P.}~\bibnamefont {Zoller}}, \bibinfo {author} {\bibfnamefont
			{R.}~\bibnamefont {Blatt}},\ and\ \bibinfo {author} {\bibfnamefont {C.~F.}\
			\bibnamefont {Roos}},\ }\bibfield  {title} {\bibinfo {title} {{Probing
				R{\'{e}}nyi entanglement entropy via randomized measurements}},\ }\href
	{https://doi.org/10.1126/science.aau4963} {\bibfield  {journal} {\bibinfo
			{journal} {Science}\ }\textbf {\bibinfo {volume} {364}},\ \bibinfo {pages}
		{260} (\bibinfo {year} {2019})}\BibitemShut {NoStop}%
	\bibitem [{\citenamefont {Krojanski}\ and\ \citenamefont
		{Suter}(2004)}]{Krojanski2004a}%
	\BibitemOpen
	\bibfield  {author} {\bibinfo {author} {\bibfnamefont {H.~G.}\ \bibnamefont
			{Krojanski}}\ and\ \bibinfo {author} {\bibfnamefont {D.}~\bibnamefont
			{Suter}},\ }\bibfield  {title} {\bibinfo {title} {{Scaling of Decoherence in
				Wide NMR Quantum Registers}},\ }\href
	{https://doi.org/10.1103/PhysRevLett.93.090501} {\bibfield  {journal}
		{\bibinfo  {journal} {Phys. Rev. Lett.}\ }\textbf {\bibinfo {volume} {93}},\
		\bibinfo {pages} {090501} (\bibinfo {year} {2004})}\BibitemShut {NoStop}%
	\bibitem [{\citenamefont {{\'{A}}lvarez}\ and\ \citenamefont
		{Suter}(2010)}]{alvarez2010nmr}%
	\BibitemOpen
	\bibfield  {author} {\bibinfo {author} {\bibfnamefont {G.~A.}\ \bibnamefont
			{{\'{A}}lvarez}}\ and\ \bibinfo {author} {\bibfnamefont {D.}~\bibnamefont
			{Suter}},\ }\bibfield  {title} {\bibinfo {title} {{NMR} quantum simulation of
			localization effects induced by decoherence},\ }\href
	{https://doi.org/10.1103/physrevlett.104.230403} {\bibfield  {journal}
		{\bibinfo  {journal} {Phys. Rev. Lett.}\ }\textbf {\bibinfo {volume} {104}},\
		\bibinfo {pages} {230403} (\bibinfo {year} {2010})}\BibitemShut {NoStop}%
	\bibitem [{\citenamefont {S\'anchez}\ \emph {et~al.}(2014)\citenamefont
		{S\'anchez}, \citenamefont {Acosta}, \citenamefont {Levstein}, \citenamefont
		{Pastawski},\ and\ \citenamefont {Chattah}}]{sanchez2014}%
	\BibitemOpen
	\bibfield  {author} {\bibinfo {author} {\bibfnamefont {C.~M.}\ \bibnamefont
			{S\'anchez}}, \bibinfo {author} {\bibfnamefont {R.~H.}\ \bibnamefont
			{Acosta}}, \bibinfo {author} {\bibfnamefont {P.~R.}\ \bibnamefont
			{Levstein}}, \bibinfo {author} {\bibfnamefont {H.~M.}\ \bibnamefont
			{Pastawski}},\ and\ \bibinfo {author} {\bibfnamefont {A.~K.}\ \bibnamefont
			{Chattah}},\ }\bibfield  {title} {\bibinfo {title} {Clustering and
			decoherence of correlated spins under double quantum dynamics},\ }\href
	{https://doi.org/10.1103/PhysRevA.90.042122} {\bibfield  {journal} {\bibinfo
			{journal} {Phys. Rev. A}\ }\textbf {\bibinfo {volume} {90}},\ \bibinfo
		{pages} {042122} (\bibinfo {year} {2014})}\BibitemShut {NoStop}%
	\bibitem [{\citenamefont {Niknam}\ \emph {et~al.}(2020)\citenamefont {Niknam},
		\citenamefont {Santos},\ and\ \citenamefont
		{Cory}}]{niknam_sensitivity_2020}%
	\BibitemOpen
	\bibfield  {author} {\bibinfo {author} {\bibfnamefont {M.}~\bibnamefont
			{Niknam}}, \bibinfo {author} {\bibfnamefont {L.~F.}\ \bibnamefont {Santos}},\
		and\ \bibinfo {author} {\bibfnamefont {D.~G.}\ \bibnamefont {Cory}},\
	}\bibfield  {title} {\bibinfo {title} {{Sensitivity of quantum information to
				environment perturbations measured with a nonlocal out-of-time-order
				correlation function}},\ }\href
	{https://doi.org/10.1103/PhysRevResearch.2.013200} {\bibfield  {journal}
		{\bibinfo  {journal} {Phys. Rev. Research}\ }\textbf {\bibinfo {volume}
			{2}},\ \bibinfo {pages} {13200} (\bibinfo {year} {2020})}\BibitemShut
	{NoStop}%
	\bibitem [{\citenamefont {Dom\'{\i}nguez}\ \emph {et~al.}(2021)\citenamefont
		{Dom\'{\i}nguez}, \citenamefont {Rodr\'{\i}guez}, \citenamefont {Kaiser},
		\citenamefont {Suter},\ and\ \citenamefont {\'Alvarez}}]{Dominguez}%
	\BibitemOpen
	\bibfield  {author} {\bibinfo {author} {\bibfnamefont {F.~D.}\ \bibnamefont
			{Dom\'{\i}nguez}}, \bibinfo {author} {\bibfnamefont {M.~C.}\ \bibnamefont
			{Rodr\'{\i}guez}}, \bibinfo {author} {\bibfnamefont {R.}~\bibnamefont
			{Kaiser}}, \bibinfo {author} {\bibfnamefont {D.}~\bibnamefont {Suter}},\ and\
		\bibinfo {author} {\bibfnamefont {G.~A.}\ \bibnamefont {\'Alvarez}},\
	}\bibfield  {title} {\bibinfo {title} {Decoherence scaling transition in the
			dynamics of quantum information scrambling},\ }\href
	{https://doi.org/10.1103/PhysRevA.104.012402} {\bibfield  {journal} {\bibinfo
			{journal} {Phys. Rev. A}\ }\textbf {\bibinfo {volume} {104}},\ \bibinfo
		{pages} {012402} (\bibinfo {year} {2021})}\BibitemShut {NoStop}%
	\bibitem [{\citenamefont {Hosur}\ \emph {et~al.}(2016)\citenamefont {Hosur},
		\citenamefont {Qi}, \citenamefont {Roberts},\ and\ \citenamefont
		{Yoshida}}]{Hosur2016}%
	\BibitemOpen
	\bibfield  {author} {\bibinfo {author} {\bibfnamefont {P.}~\bibnamefont
			{Hosur}}, \bibinfo {author} {\bibfnamefont {X.-L.}\ \bibnamefont {Qi}},
		\bibinfo {author} {\bibfnamefont {D.~A.}\ \bibnamefont {Roberts}},\ and\
		\bibinfo {author} {\bibfnamefont {B.}~\bibnamefont {Yoshida}},\ }\bibfield
	{title} {\bibinfo {title} {{Chaos in quantum channels}},\ }\href
	{https://doi.org/10.1007/JHEP02(2016)004} {\bibfield  {journal} {\bibinfo
			{journal} {J. High Energy Phys.}\ }\textbf {\bibinfo {volume} {2016}},\
		\bibinfo {pages} {4}}\BibitemShut {NoStop}%
	\bibitem [{\citenamefont {Garttner}\ \emph {et~al.}(2017)\citenamefont
		{Garttner}, \citenamefont {Bohnet}, \citenamefont {Safavi-Naini},
		\citenamefont {Wall}, \citenamefont {Bollinger},\ and\ \citenamefont
		{Rey}}]{Garttner2017}%
	\BibitemOpen
	\bibfield  {author} {\bibinfo {author} {\bibfnamefont {M.}~\bibnamefont
			{Garttner}}, \bibinfo {author} {\bibfnamefont {J.~G.}\ \bibnamefont
			{Bohnet}}, \bibinfo {author} {\bibfnamefont {A.}~\bibnamefont
			{Safavi-Naini}}, \bibinfo {author} {\bibfnamefont {M.~L.}\ \bibnamefont
			{Wall}}, \bibinfo {author} {\bibfnamefont {J.~J.}\ \bibnamefont
			{Bollinger}},\ and\ \bibinfo {author} {\bibfnamefont {A.~M.}\ \bibnamefont
			{Rey}},\ }\bibfield  {title} {\bibinfo {title} {{Measuring out-of-time-order
				correlations and multiple quantum spectra in a trapped-ion quantum magnet}},\
	}\href {https://doi.org/10.1038/NPHYS4119} {\bibfield  {journal} {\bibinfo
			{journal} {Nat. Phys.}\ }\textbf {\bibinfo {volume} {13}},\ \bibinfo {pages}
		{781} (\bibinfo {year} {2017})}\BibitemShut {NoStop}%
	\bibitem [{\citenamefont {Li}\ \emph {et~al.}(2017)\citenamefont {Li},
		\citenamefont {Fan}, \citenamefont {Wang}, \citenamefont {Ye}, \citenamefont
		{Zeng}, \citenamefont {Zhai}, \citenamefont {Peng},\ and\ \citenamefont
		{Du}}]{Li2017}%
	\BibitemOpen
	\bibfield  {author} {\bibinfo {author} {\bibfnamefont {J.}~\bibnamefont
			{Li}}, \bibinfo {author} {\bibfnamefont {R.}~\bibnamefont {Fan}}, \bibinfo
		{author} {\bibfnamefont {H.}~\bibnamefont {Wang}}, \bibinfo {author}
		{\bibfnamefont {B.}~\bibnamefont {Ye}}, \bibinfo {author} {\bibfnamefont
			{B.}~\bibnamefont {Zeng}}, \bibinfo {author} {\bibfnamefont {H.}~\bibnamefont
			{Zhai}}, \bibinfo {author} {\bibfnamefont {X.}~\bibnamefont {Peng}},\ and\
		\bibinfo {author} {\bibfnamefont {J.}~\bibnamefont {Du}},\ }\bibfield
	{title} {\bibinfo {title} {{Measuring out-of-time-order correlators on a
				nuclear magnetic resonance quantum simulator}},\ }\href
	{https://doi.org/10.1103/PhysRevX.7.031011} {\bibfield  {journal} {\bibinfo
			{journal} {Phys. Rev. X}\ }\textbf {\bibinfo {volume} {7}},\ \bibinfo {pages}
		{031011} (\bibinfo {year} {2017})}\BibitemShut {NoStop}%
	\bibitem [{\citenamefont {Roberts}\ \emph {et~al.}(2015)\citenamefont
		{Roberts}, \citenamefont {Stanford},\ and\ \citenamefont
		{Susskind}}]{Roberts2015}%
	\BibitemOpen
	\bibfield  {author} {\bibinfo {author} {\bibfnamefont {D.~A.}\ \bibnamefont
			{Roberts}}, \bibinfo {author} {\bibfnamefont {D.}~\bibnamefont {Stanford}},\
		and\ \bibinfo {author} {\bibfnamefont {L.}~\bibnamefont {Susskind}},\
	}\bibfield  {title} {\bibinfo {title} {{Localized shocks}},\ }\href
	{https://doi.org/10.1007/JHEP03(2015)051} {\bibfield  {journal} {\bibinfo
			{journal} {J. High Energy Phys.}\ }\textbf {\bibinfo {volume} {2015}},\
		\bibinfo {pages} {1}}\BibitemShut {NoStop}%
	\bibitem [{\citenamefont {Larkin}\ and\ \citenamefont
		{Ovchinnikov}(1969)}]{larkin1969quasiclassical}%
	\BibitemOpen
	\bibfield  {author} {\bibinfo {author} {\bibfnamefont {A.}~\bibnamefont
			{Larkin}}\ and\ \bibinfo {author} {\bibfnamefont {Y.~N.}\ \bibnamefont
			{Ovchinnikov}},\ }\bibfield  {title} {\bibinfo {title} {Quasiclassical method
			in the theory of superconductivity},\ }\href@noop {} {\bibfield  {journal}
		{\bibinfo  {journal} {Sov. Phys. JETP}\ }\textbf {\bibinfo {volume} {28}},\
		\bibinfo {pages} {1200} (\bibinfo {year} {1969})}\BibitemShut {NoStop}%
	\bibitem [{\citenamefont {Shenker}\ and\ \citenamefont
		{Stanford}(2014)}]{Shenker2014}%
	\BibitemOpen
	\bibfield  {author} {\bibinfo {author} {\bibfnamefont {S.~H.}\ \bibnamefont
			{Shenker}}\ and\ \bibinfo {author} {\bibfnamefont {D.}~\bibnamefont
			{Stanford}},\ }\bibfield  {title} {\bibinfo {title} {Black holes and the
			butterfly effect},\ }\href {https://doi.org/10.1007/JHEP03(2014)067}
	{\bibfield  {journal} {\bibinfo  {journal} {J. High Energy Phys.}\ }\textbf
		{\bibinfo {volume} {2014}},\ \bibinfo {pages} {67}}\BibitemShut {NoStop}%
	\bibitem [{\citenamefont {Maldacena}\ \emph {et~al.}(2016)\citenamefont
		{Maldacena}, \citenamefont {Shenker},\ and\ \citenamefont
		{Stanford}}]{maldacena2016bound}%
	\BibitemOpen
	\bibfield  {author} {\bibinfo {author} {\bibfnamefont {J.}~\bibnamefont
			{Maldacena}}, \bibinfo {author} {\bibfnamefont {S.~H.}\ \bibnamefont
			{Shenker}},\ and\ \bibinfo {author} {\bibfnamefont {D.}~\bibnamefont
			{Stanford}},\ }\bibfield  {title} {\bibinfo {title} {{A bound on chaos}},\
	}\href {https://doi.org/10.1007/jhep08(2016)106} {\bibfield  {journal}
		{\bibinfo  {journal} {J. High Energy Phys.}\ }\textbf {\bibinfo {volume}
			{2016}},\ \bibinfo {pages} {106}}\BibitemShut {NoStop}%
	\bibitem [{\citenamefont {Garc\'{\i}a-Mata}\ \emph {et~al.}(2018)\citenamefont
		{Garc\'{\i}a-Mata}, \citenamefont {Saraceno}, \citenamefont {Jalabert},
		\citenamefont {Roncaglia},\ and\ \citenamefont
		{Wisniacki}}]{garcia2018_chaos}%
	\BibitemOpen
	\bibfield  {author} {\bibinfo {author} {\bibfnamefont {I.}~\bibnamefont
			{Garc\'{\i}a-Mata}}, \bibinfo {author} {\bibfnamefont {M.}~\bibnamefont
			{Saraceno}}, \bibinfo {author} {\bibfnamefont {R.~A.}\ \bibnamefont
			{Jalabert}}, \bibinfo {author} {\bibfnamefont {A.~J.}\ \bibnamefont
			{Roncaglia}},\ and\ \bibinfo {author} {\bibfnamefont {D.~A.}\ \bibnamefont
			{Wisniacki}},\ }\bibfield  {title} {\bibinfo {title} {Chaos signatures in the
			short and long time behavior of the out-of-time ordered correlator},\ }\href
	{https://doi.org/10.1103/PhysRevLett.121.210601} {\bibfield  {journal}
		{\bibinfo  {journal} {Phys. Rev. Lett.}\ }\textbf {\bibinfo {volume} {121}},\
		\bibinfo {pages} {210601} (\bibinfo {year} {2018})}\BibitemShut {NoStop}%
	\bibitem [{\citenamefont {Wei}\ \emph {et~al.}(2018)\citenamefont {Wei},
		\citenamefont {Ramanathan},\ and\ \citenamefont {Cappellaro}}]{Wei2018}%
	\BibitemOpen
	\bibfield  {author} {\bibinfo {author} {\bibfnamefont {K.~X.}\ \bibnamefont
			{Wei}}, \bibinfo {author} {\bibfnamefont {C.}~\bibnamefont {Ramanathan}},\
		and\ \bibinfo {author} {\bibfnamefont {P.}~\bibnamefont {Cappellaro}},\
	}\bibfield  {title} {\bibinfo {title} {{Exploring Localization in Nuclear
				Spin Chains}},\ }\href {https://doi.org/10.1103/PhysRevLett.120.070501}
	{\bibfield  {journal} {\bibinfo  {journal} {Phys. Rev. Lett.}\ }\textbf
		{\bibinfo {volume} {120}},\ \bibinfo {pages} {070501} (\bibinfo {year}
		{2018})}\BibitemShut {NoStop}%
	\bibitem [{\citenamefont {Wei}\ \emph {et~al.}(2019)\citenamefont {Wei},
		\citenamefont {Peng}, \citenamefont {Shtanko}, \citenamefont {Marvian},
		\citenamefont {Lloyd}, \citenamefont {Ramanathan},\ and\ \citenamefont
		{Cappellaro}}]{Wei2019}%
	\BibitemOpen
	\bibfield  {author} {\bibinfo {author} {\bibfnamefont {K.~X.}\ \bibnamefont
			{Wei}}, \bibinfo {author} {\bibfnamefont {P.}~\bibnamefont {Peng}}, \bibinfo
		{author} {\bibfnamefont {O.}~\bibnamefont {Shtanko}}, \bibinfo {author}
		{\bibfnamefont {I.}~\bibnamefont {Marvian}}, \bibinfo {author} {\bibfnamefont
			{S.}~\bibnamefont {Lloyd}}, \bibinfo {author} {\bibfnamefont
			{C.}~\bibnamefont {Ramanathan}},\ and\ \bibinfo {author} {\bibfnamefont
			{P.}~\bibnamefont {Cappellaro}},\ }\bibfield  {title} {\bibinfo {title}
		{{Emergent Prethermalization Signatures in Out-of-Time Ordered
				Correlations}},\ }\href {https://doi.org/10.1103/PhysRevLett.123.090605}
	{\bibfield  {journal} {\bibinfo  {journal} {Phys. Rev. Lett.}\ }\textbf
		{\bibinfo {volume} {123}},\ \bibinfo {pages} {090605} (\bibinfo {year}
		{2019})}\BibitemShut {NoStop}%
	\bibitem [{\citenamefont {S{\'{a}}nchez}\ \emph {et~al.}(2020)\citenamefont
		{S{\'{a}}nchez}, \citenamefont {Chattah}, \citenamefont {Wei}, \citenamefont
		{Buljubasich}, \citenamefont {Cappellaro},\ and\ \citenamefont
		{Pastawski}}]{Sanchez2020}%
	\BibitemOpen
	\bibfield  {author} {\bibinfo {author} {\bibfnamefont {C.~M.}\ \bibnamefont
			{S{\'{a}}nchez}}, \bibinfo {author} {\bibfnamefont {A.~K.}\ \bibnamefont
			{Chattah}}, \bibinfo {author} {\bibfnamefont {K.~X.}\ \bibnamefont {Wei}},
		\bibinfo {author} {\bibfnamefont {L.}~\bibnamefont {Buljubasich}}, \bibinfo
		{author} {\bibfnamefont {P.}~\bibnamefont {Cappellaro}},\ and\ \bibinfo
		{author} {\bibfnamefont {H.~M.}\ \bibnamefont {Pastawski}},\ }\bibfield
	{title} {\bibinfo {title} {{Perturbation Independent Decay of the Loschmidt
				Echo in a Many-Body System}},\ }\href
	{https://doi.org/10.1103/PhysRevLett.124.030601} {\bibfield  {journal}
		{\bibinfo  {journal} {Phys. Rev. Lett.}\ }\textbf {\bibinfo {volume} {124}},\
		\bibinfo {pages} {030601} (\bibinfo {year} {2020})}\BibitemShut {NoStop}%
	\bibitem [{\citenamefont {Joshi}\ \emph {et~al.}(2020)\citenamefont {Joshi},
		\citenamefont {Elben}, \citenamefont {Vermersch}, \citenamefont {Brydges},
		\citenamefont {Maier}, \citenamefont {Zoller}, \citenamefont {Blatt},\ and\
		\citenamefont {Roos}}]{Joshi2020}%
	\BibitemOpen
	\bibfield  {author} {\bibinfo {author} {\bibfnamefont {M.~K.}\ \bibnamefont
			{Joshi}}, \bibinfo {author} {\bibfnamefont {A.}~\bibnamefont {Elben}},
		\bibinfo {author} {\bibfnamefont {B.}~\bibnamefont {Vermersch}}, \bibinfo
		{author} {\bibfnamefont {T.}~\bibnamefont {Brydges}}, \bibinfo {author}
		{\bibfnamefont {C.}~\bibnamefont {Maier}}, \bibinfo {author} {\bibfnamefont
			{P.}~\bibnamefont {Zoller}}, \bibinfo {author} {\bibfnamefont
			{R.}~\bibnamefont {Blatt}},\ and\ \bibinfo {author} {\bibfnamefont {C.~F.}\
			\bibnamefont {Roos}},\ }\bibfield  {title} {\bibinfo {title} {{Quantum
				Information Scrambling in a Trapped-Ion Quantum Simulator with Tunable Range
				Interactions}},\ }\href {https://doi.org/10.1103/PhysRevLett.124.240505}
	{\bibfield  {journal} {\bibinfo  {journal} {Phys. Rev. Lett.}\ }\textbf
		{\bibinfo {volume} {124}},\ \bibinfo {pages} {240505} (\bibinfo {year}
		{2020})}\BibitemShut {NoStop}%
	\bibitem [{\citenamefont {Nie}\ \emph {et~al.}(2020)\citenamefont {Nie},
		\citenamefont {Wei}, \citenamefont {Chen}, \citenamefont {Zhang},
		\citenamefont {Zhao}, \citenamefont {Qiu}, \citenamefont {Tian},
		\citenamefont {Ji}, \citenamefont {Xin}, \citenamefont {Lu},\ and\
		\citenamefont {Li}}]{Nie2020}%
	\BibitemOpen
	\bibfield  {author} {\bibinfo {author} {\bibfnamefont {X.}~\bibnamefont
			{Nie}}, \bibinfo {author} {\bibfnamefont {B.~B.}\ \bibnamefont {Wei}},
		\bibinfo {author} {\bibfnamefont {X.}~\bibnamefont {Chen}}, \bibinfo {author}
		{\bibfnamefont {Z.}~\bibnamefont {Zhang}}, \bibinfo {author} {\bibfnamefont
			{X.}~\bibnamefont {Zhao}}, \bibinfo {author} {\bibfnamefont {C.}~\bibnamefont
			{Qiu}}, \bibinfo {author} {\bibfnamefont {Y.}~\bibnamefont {Tian}}, \bibinfo
		{author} {\bibfnamefont {Y.}~\bibnamefont {Ji}}, \bibinfo {author}
		{\bibfnamefont {T.}~\bibnamefont {Xin}}, \bibinfo {author} {\bibfnamefont
			{D.}~\bibnamefont {Lu}},\ and\ \bibinfo {author} {\bibfnamefont
			{J.}~\bibnamefont {Li}},\ }\bibfield  {title} {\bibinfo {title}
		{{Experimental Observation of Equilibrium and Dynamical Quantum Phase
				Transitions via Out-of-Time-Ordered Correlators}},\ }\href
	{https://doi.org/10.1103/PhysRevLett.124.250601} {\bibfield  {journal}
		{\bibinfo  {journal} {Phys. Rev. Lett.}\ }\textbf {\bibinfo {volume} {124}},\
		\bibinfo {pages} {250601} (\bibinfo {year} {2020})}\BibitemShut {NoStop}%
	\bibitem [{\citenamefont {Mi}\ \emph {et~al.}(2021)\citenamefont {Mi},
		\citenamefont {Roushan}, \citenamefont {Quintana}, \citenamefont {Mandra},
		\citenamefont {Marshall}, \citenamefont {Neill}, \citenamefont {Arute},
		\citenamefont {Arya}, \citenamefont {Atalaya}, \citenamefont {Babbush},
		\citenamefont {Bardin}, \citenamefont {Barends}, \citenamefont {Bengtsson},
		\citenamefont {Boixo}, \citenamefont {Bourassa}, \citenamefont {Broughton},
		\citenamefont {Buckley}, \citenamefont {Buell}, \citenamefont {Burkett},
		\citenamefont {Bushnell}, \citenamefont {Chen}, \citenamefont {Chiaro},
		\citenamefont {Collins}, \citenamefont {Courtney}, \citenamefont {Demura},
		\citenamefont {Derk}, \citenamefont {Dunsworth}, \citenamefont {Eppens},
		\citenamefont {Erickson}, \citenamefont {Farhi}, \citenamefont {Fowler},
		\citenamefont {Foxen}, \citenamefont {Gidney}, \citenamefont {Giustina},
		\citenamefont {Gross}, \citenamefont {Harrigan}, \citenamefont {Harrington},
		\citenamefont {Hilton}, \citenamefont {Ho}, \citenamefont {Hong},
		\citenamefont {Huang}, \citenamefont {Huggins}, \citenamefont {Ioffe},
		\citenamefont {Isakov}, \citenamefont {Jeffrey}, \citenamefont {Jiang},
		\citenamefont {Jones}, \citenamefont {Kafri}, \citenamefont {Kelly},
		\citenamefont {Kim}, \citenamefont {Kitaev}, \citenamefont {Klimov},
		\citenamefont {Korotkov}, \citenamefont {Kostritsa}, \citenamefont
		{Landhuis}, \citenamefont {Laptev}, \citenamefont {Lucero}, \citenamefont
		{Martin}, \citenamefont {McClean}, \citenamefont {McCourt}, \citenamefont
		{McEwen}, \citenamefont {Megrant}, \citenamefont {Miao}, \citenamefont
		{Mohseni}, \citenamefont {Mruczkiewicz}, \citenamefont {Mutus}, \citenamefont
		{Naaman}, \citenamefont {Neeley}, \citenamefont {Newman}, \citenamefont
		{Niu}, \citenamefont {O'Brien}, \citenamefont {Opremcak}, \citenamefont
		{Ostby}, \citenamefont {Pato}, \citenamefont {Petukhov}, \citenamefont
		{Redd}, \citenamefont {Rubin}, \citenamefont {Sank}, \citenamefont
		{Satzinger}, \citenamefont {Shvarts}, \citenamefont {Strain}, \citenamefont
		{Szalay}, \citenamefont {Trevithick}, \citenamefont {Villalonga},
		\citenamefont {White}, \citenamefont {Yao}, \citenamefont {Yeh},
		\citenamefont {Zalcman}, \citenamefont {Neven}, \citenamefont {Aleiner},
		\citenamefont {Kechedzhi}, \citenamefont {Smelyanskiy},\ and\ \citenamefont
		{Chen}}]{Mi2021}%
	\BibitemOpen
	\bibfield  {author} {\bibinfo {author} {\bibfnamefont {X.}~\bibnamefont
			{Mi}}, \bibinfo {author} {\bibfnamefont {P.}~\bibnamefont {Roushan}},
		\bibinfo {author} {\bibfnamefont {C.}~\bibnamefont {Quintana}}, \bibinfo
		{author} {\bibfnamefont {S.}~\bibnamefont {Mandra}}, \bibinfo {author}
		{\bibfnamefont {J.}~\bibnamefont {Marshall}}, \bibinfo {author}
		{\bibfnamefont {C.}~\bibnamefont {Neill}}, \bibinfo {author} {\bibfnamefont
			{F.}~\bibnamefont {Arute}}, \bibinfo {author} {\bibfnamefont
			{K.}~\bibnamefont {Arya}}, \bibinfo {author} {\bibfnamefont {J.}~\bibnamefont
			{Atalaya}}, \bibinfo {author} {\bibfnamefont {R.}~\bibnamefont {Babbush}},
		\bibinfo {author} {\bibfnamefont {J.~C.}\ \bibnamefont {Bardin}}, \bibinfo
		{author} {\bibfnamefont {R.}~\bibnamefont {Barends}}, \bibinfo {author}
		{\bibfnamefont {A.}~\bibnamefont {Bengtsson}}, \bibinfo {author}
		{\bibfnamefont {S.}~\bibnamefont {Boixo}}, \bibinfo {author} {\bibfnamefont
			{A.}~\bibnamefont {Bourassa}}, \bibinfo {author} {\bibfnamefont
			{M.}~\bibnamefont {Broughton}}, \bibinfo {author} {\bibfnamefont {B.~B.}\
			\bibnamefont {Buckley}}, \bibinfo {author} {\bibfnamefont {D.~A.}\
			\bibnamefont {Buell}}, \bibinfo {author} {\bibfnamefont {B.}~\bibnamefont
			{Burkett}}, \bibinfo {author} {\bibfnamefont {N.}~\bibnamefont {Bushnell}},
		\bibinfo {author} {\bibfnamefont {Z.}~\bibnamefont {Chen}}, \bibinfo {author}
		{\bibfnamefont {B.}~\bibnamefont {Chiaro}}, \bibinfo {author} {\bibfnamefont
			{R.}~\bibnamefont {Collins}}, \bibinfo {author} {\bibfnamefont
			{W.}~\bibnamefont {Courtney}}, \bibinfo {author} {\bibfnamefont
			{S.}~\bibnamefont {Demura}}, \bibinfo {author} {\bibfnamefont {A.~R.}\
			\bibnamefont {Derk}}, \bibinfo {author} {\bibfnamefont {A.}~\bibnamefont
			{Dunsworth}}, \bibinfo {author} {\bibfnamefont {D.}~\bibnamefont {Eppens}},
		\bibinfo {author} {\bibfnamefont {C.}~\bibnamefont {Erickson}}, \bibinfo
		{author} {\bibfnamefont {E.}~\bibnamefont {Farhi}}, \bibinfo {author}
		{\bibfnamefont {A.~G.}\ \bibnamefont {Fowler}}, \bibinfo {author}
		{\bibfnamefont {B.}~\bibnamefont {Foxen}}, \bibinfo {author} {\bibfnamefont
			{C.}~\bibnamefont {Gidney}}, \bibinfo {author} {\bibfnamefont
			{M.}~\bibnamefont {Giustina}}, \bibinfo {author} {\bibfnamefont {J.~A.}\
			\bibnamefont {Gross}}, \bibinfo {author} {\bibfnamefont {M.~P.}\ \bibnamefont
			{Harrigan}}, \bibinfo {author} {\bibfnamefont {S.~D.}\ \bibnamefont
			{Harrington}}, \bibinfo {author} {\bibfnamefont {J.}~\bibnamefont {Hilton}},
		\bibinfo {author} {\bibfnamefont {A.}~\bibnamefont {Ho}}, \bibinfo {author}
		{\bibfnamefont {S.}~\bibnamefont {Hong}}, \bibinfo {author} {\bibfnamefont
			{T.}~\bibnamefont {Huang}}, \bibinfo {author} {\bibfnamefont {W.~J.}\
			\bibnamefont {Huggins}}, \bibinfo {author} {\bibfnamefont {L.~B.}\
			\bibnamefont {Ioffe}}, \bibinfo {author} {\bibfnamefont {S.~V.}\ \bibnamefont
			{Isakov}}, \bibinfo {author} {\bibfnamefont {E.}~\bibnamefont {Jeffrey}},
		\bibinfo {author} {\bibfnamefont {Z.}~\bibnamefont {Jiang}}, \bibinfo
		{author} {\bibfnamefont {C.}~\bibnamefont {Jones}}, \bibinfo {author}
		{\bibfnamefont {D.}~\bibnamefont {Kafri}}, \bibinfo {author} {\bibfnamefont
			{J.}~\bibnamefont {Kelly}}, \bibinfo {author} {\bibfnamefont
			{S.}~\bibnamefont {Kim}}, \bibinfo {author} {\bibfnamefont {A.}~\bibnamefont
			{Kitaev}}, \bibinfo {author} {\bibfnamefont {P.~V.}\ \bibnamefont {Klimov}},
		\bibinfo {author} {\bibfnamefont {A.~N.}\ \bibnamefont {Korotkov}}, \bibinfo
		{author} {\bibfnamefont {F.}~\bibnamefont {Kostritsa}}, \bibinfo {author}
		{\bibfnamefont {D.}~\bibnamefont {Landhuis}}, \bibinfo {author}
		{\bibfnamefont {P.}~\bibnamefont {Laptev}}, \bibinfo {author} {\bibfnamefont
			{E.}~\bibnamefont {Lucero}}, \bibinfo {author} {\bibfnamefont
			{O.}~\bibnamefont {Martin}}, \bibinfo {author} {\bibfnamefont {J.~R.}\
			\bibnamefont {McClean}}, \bibinfo {author} {\bibfnamefont {T.}~\bibnamefont
			{McCourt}}, \bibinfo {author} {\bibfnamefont {M.}~\bibnamefont {McEwen}},
		\bibinfo {author} {\bibfnamefont {A.}~\bibnamefont {Megrant}}, \bibinfo
		{author} {\bibfnamefont {K.~C.}\ \bibnamefont {Miao}}, \bibinfo {author}
		{\bibfnamefont {M.}~\bibnamefont {Mohseni}}, \bibinfo {author} {\bibfnamefont
			{W.}~\bibnamefont {Mruczkiewicz}}, \bibinfo {author} {\bibfnamefont
			{J.}~\bibnamefont {Mutus}}, \bibinfo {author} {\bibfnamefont
			{O.}~\bibnamefont {Naaman}}, \bibinfo {author} {\bibfnamefont
			{M.}~\bibnamefont {Neeley}}, \bibinfo {author} {\bibfnamefont
			{M.}~\bibnamefont {Newman}}, \bibinfo {author} {\bibfnamefont {M.~Y.}\
			\bibnamefont {Niu}}, \bibinfo {author} {\bibfnamefont {T.~E.}\ \bibnamefont
			{O'Brien}}, \bibinfo {author} {\bibfnamefont {A.}~\bibnamefont {Opremcak}},
		\bibinfo {author} {\bibfnamefont {E.}~\bibnamefont {Ostby}}, \bibinfo
		{author} {\bibfnamefont {B.}~\bibnamefont {Pato}}, \bibinfo {author}
		{\bibfnamefont {A.}~\bibnamefont {Petukhov}}, \bibinfo {author}
		{\bibfnamefont {N.}~\bibnamefont {Redd}}, \bibinfo {author} {\bibfnamefont
			{N.~C.}\ \bibnamefont {Rubin}}, \bibinfo {author} {\bibfnamefont
			{D.}~\bibnamefont {Sank}}, \bibinfo {author} {\bibfnamefont {K.~J.}\
			\bibnamefont {Satzinger}}, \bibinfo {author} {\bibfnamefont {V.}~\bibnamefont
			{Shvarts}}, \bibinfo {author} {\bibfnamefont {D.}~\bibnamefont {Strain}},
		\bibinfo {author} {\bibfnamefont {M.}~\bibnamefont {Szalay}}, \bibinfo
		{author} {\bibfnamefont {M.~D.}\ \bibnamefont {Trevithick}}, \bibinfo
		{author} {\bibfnamefont {B.}~\bibnamefont {Villalonga}}, \bibinfo {author}
		{\bibfnamefont {T.}~\bibnamefont {White}}, \bibinfo {author} {\bibfnamefont
			{Z.~J.}\ \bibnamefont {Yao}}, \bibinfo {author} {\bibfnamefont
			{P.}~\bibnamefont {Yeh}}, \bibinfo {author} {\bibfnamefont {A.}~\bibnamefont
			{Zalcman}}, \bibinfo {author} {\bibfnamefont {H.}~\bibnamefont {Neven}},
		\bibinfo {author} {\bibfnamefont {I.}~\bibnamefont {Aleiner}}, \bibinfo
		{author} {\bibfnamefont {K.}~\bibnamefont {Kechedzhi}}, \bibinfo {author}
		{\bibfnamefont {V.}~\bibnamefont {Smelyanskiy}},\ and\ \bibinfo {author}
		{\bibfnamefont {Y.}~\bibnamefont {Chen}},\ }\href@noop {} {\bibinfo {title}
		{Information scrambling in computationally complex quantum circuits}}
	(\bibinfo {year} {2021}),\ \Eprint {https://arxiv.org/abs/2101.08870}
	{arXiv:2101.08870} \BibitemShut {NoStop}%
	\bibitem [{\citenamefont {Peres}(1984)}]{peres1984stability}%
	\BibitemOpen
	\bibfield  {author} {\bibinfo {author} {\bibfnamefont {A.}~\bibnamefont
			{Peres}},\ }\bibfield  {title} {\bibinfo {title} {{Stability of quantum
				motion in chaotic and regular systems}},\ }\href
	{https://doi.org/10.1103/physreva.30.1610} {\bibfield  {journal} {\bibinfo
			{journal} {Phys. Rev. A}\ }\textbf {\bibinfo {volume} {30}},\ \bibinfo
		{pages} {1610} (\bibinfo {year} {1984})}\BibitemShut {NoStop}%
	\bibitem [{\citenamefont {Jalabert}\ and\ \citenamefont
		{Pastawski}(2001)}]{Jalabert2001}%
	\BibitemOpen
	\bibfield  {author} {\bibinfo {author} {\bibfnamefont {R.~A.}\ \bibnamefont
			{Jalabert}}\ and\ \bibinfo {author} {\bibfnamefont {H.~M.}\ \bibnamefont
			{Pastawski}},\ }\bibfield  {title} {\bibinfo {title}
		{{Environment-independent decoherence rate in classically chaotic systems}},\
	}\href {https://doi.org/10.1103/PhysRevLett.86.2490} {\bibfield  {journal}
		{\bibinfo  {journal} {Phys. Rev. Lett.}\ }\textbf {\bibinfo {volume} {86}},\
		\bibinfo {pages} {2490} (\bibinfo {year} {2001})}\BibitemShut {NoStop}%
	\bibitem [{\citenamefont {Jacquod}\ and\ \citenamefont
		{Petitjean}(2009)}]{Jacquod2009a}%
	\BibitemOpen
	\bibfield  {author} {\bibinfo {author} {\bibfnamefont {P.}~\bibnamefont
			{Jacquod}}\ and\ \bibinfo {author} {\bibfnamefont {C.}~\bibnamefont
			{Petitjean}},\ }\bibfield  {title} {\bibinfo {title} {{Decoherence,
				entanglement and irreversibility in quantum dynamical systems with few
				degrees of freedom}},\ }\href {https://doi.org/10.1080/00018730902831009}
	{\bibfield  {journal} {\bibinfo  {journal} {Adv. Phys.}\ }\textbf {\bibinfo
			{volume} {58}},\ \bibinfo {pages} {67} (\bibinfo {year} {2009})}\BibitemShut
	{NoStop}%
	\bibitem [{\citenamefont {Gorin}\ \emph {et~al.}(2006)\citenamefont {Gorin},
		\citenamefont {Prosen}, \citenamefont {Seligman},\ and\ \citenamefont
		{{\v{Z}}nidari{\v{c}}}}]{Gorin2006}%
	\BibitemOpen
	\bibfield  {author} {\bibinfo {author} {\bibfnamefont {T.}~\bibnamefont
			{Gorin}}, \bibinfo {author} {\bibfnamefont {T.}~\bibnamefont {Prosen}},
		\bibinfo {author} {\bibfnamefont {T.~H.}\ \bibnamefont {Seligman}},\ and\
		\bibinfo {author} {\bibfnamefont {M.}~\bibnamefont {{\v{Z}}nidari{\v{c}}}},\
	}\bibfield  {title} {\bibinfo {title} {{Dynamics of Loschmidt echoes and
				fidelity decay}},\ }\href {https://doi.org/10.1016/j.physrep.2006.09.003}
	{\bibfield  {journal} {\bibinfo  {journal} {Phys. Rep.}\ }\textbf {\bibinfo
			{volume} {435}},\ \bibinfo {pages} {33} (\bibinfo {year} {2006})}\BibitemShut
	{NoStop}%
	\bibitem [{\citenamefont {Goussev}\ \emph {et~al.}(2012)\citenamefont
		{Goussev}, \citenamefont {Jalabert}, \citenamefont {Pastawski},\ and\
		\citenamefont {Wisniacki}}]{Goussev:2012}%
	\BibitemOpen
	\bibfield  {author} {\bibinfo {author} {\bibfnamefont {A.}~\bibnamefont
			{Goussev}}, \bibinfo {author} {\bibfnamefont {R.~A.}\ \bibnamefont
			{Jalabert}}, \bibinfo {author} {\bibfnamefont {H.~M.}\ \bibnamefont
			{Pastawski}},\ and\ \bibinfo {author} {\bibfnamefont {D.~A.}\ \bibnamefont
			{Wisniacki}},\ }\bibfield  {title} {\bibinfo {title} {{L}oschmidt echo},\
	}\href {https://doi.org/10.4249/scholarpedia.11687} {\bibfield  {journal}
		{\bibinfo  {journal} {Scholarpedia}\ }\textbf {\bibinfo {volume} {7}},\
		\bibinfo {pages} {11687} (\bibinfo {year} {2012})}\BibitemShut {NoStop}%
	\bibitem [{\citenamefont {Suter}\ and\ \citenamefont
		{{\'{A}}lvarez}(2016)}]{Suter2016}%
	\BibitemOpen
	\bibfield  {author} {\bibinfo {author} {\bibfnamefont {D.}~\bibnamefont
			{Suter}}\ and\ \bibinfo {author} {\bibfnamefont {G.~A.}\ \bibnamefont
			{{\'{A}}lvarez}},\ }\bibfield  {title} {\bibinfo {title} {{Colloquium:
				Protecting quantum information against environmental noise}},\ }\href
	{https://doi.org/10.1103/RevModPhys.88.041001} {\bibfield  {journal}
		{\bibinfo  {journal} {Rev. Mod. Phys.}\ }\textbf {\bibinfo {volume} {88}},\
		\bibinfo {pages} {041001} (\bibinfo {year} {2016})}\BibitemShut {NoStop}%
	\bibitem [{\citenamefont {Swingle}\ and\ \citenamefont {{Yunger
				Halpern}}(2018)}]{Swingle2018a}%
	\BibitemOpen
	\bibfield  {author} {\bibinfo {author} {\bibfnamefont {B.}~\bibnamefont
			{Swingle}}\ and\ \bibinfo {author} {\bibfnamefont {N.}~\bibnamefont {{Yunger
					Halpern}}},\ }\bibfield  {title} {\bibinfo {title} {{Resilience of scrambling
				measurements}},\ }\href {https://doi.org/10.1103/PhysRevA.97.062113}
	{\bibfield  {journal} {\bibinfo  {journal} {Phys. Rev. A}\ }\textbf {\bibinfo
			{volume} {97}},\ \bibinfo {pages} {062113} (\bibinfo {year}
		{2018})}\BibitemShut {NoStop}%
	\bibitem [{\citenamefont {Syzranov}\ \emph {et~al.}(2018)\citenamefont
		{Syzranov}, \citenamefont {Gorshkov},\ and\ \citenamefont
		{Galitski}}]{Syzranov2018_out}%
	\BibitemOpen
	\bibfield  {author} {\bibinfo {author} {\bibfnamefont {S.~V.}\ \bibnamefont
			{Syzranov}}, \bibinfo {author} {\bibfnamefont {A.~V.}\ \bibnamefont
			{Gorshkov}},\ and\ \bibinfo {author} {\bibfnamefont {V.}~\bibnamefont
			{Galitski}},\ }\bibfield  {title} {\bibinfo {title} {Out-of-time-order
			correlators in finite open systems},\ }\href
	{https://doi.org/10.1103/PhysRevB.97.161114} {\bibfield  {journal} {\bibinfo
			{journal} {Phys. Rev. B}\ }\textbf {\bibinfo {volume} {97}},\ \bibinfo
		{pages} {161114} (\bibinfo {year} {2018})}\BibitemShut {NoStop}%
	\bibitem [{\citenamefont {{Gonz{\'{a}}lez Alonso}}\ \emph
		{et~al.}(2019)\citenamefont {{Gonz{\'{a}}lez Alonso}}, \citenamefont {{Yunger
				Halpern}},\ and\ \citenamefont {Dressel}}]{GonzalezAlonso2019}%
	\BibitemOpen
	\bibfield  {author} {\bibinfo {author} {\bibfnamefont {J.~R.}\ \bibnamefont
			{{Gonz{\'{a}}lez Alonso}}}, \bibinfo {author} {\bibfnamefont
			{N.}~\bibnamefont {{Yunger Halpern}}},\ and\ \bibinfo {author} {\bibfnamefont
			{J.}~\bibnamefont {Dressel}},\ }\bibfield  {title} {\bibinfo {title}
		{{Out-of-Time-Ordered-Correlator Quasiprobabilities Robustly Witness
				Scrambling}},\ }\href {https://doi.org/10.1103/PhysRevLett.122.040404}
	{\bibfield  {journal} {\bibinfo  {journal} {Phys. Rev. Lett.}\ }\textbf
		{\bibinfo {volume} {122}},\ \bibinfo {pages} {040404} (\bibinfo {year}
		{2019})}\BibitemShut {NoStop}%
	\bibitem [{\citenamefont {Tuziemski}(2019)}]{Tuziemski2019}%
	\BibitemOpen
	\bibfield  {author} {\bibinfo {author} {\bibfnamefont {J.}~\bibnamefont
			{Tuziemski}},\ }\bibfield  {title} {\bibinfo {title} {{Out-of-time-ordered
				correlation functions in open systems: A Feynman-Vernon influence functional
				approach}},\ }\href {https://doi.org/10.1103/PhysRevA.100.062106} {\bibfield
		{journal} {\bibinfo  {journal} {Phys. Rev. A}\ }\textbf {\bibinfo {volume}
			{100}},\ \bibinfo {pages} {062106} (\bibinfo {year} {2019})}\BibitemShut
	{NoStop}%
	\bibitem [{\citenamefont {Zanardi}\ and\ \citenamefont
		{Anand}(2021)}]{zanardi2021information}%
	\BibitemOpen
	\bibfield  {author} {\bibinfo {author} {\bibfnamefont {P.}~\bibnamefont
			{Zanardi}}\ and\ \bibinfo {author} {\bibfnamefont {N.}~\bibnamefont
			{Anand}},\ }\bibfield  {title} {\bibinfo {title} {Information scrambling and
			chaos in open quantum systems},\ }\href
	{https://doi.org/10.1103/PhysRevA.103.062214} {\bibfield  {journal} {\bibinfo
			{journal} {Phys. Rev. A}\ }\textbf {\bibinfo {volume} {103}},\ \bibinfo
		{pages} {062214} (\bibinfo {year} {2021})}\BibitemShut {NoStop}%
	\bibitem [{\citenamefont {S{\o}rensen}\ \emph {et~al.}(1983)\citenamefont
		{S{\o}rensen}, \citenamefont {Levitt},\ and\ \citenamefont
		{Ernst}}]{Sorensen1983}%
	\BibitemOpen
	\bibfield  {author} {\bibinfo {author} {\bibfnamefont {O.~W.}\ \bibnamefont
			{S{\o}rensen}}, \bibinfo {author} {\bibfnamefont {M.~H.}\ \bibnamefont
			{Levitt}},\ and\ \bibinfo {author} {\bibfnamefont {R.~R.}\ \bibnamefont
			{Ernst}},\ }\bibfield  {title} {\bibinfo {title} {{Uniform excitation of
				multiple-quantum coherence: Application to multiple quantum filtering}},\
	}\href {https://doi.org/10.1016/0022-2364(83)90280-9} {\bibfield  {journal}
		{\bibinfo  {journal} {J. Magn. Reson.}\ }\textbf {\bibinfo {volume} {55}},\
		\bibinfo {pages} {104} (\bibinfo {year} {1983})}\BibitemShut {NoStop}%
	\bibitem [{\citenamefont {Griesinger}\ \emph {et~al.}(1986)\citenamefont
		{Griesinger}, \citenamefont {S{\o}rensen},\ and\ \citenamefont
		{Ernst}}]{Griesinger1986}%
	\BibitemOpen
	\bibfield  {author} {\bibinfo {author} {\bibfnamefont {C.}~\bibnamefont
			{Griesinger}}, \bibinfo {author} {\bibfnamefont {O.~W.}\ \bibnamefont
			{S{\o}rensen}},\ and\ \bibinfo {author} {\bibfnamefont {R.~R.}\ \bibnamefont
			{Ernst}},\ }\bibfield  {title} {\bibinfo {title} {{Correlation of connected
				transitions by two-dimensional NMR spectroscopy}},\ }\href
	{https://doi.org/10.1063/1.451421} {\bibfield  {journal} {\bibinfo  {journal}
			{J. Chem. Phys.}\ }\textbf {\bibinfo {volume} {85}},\ \bibinfo {pages} {6837}
		(\bibinfo {year} {1986})}\BibitemShut {NoStop}%
	\bibitem [{\citenamefont {Baum}\ \emph {et~al.}(1985)\citenamefont {Baum},
		\citenamefont {Munowitz}, \citenamefont {Garroway},\ and\ \citenamefont
		{Pines}}]{Baum1985c}%
	\BibitemOpen
	\bibfield  {author} {\bibinfo {author} {\bibfnamefont {J.}~\bibnamefont
			{Baum}}, \bibinfo {author} {\bibfnamefont {M.}~\bibnamefont {Munowitz}},
		\bibinfo {author} {\bibfnamefont {A.~N.}\ \bibnamefont {Garroway}},\ and\
		\bibinfo {author} {\bibfnamefont {A.}~\bibnamefont {Pines}},\ }\bibfield
	{title} {\bibinfo {title} {{Multiple-quantum dynamics in solid state NMR}},\
	}\href {https://doi.org/10.1063/1.449344} {\bibfield  {journal} {\bibinfo
			{journal} {J. Chem. Phys.}\ }\textbf {\bibinfo {volume} {83}},\ \bibinfo
		{pages} {2015} (\bibinfo {year} {1985})}\BibitemShut {NoStop}%
	\bibitem [{\citenamefont {Munowitz}\ \emph {et~al.}(1987)\citenamefont
		{Munowitz}, \citenamefont {Pines},\ and\ \citenamefont
		{Mehring}}]{Munowitz1987}%
	\BibitemOpen
	\bibfield  {author} {\bibinfo {author} {\bibfnamefont {M.}~\bibnamefont
			{Munowitz}}, \bibinfo {author} {\bibfnamefont {A.}~\bibnamefont {Pines}},\
		and\ \bibinfo {author} {\bibfnamefont {M.}~\bibnamefont {Mehring}},\
	}\bibfield  {title} {\bibinfo {title} {{Multiple-quantum dynamics in NMR: A
				directed walk through Liouville space}},\ }\href
	{https://doi.org/10.1063/1.452028} {\bibfield  {journal} {\bibinfo  {journal}
			{J. Chem. Phys.}\ }\textbf {\bibinfo {volume} {86}},\ \bibinfo {pages} {3172}
		(\bibinfo {year} {1987})}\BibitemShut {NoStop}%
	\bibitem [{\citenamefont {Lacelle}(1991)}]{Lacelle1991}%
	\BibitemOpen
	\bibfield  {author} {\bibinfo {author} {\bibfnamefont {S.}~\bibnamefont
			{Lacelle}},\ }\bibinfo {title} {{On the Growth of Multiple Spin Coherences in
			NMR of Solids}},\ in\ \href
	{https://doi.org/10.1016/B978-0-12-025516-0.50007-3} {\emph {\bibinfo
			{booktitle} {Adv. Magn. Opt. Reson.}}},\ Vol.~\bibinfo {volume} {16}\
	(\bibinfo  {publisher} {Elsevier},\ \bibinfo {year} {1991})\ pp.\ \bibinfo
	{pages} {173--263}\BibitemShut {NoStop}%
	\bibitem [{\citenamefont {Lacelle}\ \emph {et~al.}(1993)\citenamefont
		{Lacelle}, \citenamefont {Hwang},\ and\ \citenamefont
		{Gerstein}}]{Lacelle1993a}%
	\BibitemOpen
	\bibfield  {author} {\bibinfo {author} {\bibfnamefont {S.}~\bibnamefont
			{Lacelle}}, \bibinfo {author} {\bibfnamefont {S.~J.}\ \bibnamefont {Hwang}},\
		and\ \bibinfo {author} {\bibfnamefont {B.~C.}\ \bibnamefont {Gerstein}},\
	}\bibfield  {title} {\bibinfo {title} {{Multiple quantum nuclear magnetic
				resonance of solids: A cautionary note for data analysis and
				interpretation}},\ }\href {https://doi.org/10.1063/1.465616} {\bibfield
		{journal} {\bibinfo  {journal} {J. Chem. Phys.}\ }\textbf {\bibinfo {volume}
			{99}},\ \bibinfo {pages} {8407} (\bibinfo {year} {1993})}\BibitemShut
	{NoStop}%
	\bibitem [{\citenamefont {Hughes}(2004)}]{Hughes2004}%
	\BibitemOpen
	\bibfield  {author} {\bibinfo {author} {\bibfnamefont {C.~E.}\ \bibnamefont
			{Hughes}},\ }\bibfield  {title} {\bibinfo {title} {{Spin counting}},\ }\href
	{https://doi.org/10.1016/j.pnmrs.2004.08.002} {\bibfield  {journal} {\bibinfo
			{journal} {Prog. Nucl. Magn. Reson. Spectrosc.}\ }\textbf {\bibinfo {volume}
			{45}},\ \bibinfo {pages} {301} (\bibinfo {year} {2004})}\BibitemShut
	{NoStop}%
	\bibitem [{\citenamefont {Murdoch}\ \emph {et~al.}(1984)\citenamefont
		{Murdoch}, \citenamefont {Warren}, \citenamefont {Weitekamp},\ and\
		\citenamefont {Pines}}]{Murdoch1984}%
	\BibitemOpen
	\bibfield  {author} {\bibinfo {author} {\bibfnamefont {J.~B.}\ \bibnamefont
			{Murdoch}}, \bibinfo {author} {\bibfnamefont {W.~S.}\ \bibnamefont {Warren}},
		\bibinfo {author} {\bibfnamefont {D.~P.}\ \bibnamefont {Weitekamp}},\ and\
		\bibinfo {author} {\bibfnamefont {A.}~\bibnamefont {Pines}},\ }\bibfield
	{title} {\bibinfo {title} {{Computer simulations of multiple-quantum NMR
				experiments. I. Nonselective excitation}},\ }\href
	{https://doi.org/10.1016/0022-2364(84)90327-5} {\bibfield  {journal}
		{\bibinfo  {journal} {J. Magn. Reson.}\ }\textbf {\bibinfo {volume} {60}},\
		\bibinfo {pages} {205} (\bibinfo {year} {1984})}\BibitemShut {NoStop}%
	\bibitem [{\citenamefont {Levy}\ and\ \citenamefont
		{Gleason}(1992)}]{Levy1992c}%
	\BibitemOpen
	\bibfield  {author} {\bibinfo {author} {\bibfnamefont {D.}~\bibnamefont
			{Levy}}\ and\ \bibinfo {author} {\bibfnamefont {K.}~\bibnamefont {Gleason}},\
	}\bibfield  {title} {\bibinfo {title} {{Multiple quantum nuclear magnetic
				resonance as a probe for the dimensionality of hydrogen in polycrystalline
				powders and diamond films}},\ }\href {https://doi.org/10.1021/j100199a056}
	{\bibfield  {journal} {\bibinfo  {journal} {J. Phys. Chem.}\ }\textbf
		{\bibinfo {volume} {96}},\ \bibinfo {pages} {8125} (\bibinfo {year}
		{1992})}\BibitemShut {NoStop}%
	\bibitem [{\citenamefont {Zobov}\ and\ \citenamefont
		{Lundin}(2006)}]{Zobov2006}%
	\BibitemOpen
	\bibfield  {author} {\bibinfo {author} {\bibfnamefont {V.~E.}\ \bibnamefont
			{Zobov}}\ and\ \bibinfo {author} {\bibfnamefont {A.~A.}\ \bibnamefont
			{Lundin}},\ }\bibfield  {title} {\bibinfo {title} {{Second moment of
				multiple-quantum NMR and a time-dependent growth of the number of multispin
				correlations in solids}},\ }\href {https://doi.org/10.1134/S1063776106120089}
	{\bibfield  {journal} {\bibinfo  {journal} {J. Exp. Theor. Phys.}\ }\textbf
		{\bibinfo {volume} {103}},\ \bibinfo {pages} {904} (\bibinfo {year}
		{2006})}\BibitemShut {NoStop}%
	\bibitem [{\citenamefont {Zobov}\ and\ \citenamefont
		{Lundin}(2008)}]{Zobov2008}%
	\BibitemOpen
	\bibfield  {author} {\bibinfo {author} {\bibfnamefont {V.~E.}\ \bibnamefont
			{Zobov}}\ and\ \bibinfo {author} {\bibfnamefont {A.~A.}\ \bibnamefont
			{Lundin}},\ }\bibfield  {title} {\bibinfo {title} {{On the second moment of
				the multiquantum NMR spectrum of a solid}},\ }\href
	{https://doi.org/10.1134/S1990793108050035} {\bibfield  {journal} {\bibinfo
			{journal} {Russ. J. Phys. Chem. B}\ }\textbf {\bibinfo {volume} {2}},\
		\bibinfo {pages} {676} (\bibinfo {year} {2008})}\BibitemShut {NoStop}%
	\bibitem [{\citenamefont {Mogami}\ \emph {et~al.}(2013)\citenamefont {Mogami},
		\citenamefont {Noda}, \citenamefont {Ishikawa},\ and\ \citenamefont
		{Takegoshi}}]{Mogami2013a}%
	\BibitemOpen
	\bibfield  {author} {\bibinfo {author} {\bibfnamefont {Y.}~\bibnamefont
			{Mogami}}, \bibinfo {author} {\bibfnamefont {Y.}~\bibnamefont {Noda}},
		\bibinfo {author} {\bibfnamefont {H.}~\bibnamefont {Ishikawa}},\ and\
		\bibinfo {author} {\bibfnamefont {K.}~\bibnamefont {Takegoshi}},\ }\bibfield
	{title} {\bibinfo {title} {{A statistical approach for analyzing the
				development of 1H multiple-quantum coherence in solids}},\ }\href
	{https://doi.org/10.1039/c3cp43778g} {\bibfield  {journal} {\bibinfo
			{journal} {Phys. Chem. Chem. Phys.}\ }\textbf {\bibinfo {volume} {15}},\
		\bibinfo {pages} {7403} (\bibinfo {year} {2013})}\BibitemShut {NoStop}%
	\bibitem [{\citenamefont {Zobov}\ and\ \citenamefont
		{Lundin}(2011)}]{Zobov2011}%
	\BibitemOpen
	\bibfield  {author} {\bibinfo {author} {\bibfnamefont {V.~E.}\ \bibnamefont
			{Zobov}}\ and\ \bibinfo {author} {\bibfnamefont {A.~A.}\ \bibnamefont
			{Lundin}},\ }\bibfield  {title} {\bibinfo {title} {Decay of multispin
			multiquantum coherent states in the nmr of a solid},\ }\href
	{https://doi.org/10.1134/S1063776111020129} {\bibfield  {journal} {\bibinfo
			{journal} {J. Exp. Theor. Phys.}\ }\textbf {\bibinfo {volume} {112}},\
		\bibinfo {pages} {451} (\bibinfo {year} {2011})}\BibitemShut {NoStop}%
	\bibitem [{\citenamefont {Lundin}\ and\ \citenamefont
		{Zobov}(2016)}]{Lundin2016}%
	\BibitemOpen
	\bibfield  {author} {\bibinfo {author} {\bibfnamefont {A.~A.}\ \bibnamefont
			{Lundin}}\ and\ \bibinfo {author} {\bibfnamefont {V.~E.}\ \bibnamefont
			{Zobov}},\ }\bibfield  {title} {\bibinfo {title} {{Decoherence-Induced
				Stabilization of the Multiple-Quantum NMR-Spectrum Width}},\ }\href
	{https://doi.org/10.1007/s00723-016-0770-z} {\bibfield  {journal} {\bibinfo
			{journal} {Appl. Magn. Reson.}\ }\textbf {\bibinfo {volume} {47}},\ \bibinfo
		{pages} {701} (\bibinfo {year} {2016})}\BibitemShut {NoStop}%
	\bibitem [{\citenamefont {Cho}\ and\ \citenamefont
		{Yesinowski}(1993)}]{Cho1993}%
	\BibitemOpen
	\bibfield  {author} {\bibinfo {author} {\bibfnamefont {G.}~\bibnamefont
			{Cho}}\ and\ \bibinfo {author} {\bibfnamefont {J.~P.}\ \bibnamefont
			{Yesinowski}},\ }\bibfield  {title} {\bibinfo {title} {{Multiple-quantum NMR
				dynamics in the quasi-one-dimensional distribution of protons in
				hydroxyapatite}},\ }\href {https://doi.org/10.1016/0009-2614(93)85157-J}
	{\bibfield  {journal} {\bibinfo  {journal} {Chem. Phys. Lett.}\ }\textbf
		{\bibinfo {volume} {205}},\ \bibinfo {pages} {1} (\bibinfo {year}
		{1993})}\BibitemShut {NoStop}%
	\bibitem [{\citenamefont {Cho}\ and\ \citenamefont
		{Yesinowski}(1996)}]{Cho1996}%
	\BibitemOpen
	\bibfield  {author} {\bibinfo {author} {\bibfnamefont {G.}~\bibnamefont
			{Cho}}\ and\ \bibinfo {author} {\bibfnamefont {J.~P.}\ \bibnamefont
			{Yesinowski}},\ }\bibfield  {title} {\bibinfo {title} {{1H and19F
				Multiple-quantum NMR dynamics in quasi-one-dimensional spin clusters in
				apatites}},\ }\href {https://doi.org/10.1021/jp9614815} {\bibfield  {journal}
		{\bibinfo  {journal} {J. Phys. Chem.}\ }\textbf {\bibinfo {volume} {100}},\
		\bibinfo {pages} {15716} (\bibinfo {year} {1996})}\BibitemShut {NoStop}%
	\bibitem [{\citenamefont {Cho}\ \emph {et~al.}(2005)\citenamefont {Cho},
		\citenamefont {Ladd}, \citenamefont {Baugh}, \citenamefont {Cory},\ and\
		\citenamefont {Ramanathan}}]{Cho2005}%
	\BibitemOpen
	\bibfield  {author} {\bibinfo {author} {\bibfnamefont {H.}~\bibnamefont
			{Cho}}, \bibinfo {author} {\bibfnamefont {T.~D.}\ \bibnamefont {Ladd}},
		\bibinfo {author} {\bibfnamefont {J.}~\bibnamefont {Baugh}}, \bibinfo
		{author} {\bibfnamefont {D.~G.}\ \bibnamefont {Cory}},\ and\ \bibinfo
		{author} {\bibfnamefont {C.}~\bibnamefont {Ramanathan}},\ }\bibfield  {title}
	{\bibinfo {title} {{Multispin dynamics of the solid-state NMR free induction
				decay}},\ }\href {https://doi.org/10.1103/PhysRevB.72.054427} {\bibfield
		{journal} {\bibinfo  {journal} {Phys. Rev. B}\ }\textbf {\bibinfo {volume}
			{72}},\ \bibinfo {pages} {054427} (\bibinfo {year} {2005})}\BibitemShut
	{NoStop}%
	\bibitem [{\citenamefont {{\'{A}}lvarez}\ and\ \citenamefont
		{Suter}(2011)}]{Alvarez2011}%
	\BibitemOpen
	\bibfield  {author} {\bibinfo {author} {\bibfnamefont {G.~A.}\ \bibnamefont
			{{\'{A}}lvarez}}\ and\ \bibinfo {author} {\bibfnamefont {D.}~\bibnamefont
			{Suter}},\ }\bibfield  {title} {\bibinfo {title} {{Localization effects
				induced by decoherence in superpositions of many-spin quantum states}},\
	}\href {https://doi.org/10.1103/PhysRevA.84.012320} {\bibfield  {journal}
		{\bibinfo  {journal} {Phys. Rev. A}\ }\textbf {\bibinfo {volume} {84}},\
		\bibinfo {pages} {012320} (\bibinfo {year} {2011})}\BibitemShut {NoStop}%
	\bibitem [{\citenamefont {Slichter}(1990)}]{slichter2013principles}%
	\BibitemOpen
	\bibfield  {author} {\bibinfo {author} {\bibfnamefont {C.~P.}\ \bibnamefont
			{Slichter}},\ }\href {https://doi.org/10.1007/978-3-662-09441-9} {\emph
		{\bibinfo {title} {{Principles of magnetic resonance}}}}\ (\bibinfo
	{publisher} {Springer-Verlag Berlin Heidelberg},\ \bibinfo {year}
	{1990})\BibitemShut {NoStop}%
	\bibitem [{\citenamefont {Haeberlen}\ and\ \citenamefont
		{Waugh}(1968)}]{Haeberlen1968}%
	\BibitemOpen
	\bibfield  {author} {\bibinfo {author} {\bibfnamefont {U.}~\bibnamefont
			{Haeberlen}}\ and\ \bibinfo {author} {\bibfnamefont {J.~S.}\ \bibnamefont
			{Waugh}},\ }\bibfield  {title} {\bibinfo {title} {Coherent averaging effects
			in magnetic resonance},\ }\href {https://doi.org/10.1103/PhysRev.175.453}
	{\bibfield  {journal} {\bibinfo  {journal} {Phys. Rev.}\ }\textbf {\bibinfo
			{volume} {175}},\ \bibinfo {pages} {453} (\bibinfo {year}
		{1968})}\BibitemShut {NoStop}%
	\bibitem [{\citenamefont {G{\"{a}}rttner}\ \emph {et~al.}(2018)\citenamefont
		{G{\"{a}}rttner}, \citenamefont {Hauke},\ and\ \citenamefont
		{Rey}}]{Garttner2018}%
	\BibitemOpen
	\bibfield  {author} {\bibinfo {author} {\bibfnamefont {M.}~\bibnamefont
			{G{\"{a}}rttner}}, \bibinfo {author} {\bibfnamefont {P.}~\bibnamefont
			{Hauke}},\ and\ \bibinfo {author} {\bibfnamefont {A.~M.}\ \bibnamefont
			{Rey}},\ }\bibfield  {title} {\bibinfo {title} {{Relating Out-of-Time-Order
				Correlations to Entanglement via Multiple-Quantum Coherences}},\ }\href
	{https://doi.org/10.1103/PhysRevLett.120.040402} {\bibfield  {journal}
		{\bibinfo  {journal} {Phys. Rev. Lett.}\ }\textbf {\bibinfo {volume} {120}},\
		\bibinfo {pages} {040402} (\bibinfo {year} {2018})}\BibitemShut {NoStop}%
	\bibitem [{\citenamefont {Yan}\ \emph {et~al.}(2020)\citenamefont {Yan},
		\citenamefont {Cincio},\ and\ \citenamefont {Zurek}}]{Yan2020}%
	\BibitemOpen
	\bibfield  {author} {\bibinfo {author} {\bibfnamefont {B.}~\bibnamefont
			{Yan}}, \bibinfo {author} {\bibfnamefont {L.}~\bibnamefont {Cincio}},\ and\
		\bibinfo {author} {\bibfnamefont {W.~H.}\ \bibnamefont {Zurek}},\ }\bibfield
	{title} {\bibinfo {title} {{Information Scrambling and Loschmidt Echo}},\
	}\href {https://doi.org/10.1103/PhysRevLett.124.160603} {\bibfield  {journal}
		{\bibinfo  {journal} {Phys. Rev. Lett.}\ }\textbf {\bibinfo {volume} {124}},\
		\bibinfo {pages} {160603} (\bibinfo {year} {2020})}\BibitemShut {NoStop}%
	\bibitem [{\citenamefont {Khitrin}(1997)}]{Khitrin1997}%
	\BibitemOpen
	\bibfield  {author} {\bibinfo {author} {\bibfnamefont {A.~K.}\ \bibnamefont
			{Khitrin}},\ }\bibfield  {title} {\bibinfo {title} {{Growth of NMR
				multiple-quantum coherences in quasi-one-dimensional systems}},\ }\href
	{https://doi.org/10.1016/S0009-2614(97)00661-1} {\bibfield  {journal}
		{\bibinfo  {journal} {Chem. Phys. Lett.}\ }\textbf {\bibinfo {volume}
			{274}},\ \bibinfo {pages} {217} (\bibinfo {year} {1997})}\BibitemShut
	{NoStop}%
	\bibitem [{\citenamefont {Munowitz}\ and\ \citenamefont
		{Pines}(1986)}]{Munowitz86_principles}%
	\BibitemOpen
	\bibfield  {author} {\bibinfo {author} {\bibfnamefont {M.}~\bibnamefont
			{Munowitz}}\ and\ \bibinfo {author} {\bibfnamefont {A.}~\bibnamefont
			{Pines}},\ }\bibinfo {title} {Principles and applications of multiple-quantum
		{NMR}},\ in\ \href
	{https://doi.org/https://doi.org/10.1002/9780470142929.ch1} {\emph {\bibinfo
			{booktitle} {Adv. Chem. Phys}}},\ Vol.~\bibinfo {volume} {66}\ (\bibinfo
	{publisher} {John Wiley {\&} Sons, Ltd},\ \bibinfo {year} {1986})\ pp.\
	\bibinfo {pages} {1--152}\BibitemShut {NoStop}%
	\bibitem [{\citenamefont {\'Alvarez}\ \emph {et~al.}(2008)\citenamefont
		{\'Alvarez}, \citenamefont {Danieli}, \citenamefont {Levstein},\ and\
		\citenamefont {Pastawski}}]{Alvarez2008}%
	\BibitemOpen
	\bibfield  {author} {\bibinfo {author} {\bibfnamefont {G.~A.}\ \bibnamefont
			{\'Alvarez}}, \bibinfo {author} {\bibfnamefont {E.~P.}\ \bibnamefont
			{Danieli}}, \bibinfo {author} {\bibfnamefont {P.~R.}\ \bibnamefont
			{Levstein}},\ and\ \bibinfo {author} {\bibfnamefont {H.~M.}\ \bibnamefont
			{Pastawski}},\ }\bibfield  {title} {\bibinfo {title} {Quantum {Parallelism}
			as a {Tool} for {Ensemble} {Spin} {Dynamics} {Calculations}},\ }\href
	{https://doi.org/10.1103/PhysRevLett.101.120503} {\bibfield  {journal}
		{\bibinfo  {journal} {Phys. Rev. Lett.}\ }\textbf {\bibinfo {volume} {101}},\
		\bibinfo {pages} {120503} (\bibinfo {year} {2008})}\BibitemShut {NoStop}%
	\bibitem [{\citenamefont {Scruggs}\ and\ \citenamefont
		{Gleason}(1992)}]{Scruggs1992b}%
	\BibitemOpen
	\bibfield  {author} {\bibinfo {author} {\bibfnamefont {B.~E.}\ \bibnamefont
			{Scruggs}}\ and\ \bibinfo {author} {\bibfnamefont {K.}~\bibnamefont
			{Gleason}},\ }\bibfield  {title} {\bibinfo {title} {{Computer-simulation of
				the multiple-quantum dynamics of one- , two- and three-dimensional spin
				distributions}},\ }\href {https://doi.org/10.1016/0301-0104(92)80096-e}
	{\bibfield  {journal} {\bibinfo  {journal} {Chem. Phys.}\ }\textbf {\bibinfo
			{volume} {166}},\ \bibinfo {pages} {367} (\bibinfo {year}
		{1992})}\BibitemShut {NoStop}%
	\bibitem [{\citenamefont {Doronin}\ \emph {et~al.}(2001)\citenamefont
		{Doronin}, \citenamefont {Fel{'}dman}, \citenamefont {Guinzbourg},\ and\
		\citenamefont {Maximov}}]{Doronin2001}%
	\BibitemOpen
	\bibfield  {author} {\bibinfo {author} {\bibfnamefont {S.~I.}\ \bibnamefont
			{Doronin}}, \bibinfo {author} {\bibfnamefont {E.~B.}\ \bibnamefont
			{Fel{'}dman}}, \bibinfo {author} {\bibfnamefont {I.~Y.}\ \bibnamefont
			{Guinzbourg}},\ and\ \bibinfo {author} {\bibfnamefont {I.~I.}\ \bibnamefont
			{Maximov}},\ }\bibfield  {title} {\bibinfo {title} {{Supercomputer analysis
				of one-dimensional multiple-quantum dynamics of nuclear spins in solids}},\
	}\href {https://doi.org/10.1016/s0009-2614(01)00472-9} {\bibfield  {journal}
		{\bibinfo  {journal} {Chem. Phys. Lett.}\ }\textbf {\bibinfo {volume}
			{341}},\ \bibinfo {pages} {144} (\bibinfo {year} {2001})}\BibitemShut
	{NoStop}%
	\bibitem [{\citenamefont {Munowitz}(2006)}]{Munowitz2006}%
	\BibitemOpen
	\bibfield  {author} {\bibinfo {author} {\bibfnamefont {M.}~\bibnamefont
			{Munowitz}},\ }\bibfield  {title} {\bibinfo {title} {{Exact simulation of
				multiple-quantum dynamics in solid-state NMR: implications for spin
				counting}},\ }\href {https://doi.org/10.1080/00268979000102261} {\bibfield
		{journal} {\bibinfo  {journal} {Mol. Phys.}\ }\textbf {\bibinfo {volume}
			{71}},\ \bibinfo {pages} {37} (\bibinfo {year} {2006})}\BibitemShut {NoStop}%
	\bibitem [{\citenamefont {Doronin}\ \emph {et~al.}(2009)\citenamefont
		{Doronin}, \citenamefont {Fedorova},\ and\ \citenamefont
		{Zenchuk}}]{Doronin2009}%
	\BibitemOpen
	\bibfield  {author} {\bibinfo {author} {\bibfnamefont {S.~I.}\ \bibnamefont
			{Doronin}}, \bibinfo {author} {\bibfnamefont {A.~V.}\ \bibnamefont
			{Fedorova}},\ and\ \bibinfo {author} {\bibfnamefont {A.~I.}\ \bibnamefont
			{Zenchuk}},\ }\bibfield  {title} {\bibinfo {title} {{Multiple quantum NMR
				dynamics of spin- 2 carrying molecules of a gas in nanopores}},\ }\href
	{https://doi.org/10.1063/1.3231692} {\bibfield  {journal} {\bibinfo
			{journal} {J. Chem. Phys.}\ }\textbf {\bibinfo {volume} {131}},\ \bibinfo
		{pages} {104109} (\bibinfo {year} {2009})}\BibitemShut {NoStop}%
	\bibitem [{\citenamefont {Doronin}\ \emph {et~al.}(2011)\citenamefont
		{Doronin}, \citenamefont {Fel{'}dman},\ and\ \citenamefont
		{Zenchuk}}]{Doronin2011}%
	\BibitemOpen
	\bibfield  {author} {\bibinfo {author} {\bibfnamefont {S.~I.}\ \bibnamefont
			{Doronin}}, \bibinfo {author} {\bibfnamefont {E.~B.}\ \bibnamefont
			{Fel{'}dman}},\ and\ \bibinfo {author} {\bibfnamefont {A.~I.}\ \bibnamefont
			{Zenchuk}},\ }\bibfield  {title} {\bibinfo {title} {{Numerical analysis of
				relaxation times of multiple quantum coherences in the system with a large
				number of spins}},\ }\href {https://doi.org/10.1063/1.3528040} {\bibfield
		{journal} {\bibinfo  {journal} {J. Chem. Phys.}\ }\textbf {\bibinfo {volume}
			{134}},\ \bibinfo {pages} {034102} (\bibinfo {year} {2011})}\BibitemShut
	{NoStop}%
	\bibitem [{\citenamefont {De~Raedt}\ and\ \citenamefont
		{Michielsen}(2004)}]{de2004computational}%
	\BibitemOpen
	\bibfield  {author} {\bibinfo {author} {\bibfnamefont {H.}~\bibnamefont
			{De~Raedt}}\ and\ \bibinfo {author} {\bibfnamefont {K.}~\bibnamefont
			{Michielsen}},\ }\href@noop {} {\bibinfo {title} {Computational methods for
			simulating quantum computers}} (\bibinfo {year} {2004}),\ \Eprint
	{https://arxiv.org/abs/quant-ph/0406210} {arXiv:quant-ph/0406210}
	\BibitemShut {NoStop}%
	\bibitem [{\citenamefont {Zhang}\ \emph {et~al.}(2007)\citenamefont {Zhang},
		\citenamefont {Konstantinidis}, \citenamefont {Al-Hassanieh},\ and\
		\citenamefont {Dobrovitski}}]{Zhang_2007}%
	\BibitemOpen
	\bibfield  {author} {\bibinfo {author} {\bibfnamefont {W.}~\bibnamefont
			{Zhang}}, \bibinfo {author} {\bibfnamefont {N.}~\bibnamefont
			{Konstantinidis}}, \bibinfo {author} {\bibfnamefont {K.~A.}\ \bibnamefont
			{Al-Hassanieh}},\ and\ \bibinfo {author} {\bibfnamefont {V.~V.}\ \bibnamefont
			{Dobrovitski}},\ }\bibfield  {title} {\bibinfo {title} {Modelling decoherence
			in quantum spin systems},\ }\href
	{https://doi.org/10.1088/0953-8984/19/8/083202} {\bibfield  {journal}
		{\bibinfo  {journal} {J. Condens. Matter Phys.}\ }\textbf {\bibinfo {volume}
			{19}},\ \bibinfo {pages} {083202} (\bibinfo {year} {2007})}\BibitemShut
	{NoStop}%
	\bibitem [{\citenamefont {Palma}\ \emph {et~al.}(1997)\citenamefont {Palma},
		\citenamefont {Suominen},\ and\ \citenamefont {Ekert}}]{Palma1997a}%
	\BibitemOpen
	\bibfield  {author} {\bibinfo {author} {\bibfnamefont {G.~M.}\ \bibnamefont
			{Palma}}, \bibinfo {author} {\bibfnamefont {K.-A.}\ \bibnamefont
			{Suominen}},\ and\ \bibinfo {author} {\bibfnamefont {A.~K.}\ \bibnamefont
			{Ekert}},\ }\bibfield  {title} {\bibinfo {title} {{Quantum Computers and
				Dissipation}},\ }\href {https://doi.org/10.1098/rspa.1996.0029} {\bibfield
		{journal} {\bibinfo  {journal} {Proc. Math. Phys. Eng. Sci.}\ }\textbf
		{\bibinfo {volume} {452}},\ \bibinfo {pages} {20} (\bibinfo {year}
		{1997})}\BibitemShut {NoStop}%
	\bibitem [{\citenamefont {Duan}\ and\ \citenamefont {Guo}(1998)}]{Duan1998}%
	\BibitemOpen
	\bibfield  {author} {\bibinfo {author} {\bibfnamefont {L.~M.}\ \bibnamefont
			{Duan}}\ and\ \bibinfo {author} {\bibfnamefont {G.~C.}\ \bibnamefont {Guo}},\
	}\bibfield  {title} {\bibinfo {title} {{Reducing decoherence in
				quantum-computer memory with all quantum bits coupling to the same
				environment}},\ }\href {https://doi.org/10.1103/PhysRevA.57.737} {\bibfield
		{journal} {\bibinfo  {journal} {Phys. Rev. A}\ }\textbf {\bibinfo {volume}
			{57}},\ \bibinfo {pages} {737} (\bibinfo {year} {1998})}\BibitemShut
	{NoStop}%
	\bibitem [{\citenamefont {Reina}\ \emph {et~al.}(2002)\citenamefont {Reina},
		\citenamefont {Quiroga},\ and\ \citenamefont {Johnson}}]{Reina2002}%
	\BibitemOpen
	\bibfield  {author} {\bibinfo {author} {\bibfnamefont {J.~H.}\ \bibnamefont
			{Reina}}, \bibinfo {author} {\bibfnamefont {L.}~\bibnamefont {Quiroga}},\
		and\ \bibinfo {author} {\bibfnamefont {N.~F.}\ \bibnamefont {Johnson}},\
	}\bibfield  {title} {\bibinfo {title} {{Decoherence of quantum registers}},\
	}\href {https://doi.org/10.1103/PhysRevA.65.032326} {\bibfield  {journal}
		{\bibinfo  {journal} {Phys. Rev. A}\ }\textbf {\bibinfo {volume} {65}},\
		\bibinfo {pages} {032326} (\bibinfo {year} {2002})}\BibitemShut {NoStop}%
	\bibitem [{\citenamefont {Jing}\ and\ \citenamefont {Hu}(2015)}]{Jing2015}%
	\BibitemOpen
	\bibfield  {author} {\bibinfo {author} {\bibfnamefont {J.}~\bibnamefont
			{Jing}}\ and\ \bibinfo {author} {\bibfnamefont {X.}~\bibnamefont {Hu}},\
	}\bibfield  {title} {\bibinfo {title} {{Scaling of decoherence for a system
				of uncoupled spin qubits}},\ }\href {https://doi.org/10.1038/srep17013}
	{\bibfield  {journal} {\bibinfo  {journal} {Sci. Rep.}\ }\textbf {\bibinfo
			{volume} {5}},\ \bibinfo {pages} {17013} (\bibinfo {year}
		{2015})}\BibitemShut {NoStop}%
	\bibitem [{\citenamefont {{\'{A}}lvarez}\ \emph {et~al.}(2013)\citenamefont
		{{\'{A}}lvarez}, \citenamefont {Kaiser},\ and\ \citenamefont
		{Suter}}]{Alvarez2013}%
	\BibitemOpen
	\bibfield  {author} {\bibinfo {author} {\bibfnamefont {G.~A.}\ \bibnamefont
			{{\'{A}}lvarez}}, \bibinfo {author} {\bibfnamefont {R.}~\bibnamefont
			{Kaiser}},\ and\ \bibinfo {author} {\bibfnamefont {D.}~\bibnamefont
			{Suter}},\ }\bibfield  {title} {\bibinfo {title} {{Quantum simulations of
				localization effects with dipolar interactions}},\ }\href
	{https://doi.org/10.1002/andp.201300096} {\bibfield  {journal} {\bibinfo
			{journal} {Ann. Phys.}\ }\textbf {\bibinfo {volume} {525}},\ \bibinfo {pages}
		{833} (\bibinfo {year} {2013})}\BibitemShut {NoStop}%
	\bibitem [{\citenamefont {{\'{A}}lvarez}\ \emph {et~al.}(2006)\citenamefont
		{{\'{A}}lvarez}, \citenamefont {Danieli}, \citenamefont {Levstein},\ and\
		\citenamefont {Pastawski}}]{Alvarez2006}%
	\BibitemOpen
	\bibfield  {author} {\bibinfo {author} {\bibfnamefont {G.~A.}\ \bibnamefont
			{{\'{A}}lvarez}}, \bibinfo {author} {\bibfnamefont {E.~P.}\ \bibnamefont
			{Danieli}}, \bibinfo {author} {\bibfnamefont {P.~R.}\ \bibnamefont
			{Levstein}},\ and\ \bibinfo {author} {\bibfnamefont {H.~M.}\ \bibnamefont
			{Pastawski}},\ }\bibfield  {title} {\bibinfo {title} {{Environmentally
				induced quantum dynamical phase transition in the spin swapping operation}},\
	}\href {https://doi.org/10.1063/1.2193518} {\bibfield  {journal} {\bibinfo
			{journal} {J. Chem. Phys.}\ }\textbf {\bibinfo {volume} {124}},\ \bibinfo
		{pages} {194507} (\bibinfo {year} {2006})}\BibitemShut {NoStop}%
	\bibitem [{\citenamefont {Danieli}\ \emph {et~al.}(2007)\citenamefont
		{Danieli}, \citenamefont {\'Alvarez}, \citenamefont {Levstein},\ and\
		\citenamefont {Pastawski}}]{Danieli2007}%
	\BibitemOpen
	\bibfield  {author} {\bibinfo {author} {\bibfnamefont {E.}~\bibnamefont
			{Danieli}}, \bibinfo {author} {\bibfnamefont {G.}~\bibnamefont {\'Alvarez}},
		\bibinfo {author} {\bibfnamefont {P.}~\bibnamefont {Levstein}},\ and\
		\bibinfo {author} {\bibfnamefont {H.}~\bibnamefont {Pastawski}},\ }\bibfield
	{title} {\bibinfo {title} {Quantum dynamical phase transition in a system
			with many-body interactions},\ }\href
	{https://doi.org/https://doi.org/10.1016/j.ssc.2006.11.001} {\bibfield
		{journal} {\bibinfo  {journal} {Solid State Commun.}\ }\textbf {\bibinfo
			{volume} {141}},\ \bibinfo {pages} {422} (\bibinfo {year}
		{2007})}\BibitemShut {NoStop}%
	\bibitem [{\citenamefont {Rotter}(2009)}]{Rotter2009}%
	\BibitemOpen
	\bibfield  {author} {\bibinfo {author} {\bibfnamefont {I.}~\bibnamefont
			{Rotter}},\ }\bibfield  {title} {\bibinfo {title} {{A non-hermitian hamilton
				operator and the physics of open quantum systems}},\ }\href
	{https://doi.org/10.1088/1751-8113/42/15/153001} {\bibfield  {journal}
		{\bibinfo  {journal} {J. Phys. A Math. Theor.}\ }\textbf {\bibinfo {volume}
			{42}},\ \bibinfo {pages} {153001} (\bibinfo {year} {2009})}\BibitemShut
	{NoStop}%
	\bibitem [{\citenamefont {Rotter}(2010)}]{Rotter2010}%
	\BibitemOpen
	\bibfield  {author} {\bibinfo {author} {\bibfnamefont {I.}~\bibnamefont
			{Rotter}},\ }\bibfield  {title} {\bibinfo {title} {{Environmentally induced
				effects and dynamical phase transitions in quantum systems}},\ }\href
	{https://doi.org/10.1088/2040-8978/12/6/065701} {\bibfield  {journal}
		{\bibinfo  {journal} {J. Opt.}\ }\textbf {\bibinfo {volume} {12}},\ \bibinfo
		{pages} {065701} (\bibinfo {year} {2010})}\BibitemShut {NoStop}%
	\bibitem [{\citenamefont {Rotter}\ and\ \citenamefont
		{Bird}(2015)}]{Rotter2015}%
	\BibitemOpen
	\bibfield  {author} {\bibinfo {author} {\bibfnamefont {I.}~\bibnamefont
			{Rotter}}\ and\ \bibinfo {author} {\bibfnamefont {J.~P.}\ \bibnamefont
			{Bird}},\ }\bibfield  {title} {\bibinfo {title} {{A review of progress in the
				physics of open quantum systems: Theory and experiment}},\ }\href
	{https://doi.org/10.1088/0034-4885/78/11/114001} {\bibfield  {journal}
		{\bibinfo  {journal} {Rep. Prog. Phys.}\ }\textbf {\bibinfo {volume} {78}},\
		\bibinfo {pages} {114001} (\bibinfo {year} {2015})}\BibitemShut {NoStop}%
	\bibitem [{\citenamefont {Martinez~Alvarez}\ \emph {et~al.}(2018)\citenamefont
		{Martinez~Alvarez}, \citenamefont {Barrios~Vargas},\ and\ \citenamefont
		{Foa~Torres}}]{Martinez2018}%
	\BibitemOpen
	\bibfield  {author} {\bibinfo {author} {\bibfnamefont {V.~M.}\ \bibnamefont
			{Martinez~Alvarez}}, \bibinfo {author} {\bibfnamefont {J.~E.}\ \bibnamefont
			{Barrios~Vargas}},\ and\ \bibinfo {author} {\bibfnamefont {L.~E.~F.}\
			\bibnamefont {Foa~Torres}},\ }\bibfield  {title} {\bibinfo {title}
		{Non-hermitian robust edge states in one dimension: Anomalous localization
			and eigenspace condensation at exceptional points},\ }\href
	{https://doi.org/10.1103/PhysRevB.97.121401} {\bibfield  {journal} {\bibinfo
			{journal} {Phys. Rev. B}\ }\textbf {\bibinfo {volume} {97}},\ \bibinfo
		{pages} {121401} (\bibinfo {year} {2018})}\BibitemShut {NoStop}%
	\bibitem [{\citenamefont {Metzler}\ and\ \citenamefont
		{Klafter}(2000)}]{METZLER20001}%
	\BibitemOpen
	\bibfield  {author} {\bibinfo {author} {\bibfnamefont {R.}~\bibnamefont
			{Metzler}}\ and\ \bibinfo {author} {\bibfnamefont {J.}~\bibnamefont
			{Klafter}},\ }\bibfield  {title} {\bibinfo {title} {The random walk's guide
			to anomalous diffusion: a fractional dynamics approach},\ }\href
	{https://doi.org/https://doi.org/10.1016/S0370-1573(00)00070-3} {\bibfield
		{journal} {\bibinfo  {journal} {Phys. Rep.}\ }\textbf {\bibinfo {volume}
			{339}},\ \bibinfo {pages} {1} (\bibinfo {year} {2000})}\BibitemShut {NoStop}%
	\bibitem [{\citenamefont {Mercadier}\ \emph {et~al.}(2009)\citenamefont
		{Mercadier}, \citenamefont {Guerin}, \citenamefont {Chevrollier},\ and\
		\citenamefont {Kaiser}}]{mercadier2009levy}%
	\BibitemOpen
	\bibfield  {author} {\bibinfo {author} {\bibfnamefont {N.}~\bibnamefont
			{Mercadier}}, \bibinfo {author} {\bibfnamefont {W.}~\bibnamefont {Guerin}},
		\bibinfo {author} {\bibfnamefont {M.}~\bibnamefont {Chevrollier}},\ and\
		\bibinfo {author} {\bibfnamefont {R.}~\bibnamefont {Kaiser}},\ }\bibfield
	{title} {\bibinfo {title} {L{\'e}vy flights of photons in hot atomic
			vapours},\ }\href {https://doi.org/10.1038/nphys1286} {\bibfield  {journal}
		{\bibinfo  {journal} {Nat. Phys}\ }\textbf {\bibinfo {volume} {5}},\ \bibinfo
		{pages} {602} (\bibinfo {year} {2009})}\BibitemShut {NoStop}%
	\bibitem [{\citenamefont {Maziero}\ \emph {et~al.}(2009)\citenamefont
		{Maziero}, \citenamefont {C\'eleri}, \citenamefont {Serra},\ and\
		\citenamefont {Vedral}}]{Maziero2009_classical}%
	\BibitemOpen
	\bibfield  {author} {\bibinfo {author} {\bibfnamefont {J.}~\bibnamefont
			{Maziero}}, \bibinfo {author} {\bibfnamefont {L.~C.}\ \bibnamefont
			{C\'eleri}}, \bibinfo {author} {\bibfnamefont {R.~M.}\ \bibnamefont
			{Serra}},\ and\ \bibinfo {author} {\bibfnamefont {V.}~\bibnamefont
			{Vedral}},\ }\bibfield  {title} {\bibinfo {title} {Classical and quantum
			correlations under decoherence},\ }\href
	{https://doi.org/10.1103/PhysRevA.80.044102} {\bibfield  {journal} {\bibinfo
			{journal} {Phys. Rev. A}\ }\textbf {\bibinfo {volume} {80}},\ \bibinfo
		{pages} {044102} (\bibinfo {year} {2009})}\BibitemShut {NoStop}%
	\bibitem [{\citenamefont {Xu}\ \emph {et~al.}(2010)\citenamefont {Xu},
		\citenamefont {Xu}, \citenamefont {Li}, \citenamefont {Zhang}, \citenamefont
		{Zou},\ and\ \citenamefont {Guo}}]{Xu2010_experimental}%
	\BibitemOpen
	\bibfield  {author} {\bibinfo {author} {\bibfnamefont {J.-S.}\ \bibnamefont
			{Xu}}, \bibinfo {author} {\bibfnamefont {X.-Y.}\ \bibnamefont {Xu}}, \bibinfo
		{author} {\bibfnamefont {C.-F.}\ \bibnamefont {Li}}, \bibinfo {author}
		{\bibfnamefont {C.-J.}\ \bibnamefont {Zhang}}, \bibinfo {author}
		{\bibfnamefont {X.-B.}\ \bibnamefont {Zou}},\ and\ \bibinfo {author}
		{\bibfnamefont {G.-C.}\ \bibnamefont {Guo}},\ }\bibfield  {title} {\bibinfo
		{title} {Experimental investigation of classical and quantum correlations
			under decoherence},\ }\href {https://doi.org/10.1038/ncomms1005} {\bibfield
		{journal} {\bibinfo  {journal} {Nat. Commun.}\ }\textbf {\bibinfo {volume}
			{1}},\ \bibinfo {pages} {7} (\bibinfo {year} {2010})}\BibitemShut {NoStop}%
	\bibitem [{\citenamefont {Touil}\ and\ \citenamefont
		{Deffner}(2021)}]{Touil2021_information}%
	\BibitemOpen
	\bibfield  {author} {\bibinfo {author} {\bibfnamefont {A.}~\bibnamefont
			{Touil}}\ and\ \bibinfo {author} {\bibfnamefont {S.}~\bibnamefont
			{Deffner}},\ }\bibfield  {title} {\bibinfo {title} {Information scrambling
			versus decoherence---two competing sinks for entropy},\ }\href
	{https://doi.org/10.1103/PRXQuantum.2.010306} {\bibfield  {journal} {\bibinfo
			{journal} {PRX Quantum}\ }\textbf {\bibinfo {volume} {2}},\ \bibinfo {pages}
		{010306} (\bibinfo {year} {2021})}\BibitemShut {NoStop}%
	\bibitem [{\citenamefont {Xu}\ \emph {et~al.}(2021)\citenamefont {Xu},
		\citenamefont {Chenu}, \citenamefont {Prosen},\ and\ \citenamefont {del
			Campo}}]{Xu2021_thermofield}%
	\BibitemOpen
	\bibfield  {author} {\bibinfo {author} {\bibfnamefont {Z.}~\bibnamefont
			{Xu}}, \bibinfo {author} {\bibfnamefont {A.}~\bibnamefont {Chenu}}, \bibinfo
		{author} {\bibfnamefont {T.~c.~v.}\ \bibnamefont {Prosen}},\ and\ \bibinfo
		{author} {\bibfnamefont {A.}~\bibnamefont {del Campo}},\ }\bibfield  {title}
	{\bibinfo {title} {Thermofield dynamics: Quantum chaos versus decoherence},\
	}\href {https://doi.org/10.1103/PhysRevB.103.064309} {\bibfield  {journal}
		{\bibinfo  {journal} {Phys. Rev. B}\ }\textbf {\bibinfo {volume} {103}},\
		\bibinfo {pages} {064309} (\bibinfo {year} {2021})}\BibitemShut {NoStop}%
	\bibitem [{\citenamefont {Styliaris}\ \emph {et~al.}(2021)\citenamefont
		{Styliaris}, \citenamefont {Anand},\ and\ \citenamefont
		{Zanardi}}]{Styliaris2021_information}%
	\BibitemOpen
	\bibfield  {author} {\bibinfo {author} {\bibfnamefont {G.}~\bibnamefont
			{Styliaris}}, \bibinfo {author} {\bibfnamefont {N.}~\bibnamefont {Anand}},\
		and\ \bibinfo {author} {\bibfnamefont {P.}~\bibnamefont {Zanardi}},\
	}\bibfield  {title} {\bibinfo {title} {Information scrambling over
			bipartitions: Equilibration, entropy production, and typicality},\ }\href
	{https://doi.org/10.1103/PhysRevLett.126.030601} {\bibfield  {journal}
		{\bibinfo  {journal} {Phys. Rev. Lett.}\ }\textbf {\bibinfo {volume} {126}},\
		\bibinfo {pages} {030601} (\bibinfo {year} {2021})}\BibitemShut {NoStop}%
\end{thebibliography}
\end{document}